\documentclass[12pt]{iopart}

\usepackage{threeparttable}    
\usepackage{dcolumn}           
\newcolumntype{d}{D{.}{.}{-1}}
\usepackage{nomencl}           
\makenomenclature
\usepackage{graphicx}          
\usepackage{caption}          
\usepackage{fancyvrb}          
\fvset{fontsize=\footnotesize,xleftmargin=2em}  
\usepackage{lettrine}          
\expandafter\let\csname equation*\endcsname\relax
\expandafter\let\csname endequation*\endcsname\relax
\usepackage{amsmath}           
\usepackage{hyperref}          
\graphicspath{{figs/}}         
\usepackage{float}             
\usepackage{upquote}
\usepackage{placeins}
\usepackage[retainorgcmds]{IEEEtrantools}
\usepackage{subfig}
\usepackage{siunitx}
\usepackage[export]{adjustbox}
\usepackage[title]{appendix}
\newcommand{\be}{\begin{equation}}
\newcommand{\ee}{\end{equation}} 

\usepackage{cite}
\usepackage[nameinlink,capitalise]{cleveref}
\usepackage{tikz}
\usetikzlibrary{positioning,shadows,arrows,shapes.multipart,decorations.pathmorphing}

\newcommand{\colormarkerempty}[1]{\raisebox{0.5pt}{\protect\tikz{\protect\node[draw,scale=0.3,circle,fill=white](){};}}}

\begin{document}

\title[]{Multi-physics modeling of non-equilibrium phenomena in inductively coupled plasma discharges:  Part I. A state-to-state approach }

\author{Sanjeev Kumar, Alessandro Munaf\`{o}, Sung Min Jo, and Marco Panesi\footnote{\label{footnote_2}Corresponding author (mpanesi@illinois.edu).}  }

\address{Center for Hypersonics and Entry Systems Studies (CHESS), \\
Department of Aerospace Engineering, \\
University of Illinois at Urbana-Champaign, Urbana, IL 61801, USA}
\vspace{10pt}



\begin{abstract}
This work presents a vibrational and electronic state-to-state model for nitrogen plasma implemented within a multi-physics modular computational framework to study non-equilibrium effects in inductively coupled plasma (ICP) discharges. Within the computational framework, the set of vibronic (i.e., vibrational and electronic) master equations are solved in a tightly coupled fashion with the flow governing equations. This tight coupling eliminates the need for invoking any simplifying assumptions when computing the state of the plasma, thereby ensuring a higher degree of physical fidelity. To mitigate computational complexity, a maximum entropy coarse-graining strategy is deployed, effectively truncating the internal state space. The efficacy of this reduced StS model is empirically substantiated through zero-dimensional isochoric simulations. In these simulations, the results obtained from the reduced-order model are rigorously compared against those obtained from the full StS model, thereby confirming the accuracy of the reduced StS framework. The developed Coarse-grained StS model was employed to study the plasma discharge within the VKI Plasmatron facility. Our results reveal pronounced discrepancies between the plasma flow fields obtained from StS simulations and those derived from Local Thermodynamic Equilibrium (LTE) models, which are conventionally used in the simulation of such facilities. The analysis demonstrates a substantial departure of the internal state populations of atoms and molecules from the Boltzmann distribution. These non-equilibrium effects have important consequences on the energy coupling dynamics, thereby impacting the overall morphology of the plasma discharge. A deeper analysis of the results demonstrates that the population distribution is in a Quasi-Steady-State in the hot plasma core. This insight allows for the determination of macroscopic global rates, offering a computationally efficient avenue for the construction of multi-temperature models. The implementation and ramifications of these models constitute the subject matter for Part II of this ongoing research series.

\end{abstract}

\section{Introduction}\label{sec:intro}
The extreme heat loads experienced by the thermal protection shield of atmospheric re-entry vehicles (\emph{e.g.,} space shuttle, re-entry capsules) are often reproduced in the ground testing facilities by placing a sample of thermal protection material (TPM) in a hot jet of plasma. Historically, two different types of plasma-wind tunnels have been developed: the arc-jet facilities in which the test gas runs between electrodes and gets heated by the Ohmic dissipation of the powerful electric current running between the electrodes, and the inductively coupled plasma (ICP) facilities where the plasma is generated in an electrodeless manner using electro-magnetic induction\cite{boulos1985inductively}. One of the major advantages of ICP facilities is that there are no metallic impurities being eroded from the electrodes and polluting the plasma. Moreover, the ease and longer duration of operation and its versatility in terms of sizes have made ICPs widely used not only for testing of TPS materials but also for other industrial applications like plasma spray processes\cite{fauchais2004understanding,meillot2015numerical}, synthesis of nano-particles\cite{shigeta2011thermal}, possible electric propulsion devices for very low earth orbit satellites\cite{zheng2023atmosphere}, \emph{etc.}. 

Numerical simulation of ICPs requires physico-chemical modeling of the plasma flow field and electromagnetic phenomena inside the torch by solving the coupled system of Navier-Stokes and Maxwell equations. The earliest attempts to model the ICP torches were published in the 1960-1970s\cite{freeman1968energy,keefer1973electrodeless,eckert1970analysis,eckert1970analytical,eckert1972analysis,eckert1977two}, where the torch was approximated as an infinite solenoid and the plasma was assumed to be in Local Thermodynamic Equilibrium (LTE) conditions. These assumptions simplified the problem to the coupled solution of the energy equation for the gas and an induction equation for the electric field. The developments in the field of Computational Fluid Dynamics (CFD) led to the possibility of solving the magneto-hydrodynamic equations in multi-dimensional configurations\cite{Boulos_1976,mostaghimi1984parametric,mostaghimi1985analysis,mostaghimi1987two,proulx1987heating,mostaghimi1990effect,chen1991modeling,panesi2007analysis,abeele2000efficient,utyuzhnikov2004simulation}. Majority of the simulations available in the literature, however, assume LTE conditions for the plasma inside the ICP torch, which holds good for relatively high-pressure values (\emph{e.g.,} $\approx$ $10^4$ Pa and above) at which the collisional frequency between the gas particles is large enough to maintain local equilibrium. This assumption greatly reduces the computational cost as the thermodynamic and transport properties can be tabulated as a function of two independent state variables. However, recent non-LTE simulations have shown that the use of the LTE assumption does not always hold, especially at lower pressures\cite{kumar2022high,kumar2022self,munafoRGD2022,zhang2016analysis,mostaghimi1987two,mostaghimi1990effect,munafo2015tightly}. 

The most physically consistent approach to model NLTE plasmas is the state-to-state (StS)\cite{laux2012state,bultel2013elaboration,munafo2015tightly,panesi2013rovibrational,colonna2015non,heritier2014energy,laporta2013electron,laporta2016electron,capitelli2013plasma,bultel2002influence,magin2006nonequilibrium, capitelli2007non,pietanza2010kinetic,munafo2012qct,munafo2013modeling,venturi2020bayesian,priyadarshini2022comprehensive,esposito1999quasiclassical,kustova2014chemical} approach where each internal energy state is treated as a separate pseudo-species, thus allowing for non-Boltzmann distributions. However, the StS approach is computationally expensive and requires the knowledge of a large number of reliable elementary rate coefficients which might not be available for complex gas mixtures. To overcome these issues, reduced order models have been developed over the years to model non-equilibrium flows where the population of internal levels of chemical species is assumed to be in Maxwell-Boltzmann distribution\cite{park1989nonequilibrium,park1993review,gnoffo1989conservation}. This assumption drastically reduces the number of equations to be solved making the computation practical. One of the most widely used non-equilibrium models for ICP discharges is the two temperature (2-T) model\cite{park1989nonequilibrium,park1993review,park2001chemical} which further assumes fast equilibrium between the heavy-species translational and the rotational energy modes, while the electronic and vibrational energy modes are assumed to be in equilibrium with the translational energy of the free-electrons. Simulations of ICP discharges using the two-temperature NLTE model have been recently reported in several literatures\cite{mostaghimi1987two,Most_1989,atsuchi2006modeling,panesi2007analysis,kumar2022high,munafo2022multi,munafoRGD2022} revealing the non-equilibrium effect in the plasma and its impact on the plasma thermal and flow field. However, as per the authors' knowledge, there is hardly any literature presenting a comparison of the 2-T simulations with experiments to conclude how accurately the 2-T NLTE models predict the plasma flow field inside the ICPs and how accurate is the assumption of the Boltzmann distribution for the internal state populations of the chemical species. Recent simulations of hypersonic flows using electronic state-to-state models \cite{jo2019electronic,jo2020stagnation,johnston2018impact,panesi2009fire,panesi2011electronic} (where the electronic energy levels are considered as separate pseudo-species) have shown significant departure of the electronic energy levels from Boltzmann distribution and results obtained using the electronic StS simulations show better agreements with the experiments. Similar attempts have been reported for ICP discharges also, where electronic StS simulations show large deviation of the electronic levels from the Boltzmann distribution, especially in regions near the cold walls which are dominated by recombination and regions where sudden change in plasma conditions occur\cite{munafo2015tightly,Kumar_RGD32}. Moreover, departures from Boltzmann distribution in the vibrational levels were also observed in recombining nitrogen plasma experiments conducted at Stanford University\cite{gessman1997experimental} using an ICP facility. The experiment was further modeled using a vibrational state-specific collisional radiative (CR) model\cite{laux2012state} \emph{i.e.} the CR model was used to compute and compare the populations of the vibrational levels at the point of interest by making the quasi-steady-state (QSS) assumption. These observations motivate the objective of this paper, which is to study the non-equilibrium effects in an ICP torch using a vibronic state-to-state model, where the set of vibronic master equations is solved in a fully coupled manner with the flow governing equations without resorting to any simplifying assumptions.

As mentioned before, the computationally expensive nature of the StS models makes it almost impossible to perform CFD calculations. Hence, to make the StS problem tractable, a common approach used in the hypersonics community is to use coarse-graining, which reduces the number of internal energy levels by appropriately grouping the levels based on either simpler strategies like energy-based binning, or more advanced and accurate adaptive-binning strategies which leverage the details of the state-specific chemical kinetics to group the energy levels\cite{johnston2018impact,sahai2017adaptive,macdonald2018construction,macdonald2018construction2,munafo2014boltzmann,liu2015general,panesi2013collisional,venturi2020data,sahai2019flow,magin2012coarse,sharma2020coarse,kosareva2021four,zanardi2023adaptive,kuppa2023uncertainty}. The coarse-grain approach is based on the idea of grouping the individual states into a smaller number of macroscopic bins. A bin-wise distribution function based on macroscopic quantities (bin population, energy, etc.) and the maximum entropy principle is used to reconstruct the population of the grouped levels. The use of the bin-wise distribution function allows the temporal evolution of the state populations to be modeled accurately without needing to solve the master equations for the individual levels. This reduces the number of levels and the number of reactions in the StS model drastically, making it possible to conduct CFD calculations at a relatively cheaper cost without compromising accuracy. Hence, this work aims to develop a coarse-grained model from a state-of-the-art vibronic StS model for nitrogen gas and integrate it with the multi-physics computational framework for studying non-equilibrium plasma dynamics inside ICP discharges. To the authors' knowledge, this work is the first instance of modeling ICP discharges using a vibronic StS model fully coupled with the flow governing equations. 

The paper is organized as follows: \cref{sec:physical_modeling} discusses the mathematical model used in the computational framework to describe the plasma inside the ICP facility. This section also describes the vibronic StS model for nitrogen gas used in this work, along with the binning strategy to reduce the model. \cref{sec:numerical_method} describes the implementation of the plasma model in the CFD and the electromagnetic (EM) solver, along with the coupling strategy for the solvers for performing magneto-hydrodynamic (MHD) simulations. Results are presented in \cref{sec:results} which presents the verification of the binning strategy followed by the vibronic StS results for an ICP torch. Further, LTE results have also been presented and compared against the vibronic StS results to highlight the discrepancies in the results due to LTE assumption for the given conditions. Finally, the conclusions are summarized in \cref{sec:conclusions}.

\section{Physical Modeling}\label{sec:physical_modeling}
A complete description of a non-equilibrium plasma flow field inside an inductively coupled plasma (ICP) requires modeling of the flow and the electromagnetic field. This sub-section describes the mathematical models used for the above-mentioned aspects of ICP modeling.
\subsection{\label{sec:plasma Field}Plasma Field}
The plasmas treated in this work are modeled under the following assumptions:
\begin{enumerate}
    \item The gas is a collection of neutral and charged components/species, each behaving as an ideal gas. $\mathcal{S}$ = \{$\mathrm{e}^-$, $\mathrm{N}$, $\mathrm{N}_2$, $\mathrm{N}_2^+$, $\mathrm{N}^+$\}.
    \item The plasma is quasi-neutral (since the Debye length scale is much smaller than the flow length scale) as well as collision-dominated such that the use of a hydrodynamics description is appropriate \cite{boulos1994thermal}. 
\end{enumerate}  
Under the above assumptions, the non-equilibrium plasma hydrodynamics are governed by the set of mass continuity, global momentum and energy, and free-electron energy equations \cite{Munafo_JCP_2020,Mitchner_book,gnoffo1989conservation,capitelli2013fundamantal,magin2004transport}:
\begin{align}
&\frac{\partial \rho_{s}}{\partial t}+ {\nabla}_{\mathbf{r}}  \cdot\left[\rho_{s}\left(\mathbf{v}+ \mathbf{U}_s \right)\right] = \dot{\omega}_{s}, \quad s \in \mathcal{S}, \label{eq:cont} \\
&\frac{\partial \rho \mathbf{v}}{\partial t}+ {\nabla}_{\mathbf{r}} \cdot(\rho \mathbf{v} \mathbf{v} + p \mathsf{I}) =  {\nabla}_{\mathbf{r}} \cdot \mathsf{\tau} + \mathbf{J} \times \mathbf{B}, \label{eq:momentum}\\
 &\frac{\partial \rho E}{\partial t}+ {\nabla}_{\mathbf{r}} \cdot(\rho H \mathbf{v}) = {\nabla}_{\mathbf{r}} \cdot \left( \mathsf{\tau} \mathbf{v} - \mathbf{q}\right) + \mathbf{J} \cdot \mathbf{E^{\prime}}, \label{eq:global_E}\\
&\frac{\partial \rho e_{\mathrm{e}}}{\partial t}+ {\nabla}_{\mathbf{r}} \cdot\left(\rho e_{\mathrm{e}} \mathbf{v} \right)  =  - {\nabla}_{\mathbf{r}} \cdot \mathbf{q}_{\mathrm{e}}  -p_{\mathrm{e}} {\nabla}_{\mathbf{r}} \cdot \mathbf{v} + \Omega_{\mathrm{e}}^{\textsc{c}} + \mathbf{J} \cdot \mathbf{E^{\prime}},\label{eq:ve_eq}
\end{align}

where $\mathcal{S}$ denotes the set of species, and the e lower-script denotes the contributions from \emph{free-electrons}. The various symbols in the governing equations \cref{eq:cont,eq:momentum,eq:global_E,eq:ve_eq} have their usual meaning: $t$ denotes time, $\mathbf{r}$ the position; $\rho$ and $\mathbf{v}$ the mass density and mass-averaged velocity, respectively; $\rho_s$ and $\mathbf{U}_s$ the partial density and diffusion velocity of species $s$; $p_{\mathrm{e}}$ the pressure of free-electrons; $e$ and $H$ the total energy and enthalpy per unit-mass, respectively; $\mathsf{\tau}$ the stress tensor; $\mathbf{q}$ the heat flux vector; $\dot{\omega}_s$ the mass production rates due to collisional processes; the $\Omega^{\textsc{c}}$ term the energy exchange terms due to collisional processes; $\mathbf{J}$ the conduction current density; $\mathbf{E}$ and $\mathbf{B}$ the electric field and the magnetic induction, respectively; $\mathbf{E^{\prime}} = \mathbf{E} + \mathbf{v} \times \mathbf{B}$ the electric field in the hydrodynamic frame (non-relativistic approximation). 

For vibronic state-to-state simulations, the vibronic-state resolved master equations are solved in a coupled manner with the flow governing equations. Hence,  the index $s$ in \cref{eq:cont} denotes all the internal states which are considered state-to-state \emph{i.e.} the continuity equation is solved for all the vibronic states. The rotational modes are assumed to be in equilibrium with the translational modes and a rigid-rotor approximation is used to compute the corresponding internal energies and partition functions. This assumption generally holds as electrons do not excite the rotational modes efficiently.

Under LTE conditions with no elemental de-mixing, the governing equations reduce to the global mass, momentum, and energy equations (\emph{i.e.}, Navier-Stokes) where the effects of the \emph{fast} chemical reactions are incorporated via the definition of state-equations (\emph{e.g.} $p = p(\rho, \, T)$), thermodynamic and transport properties.

\subsubsection{Kinetics}
The mass production terms in the continuity equations are expressed as $\dot{\omega}_s=\sum_{r=1}^{N r} \dot{\omega}_{s r}$ with \begin{equation}
\dot{\omega}_{s r}=\mathcal{M}_s\left(\nu_{s r}^{\prime \prime}-\nu_{s r}^{\prime}\right)\left\{k_{f r} \prod_{j \in \mathcal{S}}\left(\frac{\rho_j}{M_j}\right)^{\nu_{j r}^{\prime r}}-k_{b r} \prod_{j \in \mathcal{S}}\left(\frac{\rho_j}{M_j}\right)^{\nu_{j r}^{\prime \prime}}\right\}
\end{equation}
where $\dot{\omega}_{s r}$ is the mass production term of species s due to the
$\mathrm{r}^\mathrm{th}$ elementary reaction and $\dot{\omega}_s=\left(d \rho_s / d t\right)_{c h e m}$ is the global mass production term for species s. $k_{f r}$ and $k_{b r}$ are the
the forward and the backward reaction rate constants, respectively for the $\mathrm{r}^\mathrm{th}$ reaction.

The source term $\Omega_{\mathrm{e}}^{\textsc{c}}$ in the free-electron energy equation is given by:
\begin{equation}
\Omega_{\mathrm{e}}^{\textsc{c}}=\Omega^{\mathrm{TE}}+\Omega^{\mathrm{DE}}+\Omega^{\mathrm{IE}} 
\end{equation}
\\
$\Omega^{\mathrm{TE}}$ denotes the elastic energy exchange in collisions between free-electrons and heavy particles and is evaluated using kinetic theory\cite{petschek1957approach}:
\begin{equation}
\Omega^{\mathrm{TE}}=\frac{3}{2} n_{\mathrm{e}} k_{\mathrm{B}} \frac{\left(T_{\mathrm{h}}-T_{\mathrm{e}}\right)}{\tau_{\mathrm{eh}}^{\mathrm{TE}}}
\end{equation}
where the elastic energy transfer relaxation time is given by
\begin{equation}
\frac{1}{\tau_{\mathrm{eh}}^{\mathrm{TE}}}=\frac{8}{3} \sqrt{\frac{8 k_{\mathrm{B}} T_{\mathrm{e}}}{\pi m_{\mathrm{e}}}} \sum_{s \in \mathcal{S}_{\mathrm{h}}}\left(\frac{m_{\mathrm{e}}}{m_s}\right) n_s \bar{Q}_{\mathrm{e} s}^{(1,1)}
\end{equation}
where $\bar{Q}_{\mathrm{e} s}^{(1,1)}$ are the reduced collision integrals for electron-heavy interactions (refer \cite{Munafo_JCP_2020}).

$\Omega^{\mathrm{DE}}$ and $\Omega^{\mathrm{IE}}$ terms represent the volumetric energy loss of free-electrons due to electron-impact dissociation and ionization reactions respectively. Disregarding the effects of excitation and assuming that secondary electrons have zero energy, the terms can be given as:
\begin{equation}
\Omega^{\mathrm{DE}}=\sum_{r \in \mathcal{R}^{\mathrm{DE}}} \Delta E_r\left(\frac{\omega_{\mathrm{e}}^r}{m_{\mathrm{e}}}\right) \quad \text { and } \quad \Omega^{\mathrm{IE}}=\sum_{r \in \mathcal{R}^{\mathrm{IE}}} \Delta E_r\left(\frac{\omega_{\mathrm{e}}^r}{m_{\mathrm{e}}}\right)
\end{equation}
where $\mathcal{R}^{\mathrm{DE}}$ and $\mathcal{R}^{\mathrm{IE}}$ represent the dissociation and ionization by electron impact reactions, respectively and $\omega_{\mathrm{e}}^r$ denotes the free-electron mass production term for reaction $r$. $\Delta E_r$ corresponds to the ground-state dissociation/ionization energy of the reactant heavy-particle.

\subsubsection{Thermodynamic and transport properties}
Under the assumptions introduced above, the gas pressure is given by Dalton's law, $p = p_h + p_e$, where the heavy-particle and free-electron pressures are given as $p_h = n_h k_B T_h$ and $p_e = n_e k_B T_e$, respectively. The mixture internal energy is given by $e=\sum_{j \in \mathrm{S}} \chi_j e_j$, with the mass fraction given as $\chi_i=\rho_i / \rho$. The species internal energy $e_i$ consists of the translational and formation contributions $\left[e_e=e_e^T\left(T_{\mathrm{e}}\right)+e_e^F\right]$ for electrons; the translational and formation contributions $\left[e_i=e_i^T\left(T\right)+e_i^F\right]$ for the atoms; and the translational, rotational, and formation contributions $\left[e_i=e_i^T(T)+e_i^R(T)+e_i^F\right]$ for all the molecules, where $T_{\mathrm{e}}$ is the free-electron temperature. The internal energy has been defined in this way because the electronic and vibrational states are treated state-to-state. Gurvich tables \cite{hildenbrand1995thermodynamic} give the formation energies of neutral and the characteristic vibrational and rotational temperatures of molecules.

Transport properties are computed using the modified Chapman-Enskog perturbative analysis for partially ionized plasmas\cite{hirschfelder1954molecular,ferziger1973mathematical,bruno2010transport}, neglecting the effect of inelastic and reactive collisions on transport properties, and assuming that the collision cross-sections for elastic scattering do not depend on the internal quantum states. A detailed explanation of the computation of transport properties can be found in Ref. \cite{Munafo_JCP_2020}. The diffusion fluxes have been computed solving the Stefan-Maxwell system of equations\cite{hirschfelder1954molecular,ferziger1973mathematical,magin2004transport}, which consists of a linear system (in the diffusion fluxes) of as many equations as the chemical species in the mixture. The system is further supplemented by the auxiliary condition that the sum of the diffusion fluxes is zero plus the ambipolar constraint. 

For LTE simulations, the thermodynamic and transport properties may be evaluated as a function of two state variables, say pressure $p$, and temperature $T$, or density $\rho$ and temperature $T$. Here the $(\rho, \, T)$ pair is chosen which translates into the following functional dependencies in the mechanical and thermal equations of state:
\be \label{eq:eos}
e = e(\rho,\, T) \quad \text{and} \quad p = \rho R T,
\ee
where $R = R(\rho, \, T)$ is the specific gas constant. Similar relations may be written for the relevant transport properties, which are the dynamic/shear viscosity, $\eta = \eta (\rho,\, T)$, and the (total) thermal conductivity, $\lambda = \lambda (\rho, \, T)$.  
The direct evaluation of the \cref{eq:eos} during a calculation is expensive since these often involve solving a non-linear set of equations. To circumvent this issue, two-dimensional LTE look-up tables are generated and loaded when starting a simulation. The evaluation of quantities of interest (\emph{e.g.}, energy, pressure) at non-grid locations may be achieved based on several interpolation techniques (\emph{e.g.}, spline, Lagrange). Here bi-linear interpolation is used since this approach, as shown in the study by Rinaldi \emph{et al.} \cite{rinaldi2014exact}, provides the best trade-off between accuracy and computational cost. More details about the computation of LTE properties can be found in \cite{Munafo2022}.

\subsection{\label{sec:EM field} Electromagnetic Field}
Electromagnetic phenomena are all governed by Maxwell's equations \cite{Mitchner_book}. To reduce the complexity of the model, the following assumptions are hereby introduced \cite{david_thesis}:
\begin{enumerate}
    \item Low frequency approximation ($f / f_p \ll 1$): the inductor frequency ($f$) is very small compared to the plasma frequency ($f_p$). This allows us to rule out electrostatic and electromagnetic waves.
    \item Low magnetic Reynolds number ($R_m=\mu_0 \sigma u L \ll 1$): for high-pressure ICPs ($>$ \SI{0.01}{atm}) considered in this work, the magnetic Reynolds number ($R_m$) is very small.
    \item Ambipolar diffusion (\emph{i.e.}, no current in the poloidal plane).
    \item Plasma is unmagnetized: this assumption is valid for the pressures ($>$ \SI{0.01}{atm}) considered in this work.
\end{enumerate}

For unmagnetized plasmas, the conduction current density is given by the generalized Ohm's law \cite{Mitchner_book}:
\begin{equation}\label{eq:Ohm}
\mathbf{J} = \sigma \left(\mathbf{E} + \mathbf{v} \times \mathbf{B}  - \dfrac{{\nabla}_{\mathbf{r}} p_{\mathrm{e}}}{n_{\mathrm{e}} q_{\mathrm{e}}} \right),    
\end{equation}
where $\sigma$ is the plasma scalar electrical conductivity, whereas $n_{\mathrm{e}}$ and $q_{\mathrm{e}}$ stand, respectively, for the number density of free-electrons and the electron charge. The electric field may be further decomposed as the sum of an electrostatic irrotational component and a divergence-free induced part: $\mathbf{E} = \mathbf{E}_s + \mathbf{E}_i$. In light of the ambipolar diffusion assumption in the poloidal plane, the electrostatic contribution reads $\mathbf{E}_s = {\nabla}_{\mathbf{r}} p_{\mathrm{e}}  /n_{\mathrm{e}} q_{\mathrm{e}}$ \cite{david_thesis}.
Using this result in Ohm's law \cref{eq:Ohm} gives:
\begin{equation}
 \mathbf{J} = \sigma \left(\mathbf{E} + \mathbf{v} \times \mathbf{B} \right),     
\end{equation}
where the electric field $\mathbf{E}$ now stands for sole the induced part. A dimensional analysis shows that the $\mathbf{v} \times \mathbf{B}$ term relates to the magnetic Reynolds number which is very small at the operating pressures ($>$ \SI{0.01}{atm}) considered in this work\cite{david_thesis}, yielding:
\begin{equation} \label{eq:Ohm_simpl}
\mathbf{J} = \sigma \mathbf{E}.    
\end{equation}
As a further consequence, the Joule heating term in the free-electron energy equation \cref{eq:ve_eq} simplifies to $\mathbf{J} \cdot \mathbf{E^{\prime}} \simeq \mathbf{J} \cdot \mathbf{E}$.

The equation governing the induced electric field is obtained assuming that, at steady-state, all quantities will undergo harmonic oscillations \cite{Most_1989}:
\begin{align} 
\mathbf{E} (\mathbf{r}, \, t)  &=  \mathbf{\tilde{E} (\mathbf{r})} \exp\left({i \omega t}\right), \label{eq:E_time_space1} \\
\mathbf{B} (\mathbf{r}, \, t)  &= \mathbf{\tilde{B} (\mathbf{r})} \exp\left({i \omega t}\right), \label{eq:E_time_space2} \\
\mathbf{J} (\mathbf{r}, \, t)  &= \mathbf{\tilde{J} (\mathbf{r})} \exp\left({i \omega t}\right) \label{eq:E_time_space3}, 
\end{align}
where the angular frequency of the current running through the inductor is $\omega = 2\pi f$, whereas $\imath = \sqrt{-1}$ denotes the imaginary unit. In the above relations amplitudes are assumed complex (\emph{e.g.}, $\mathbf{\tilde{E}} = \mathbf{\tilde{E}}_{\mathrm{R}} + \imath \mathbf{\tilde{E}}_{\mathrm{I}}$) to account for possible phase differences between electric and magnetic fields. The use of equations \cref{eq:E_time_space1,eq:E_time_space2,eq:E_time_space3} in Faraday's law leads, with the aid of Maxwell-Amp\`{e}re law with no displacement currents, to a Helmholtz-like vector equation for the electric field phasor $\mathbf{\tilde{E}}$: 
\begin{equation}\label{eq:Helm}
{\nabla}_{\mathbf{r}} \times {\nabla}_{\mathbf{r}} \times \mathbf{\tilde{E}} = - \imath \mu_0 \,   \omega \, (\mathbf{\tilde{J}} + \mathbf{\tilde{J_s}}),    
\end{equation}
where $\mu_0$ is the vacuum magnetic permeability. The current density phasor is given by $\mathbf{\tilde{J}} = \sigma \mathbf{\tilde{E}}$ in the plasma region (\emph{e.g.}, torch), whereas in the inductor coils it is prescribed as discussed later in this Section. The $\mathbf{\tilde{J_s}}$ term on the right-hand side of \cref{eq:Helm} is the current density contribution from external sources (i.e. inductor coils). Once $\mathbf{\tilde{E}}$ found, the magnetic induction may be retrieved from Faraday's law:
\begin{equation}\label{eq:Bfield}
\mathbf{\tilde{B}} = \left(\dfrac{\imath}{\omega}\right) {\nabla}_{\mathrm{r}} \times \mathbf{\tilde{E}}.
\end{equation}

In general, ICP facilities are operated at frequencies of the order of \si{\mega\hertz}. In light of this, it is reasonable to assume that, over the inductor period $1/f$, the plasma experiences a time-averaged  Lorentz force and Joule heating:
\begin{IEEEeqnarray}{rCl}\label{Lorentz_Joule_terms}
\left<\mathbf{J}\times \mathbf{B}\right>&= & \dfrac{1}{2} \left(\dfrac{\sigma}{\omega}\right) \Re{\left[
\mathbf{\tilde{E}} \times\left( \imath {\nabla}_{\mathbf{r}} \times\mathbf{ \tilde{E}}\right)^{*}\right]}, \\
\left<\mathbf{J}\cdot \mathbf{E}\right>&= &
    \frac{1}{2}\sigma\mathbf{\tilde{E}}\cdot \mathbf{\tilde{E}^*},
\end{IEEEeqnarray}
where the $*$ upperscript denotes the complex conjugate, whereas $\Re{\left[z\right]}$ indicates the real part of $z$.

For a two-dimensional axi-symmetric configuration, \cref{eq:Helm} reduces to a scalar equation for the toroidal component of the electric field phasor:
\begin{equation}
 \dfrac{\partial}{\partial r} \left(\dfrac{1}{r} \dfrac{\partial r \tilde{E}}{\partial r}  \right) + \dfrac{\partial^2 \tilde{E}}{\partial z^2} = \imath \omega \mu_0 (\tilde{J} + \tilde{J_s}),
\end{equation}
where $\tilde{E} =\tilde{E}_{\mathrm{R}} + \imath \tilde{E}_{\mathrm{I}}$, and with $r$ and $z$ being the radial and axial coordinates, respectively. The magnetic induction is always given by \cref{eq:Bfield}. The time-averaged Lorentz force and the Joule heating reduce to \cite{Most_1989}:
\begin{IEEEeqnarray}{rCl}
\left<\mathbf{J}\times \mathbf{B}\right>_z&=&-\frac{1}{2} \sigma \Re{\left[ \tilde{E} \tilde{B}^{*}_r  \right]}, \\
\left<\mathbf{J}\times \mathbf{B}\right>_r&=&\frac{1}{2} \sigma \Re{\left[ \tilde{E} \tilde{B}^{*}_z \right]}, \\
\left<\mathbf{J}\cdot \mathbf{E}\right>&=&\frac{1}{2} \sigma \tilde{E} \tilde{E}^*,
\end{IEEEeqnarray}
where $\tilde{B}_r$ and $\tilde{B}_r$ are, respectively, the radial and axial components of the magnetic induction phasor.

Following Boulos \cite{Boulos_1976}, during a simulation, the current running through the inductor coils is updated to match a prescribed value of the power ($P$) dissipated by Joule heating:
\be
P = \!\! \int \!\! \left<  \mathbf{J}\cdot \mathbf{E^{\prime}} \right> dv \simeq \!\! \int \!\! \left<  \mathbf{J}\cdot \mathbf{E} \right> dv.
\ee

\subsection{Nitrogen StS model}\label{nitrogen_cr_model}
A vibrational-specific state-to-state model for nitrogen gas has been incorporated into the current ICP framework to study non-equilibrium and non-Boltzmann effects in ICPs. The StS model has been taken from the work of \textit{Pereira} \emph{et al.} \cite{elio_thesis} where state-of-the-art microscopic kinetic rates were collected from several literatures as well as a Forced Harmonic Oscillator (FHO) model was used to determine the rates for vibrational transitions and dissociation of N\textsubscript{2} and $\mathrm{N}_2^+$ by collision with heavy-particles. Also, a vibrational redistribution procedure (VRP) was used in \cite{elio_thesis} to compute the rates for a full set of vibrational quantum numbers for cases where the rates were available only for some particular vibrational quantum numbers. Further, the rates for N\textsubscript{2}(X)-N and N\textsubscript{2}(X)-N\textsubscript{2} systems have been taken from the work of \textit{Panesi} and \textit{Macdonald} \cite{panesi2013rovibrational,Macdonald2020_N3,macdonald2018construction} where first, a potential energy surface (PES) for the system of interest is constructed from first principles and then quasi-classical trajectory (QCT) calculations are done on this PES to obtain the reaction rate coefficients. These being comparatively simple systems allow accurate construction of kinetic databases based on \textit{ab initio} quantum chemistry calculations which are much more accurate than the FHO model for a wide range of temperatures. Also, these systems play an important role in governing the plasma dynamics as will be discussed in Part II of this work, and hence getting accurate rates for the same is crucial. 

\subsubsection{Species and energy levels considered}
The species considered are N, N\textsuperscript{+}, N\textsubscript{2}, $\mathrm{N}_2^+$, and electrons. The internal structure of N atom and N\textsuperscript{+} ion consists of 131 and 81 electronic levels, respectively taken from NIST\cite{kramida2015atomic}. The X, A, B, W, B\textsuperscript{'} and C electronic states are considered for N\textsubscript{2}, while the X, A, B, D, and C electronic states are considered for $\mathrm{N}_2^+$. All the vibrational levels belonging to these electronic states are considered up to the dissociation limit, while the rotational levels have been disregarded and rigid rotor approximation was used for the molecules for vibronic StS simulations. \cref{tab:species_full} lists the energy levels considered for each component in the full vibronic StS kinetics.

\begin{table}[htb]
\centering
\caption{Components and corresponding energy levels in the full vibronic StS model}\vspace{2mm}
\begin{tabular}{lcccc}
\hline\hline
 Components & Energy levels \\
\hline
$\mathrm{N_2}$ & $\mathrm{X}^{1} \Sigma_{\mathrm{g}}^{+}([62]), \mathrm{A}^{3} \Sigma_{\mathrm{u}}^{+}([32]), \mathrm{B}^{3} \Pi_{\mathrm{g}}([33]),$\\
& $\mathrm{W}^{3} \Delta_{\mathrm{u}}([45]), \mathrm{B}^{\prime 3} \Sigma_{\mathrm{u}}^{-}([48]), \mathrm{C}^{3} \Pi_{\mathrm{u}}([5])$  \\

$\mathrm{N}_2^+$ & $\mathrm{X}^{2} \Sigma_{\mathrm{g}}^{+}([66]), \mathrm{A}^{2} \Pi_{\mathrm{u}}([67]), \mathrm{B}^{2} \Sigma_{\mathrm{u}}^{+}([39]),$\\
&$ \mathrm{D}^{2} \Pi_{\mathrm{g}}([39]), \mathrm{C}^{2} \Sigma_{\mathrm{u}}^{+}([14])$  \\

$\mathrm{N}$ & 131 levels \\
$\mathrm{N^+}$ & 81 levels \\
$\mathrm{e^-}$ & - \\
\hline\hline
\end{tabular}
\label{tab:species_full}
\end{table}

\subsubsection{Reactions considered}
The vibronic StS kinetics consists of approximately \SI{150000}{} reactions as summarized in \cref{tab:heavy_impact_reactions,tab:e_impact_reactions}. More details about the same can be found in Ref. \cite{elio_thesis}.

\begin{table}[htbp]
\centering
\caption{Heavy-impact reactions considered in the StS model}\vspace{2mm}
\footnotesize	
\begin{tabular}{p{0.35\paperwidth}p{0.25\paperwidth}p{0.075\paperwidth}}
\hline\hline
 Reaction & Remarks & Reference \\
\hline
Heavy-impact vibrational excitation\\
$\mathrm{N}_2(e, v)+\mathrm{M} \rightleftharpoons \mathrm{N}_2\left(e, v^{\prime}\right)+\mathrm{M}$ & $e \in\left\{\mathrm{X}\right\}$, $\forall v, \forall v^{\prime}>v$ and $\mathrm{M}\in\left\{\mathrm{N}_2,\mathrm{N} \right\}$ & \cite{macdonald2018construction,Macdonald2020_N3}\\
 & $e \in\left\{\mathrm{X}\right\}$, $\forall v, \forall v^{\prime}>v$ and $\mathrm{M}\in\left\{\mathrm{N}_2^{+},\mathrm{N}^{+} \right\}$ & \cite{elio_thesis}\\
 & $e \in\left\{\mathrm{A}, \mathrm{B}, \mathrm{W}, \mathrm{B}^{\prime}, \mathrm{C}\right\}$, $\forall v, \forall v^{\prime}>v$ and $\mathrm{M}\in\left\{\mathrm{N}_2,\mathrm{N}_2^{+},\mathrm{N},\mathrm{N}^{+} \right\}$ & \cite{elio_thesis}\\
$\mathrm{N}_2^{+}(e, v)+\mathrm{M} \rightleftharpoons \mathrm{N}_2^{+}\left(e, v^{\prime}\right)+\mathrm{M}$ & $e \in\left\{\mathrm{X}, \mathrm{A}, \mathrm{B}, \mathrm{D},\mathrm{C}\right\}$, $\forall v, \forall v^{\prime}>v$ and $\mathrm{M}\in\left\{\mathrm{N}_2,\mathrm{N}_2^{+},\mathrm{N},\mathrm{N}^{+} \right\}$ & \cite{elio_thesis}\\
\\

\hline
Heavy-impact vibrational-electronic excitation\\
$\mathrm{N}_2(\mathrm{~A}, v)+\mathrm{N}\left({ }^4 \mathrm{~S}_{\mathrm{u}}\right) \rightleftharpoons \mathrm{N}_2\left(\mathrm{X}, v^{\prime}\right)+\mathrm{N}\left({ }^2 \mathrm{P}_{\mathrm{u}}\right)$ & $\forall v$, and $\forall v^{\prime}$ & \cite{piper1989excitation}\\
$\mathrm{N}_2(\mathrm{~A}, v)+\mathrm{N}\left({ }^4 \mathrm{~S}_{\mathrm{u}}\right) \rightleftharpoons \mathrm{N}_2\left(\mathrm{~B}, v^{\prime}\right)+\mathrm{N}\left({ }^4 \mathrm{~S}_{\mathrm{u}}\right)$ & $\forall v$, and $\forall v^{\prime}$ & \cite{bachmann1993vibrational}\\
$\mathrm{N}_2(\mathrm{~W}, v)+\mathrm{N}\left({ }^4 \mathrm{~S}_{\mathrm{u}}\right) \rightleftharpoons \mathrm{N}_2\left(\mathrm{~B}, v^{\prime}\right)+\mathrm{N}\left({ }^4 \mathrm{~S}_{\mathrm{u}}\right)$ & $\forall v$, and $\forall v^{\prime}$ & \cite{bachmann1993vibrational}\\
$\mathrm{N}_2\left(\mathrm{~A}, v_1\right)+\mathrm{N}_2\left(\mathrm{X}, v_2\right) \rightleftharpoons \mathrm{N}_2\left(\mathrm{X}, v_1^{\prime}\right)+\mathrm{N}_2\left(\mathrm{X}, v_2^{\prime}\right)$ & $\forall v_1, \forall v_2, \forall v_1^{\prime}$ and $\forall v_2^{\prime}$ & \cite{levron1978quenching}\\
$\mathrm{N}_2\left(\mathrm{~A}, v_1\right)+\mathrm{N}_2\left(\mathrm{X}, v_2\right) \rightleftharpoons \mathrm{N}_2\left(\mathrm{~B}, v_1^{\prime}\right)+\mathrm{N}_2\left(\mathrm{X}, v_2^{\prime}\right)$ & $\forall v_1, \forall v_2, \forall v_1^{\prime}$ and $\forall v_2^{\prime}$ & \cite{bachmann1993vibrational}\\
$\mathrm{N}_2\left(\mathrm{~A}, v_1\right)+\mathrm{N}_2\left(\mathrm{~A}, v_2\right) \rightleftharpoons \mathrm{N}_2\left(\mathrm{~B}, v_1^{\prime}\right)+\mathrm{N}_2\left(\mathrm{X}, v_2^{\prime}\right)$ & $\forall v_1, \forall v_2, \forall v_1^{\prime}$ and $\forall v_2^{\prime}$ & \cite{piper1988state}\\
$\mathrm{N}_2\left(\mathrm{~A}, v_1\right)+\mathrm{N}_2\left(\mathrm{~A}, v_2\right) \rightleftharpoons \mathrm{N}_2\left(\mathrm{C}, v_1^{\prime}\right)+\mathrm{N}_2\left(\mathrm{X}, v_2^{\prime}\right)$ & $\forall v_1, \forall v_2, \forall v_1^{\prime}$ and $\forall v_2^{\prime}$ & \cite{piper1988state2}\\
$\mathrm{N}_2\left(\mathrm{~W}, v_1\right)+\mathrm{N}_2\left(\mathrm{X}, v_2\right) \rightleftharpoons \mathrm{N}_2\left(\mathrm{~B}, v_1^{\prime}\right)+\mathrm{N}_2\left(\mathrm{X}, v_2^{\prime}\right)$ & $\forall v_1, \forall v_2, \forall v_1^{\prime}$ and $\forall v_2^{\prime}$ & \cite{bachmann1993vibrational}\\
\\
\hline
Heavy-impact electronic excitation\\
$\mathrm{N}(e)+\mathrm{M} \rightleftharpoons \mathrm{N}\left(e^{\prime}\right)+\mathrm{M}$ & $\forall e, \forall e^{\prime}>e$ and $\mathrm{M} \in\left\{\mathrm{N}, \mathrm{N}_2\right\}$ & \cite{annaloro2014vibrational}\\
$\mathrm{N}^{+}(e)+\mathrm{M} \rightleftharpoons \mathrm{N}^{+}\left(e^{\prime}\right)+\mathrm{M}$ & $\forall e, \forall e^{\prime}>e$ and $\mathrm{M} \in\left\{\mathrm{N}, \mathrm{N}_2\right\}$ & \cite{annaloro2014vibrational}\\
\\
\hline
Heavy-impact dissociation\\
$\mathrm{N}_2(e, v)+\mathrm{M} \rightleftharpoons \mathrm{N}\left(e_1^{\prime}\right)+\mathrm{N}\left(e_2^{\prime}\right)+\mathrm{M}$ & $e \in\left\{\mathrm{X}\right\}$, $\forall v$ and $\mathrm{M}\in\left\{\mathrm{N}_2,\mathrm{N} \right\}$ & \cite{macdonald2018construction,Macdonald2020_N3}\\
 & $e \in\left\{\mathrm{X}\right\}$, $\forall v$ and $\mathrm{M}\in\left\{\mathrm{N}_2^{+},\mathrm{N}^{+} \right\}$ & \cite{elio_thesis}\\
& $e \in\left\{\mathrm{A}, \mathrm{B}, \mathrm{W}, \mathrm{B}^{\prime}, \mathrm{C}\right\}$, $\forall v$ and $\mathrm{M}\in\left\{\mathrm{N}_2,\mathrm{N}_2^{+},\mathrm{N},\mathrm{N}^{+} \right\}$ & \cite{elio_thesis}\\
$\mathrm{N}_2^{+}(e, v)+\mathrm{M} \rightleftharpoons \mathrm{N}\left(e_1^{\prime}\right)+\mathrm{N}^{+}\left(e_2^{\prime}\right)+\mathrm{M}$ & $e \in\left\{\mathrm{X}, \mathrm{A}, \mathrm{B}, \mathrm{D},\mathrm{C}\right\}$, $\forall v$ and $\mathrm{M}\in\left\{\mathrm{N}_2,\mathrm{N}_2^{+},\mathrm{N},\mathrm{N}^{+} \right\}$ & \cite{elio_thesis}\\
\\
\hline
Heavy-impact ionization\\
$\mathrm{N}(e)+\mathrm{M} \rightleftharpoons \mathrm{N}^{+}\left({ }^3 \mathrm{P}\right)+\mathrm{M}+\mathrm{e}^{-}$ & $\forall e$ and $\mathrm{M} \in\left\{\mathrm{N}, \mathrm{N}_2\right\}$ & \cite{annaloro2014vibrational}\\
\\
\hline
Ionization-recombination\\
$\mathrm{N}_2(\mathrm{X}, v)+\mathrm{N}^{+}\left({ }^3 \mathrm{P}\right) \rightleftharpoons \mathrm{N}_2^{+}\left(\mathrm{X}, v^{\prime}\right)+\mathrm{N}\left({ }^4 \mathrm{~S}_{\mathrm{u}}\right)$ & $\forall v$ and $\forall v^{\prime}$ & \cite{freysinger1994charge}\\

\hline\hline
\end{tabular}
\label{tab:heavy_impact_reactions}
\end{table}

\begin{table}[htbp]
\centering
\caption{Electron-impact reactions considered in the StS model}\vspace{2mm}
\footnotesize	
\begin{tabular}{p{0.35\paperwidth}p{0.25\paperwidth}p{0.075\paperwidth}}
\hline\hline
 Reaction & Remarks & Reference \\
\hline
Electron-impact vibrational excitation\\
$\mathrm{N}_2\left(\mathrm{X}^1 \Sigma_{\mathrm{g}}^{+}, v\right)+\mathrm{e}^{-} \rightleftharpoons \mathrm{N}_2\left(\mathrm{X}^1 \Sigma_{\mathrm{g}}^{+}, v^{\prime}\right)+\mathrm{e}^{-}$ & $\forall v, \forall v^{\prime}>v$ & \cite{laporta2014electron}\\
\\
\hline
Electron-impact vibrational-electronic excitation\\
$\mathrm{N}_2(\mathrm{X}, v)+\mathrm{e}^{-} \rightleftharpoons \mathrm{N}_2\left(e^{\prime}, v^{\prime}\right)+\mathrm{e}^{-}$ & $e \in\left\{\mathrm{A}, \mathrm{B}, \mathrm{W},\mathrm{B^{\prime}},\mathrm{C}\right\}$, $\forall v$, and $\forall v^{\prime}$ & \cite{brunger20036}\\
$\mathrm{N}_2^{+}(\mathrm{X}, v)+\mathrm{e}^{-} \rightleftharpoons \mathrm{N}_2^{+}\left(e^{\prime}, v^{\prime}\right)+\mathrm{e}^{-}$ & $e \in\left\{\mathrm{A}, \mathrm{B}, \mathrm{D},\mathrm{C}\right\}$, $\forall v$, and $\forall v^{\prime}$ & \cite{crandall1974absolute}\\
\\
\hline
Electron-impact electronic excitation\\
$\mathrm{N}(e)+\mathrm{e}^{-} \rightleftharpoons \mathrm{N}\left(e^{\prime}\right)+\mathrm{e}^{-}$ & $\left(e, e^{\prime}\right) \in\left\{\left({ }^4 \mathrm{~S}_{\mathrm{u}},{ }^2 \mathrm{D}_{\mathrm{u}}\right),\left({ }^4 \mathrm{~S}_{\mathrm{u}},{ }^2 \mathrm{P}_{\mathrm{u}}\right),\left({ }^2 \mathrm{D}_{\mathrm{u}},{ }^2 \mathrm{P}_{\mathrm{u}}\right)\right\}$ & \cite{berrington1975scattering}\\
& Remainder of $\left(e, e^{\prime}\right)$, with $e^{\prime}>e$ & \cite{panesi2009fire}\\
$\mathrm{N}^{+}(e)+\mathrm{e}^{-} \rightleftharpoons \mathrm{N}^{+}\left(e^{\prime}\right)+\mathrm{e}^{-}$ & $\forall e, \forall e^{\prime}>e$ & \cite{panesi2009fire}\\
\\
\hline
Electron-impact dissociation\\
$\mathrm{N}_2\left(\mathrm{X}^1 \Sigma_{\mathrm{g}}^{+}, v\right)+\mathrm{e}^{-} \rightleftharpoons \mathrm{N}\left(e_1^{\prime}\right)+\mathrm{N}\left(e_2^{\prime}\right)+\mathrm{e}^{-}$ & $\forall v,\left(e_1^{\prime}, e_2^{\prime}\right)=\left({ }^4 \mathrm{~S}_{\mathrm{u}},{ }^4 \mathrm{~S}_{\mathrm{u}}\right)$ & \cite{laporta2014electron}\\
& $\forall v,\left(e_1^{\prime}, e_2^{\prime}\right)=\left({ }^4 \mathrm{~S}_{\mathrm{u}},{ }^2 \mathrm{~D}_{\mathrm{u}}\right)$ & \cite{capitelli2001electron}\\
\\
\hline
Dissociation-recombination\\
$\mathrm{N}_2^{+}\left(\mathrm{X}^2 \Sigma_{\mathrm{g}}^{+}, v\right)+\mathrm{e}^{-} \rightleftharpoons \mathrm{N}\left(e_1^{\prime}\right)+\mathrm{N}\left(e_2^{\prime}\right)$ & - & \cite{guberman2014vibrational}\\
\\
\hline
Electron-impact ionization\\
$\mathrm{N}_2\left(\mathrm{X}^1 \Sigma_{\mathrm{g}}^{+}, v\right)+\mathrm{e}^{-} \rightleftharpoons \mathrm{N}_2^{+}\left(e^{\prime}, v^{\prime}\right)+2 \mathrm{e}^{-}$ & $\forall v$ and $e^{\prime} \in\{\mathrm{X}, \mathrm{A}, \mathrm{B}\}$ & \cite{laricchiuta2006dissociation}\\
$\mathrm{N}(e)+\mathrm{e}^{-} \rightleftharpoons \mathrm{N}^{+}\left({ }^3 \mathrm{P}\right)+2 \mathrm{e}^{-}$ & $e={ }^4 \mathrm{~S}_{\mathrm{u}}$ & \cite{brook1978measurements}\\
& $e \in\left\{{ }^2 \mathrm{D}_{\mathrm{u}},{ }^2 \mathrm{P}_{\mathrm{u}}\right\}$ & \cite{wang2014b}\\
& remainder of $e$ & \cite{panesi2009fire}\\

\hline\hline
\end{tabular}
\label{tab:e_impact_reactions}
\end{table}

\subsubsection{Complexity reduction of the StS model}
The full StS model consisting of \SI{663}{} species and around \SI{150000}{} reactions makes it almost impossible to do a 2-dimensional CFD calculation. To make the model computationally cheaper, the full StS model was reduced by energy-based binning of the vibronic states following Ref. \cite{munafo2014boltzmann,johnston2018impact}. For molecules, the vibrational states within each electronic level were grouped based on their energies reducing the number of vibrational states within each electronic level. In no case, a vibrational state of an electronic level was grouped with a vibrational state of another electronic level. Similarly, for N and N\textsuperscript{+}, the number of electronic levels was reduced by grouping them. To reduce the complexity of the grouping strategy, it was further assumed that the population of the actual states within each bin obeyed Boltzmann distribution at local translational temperature. The energy of a grouped level is given as: 
\begin{equation}
E_{i^{\prime}}=\frac{\sum_{i \in G_{i^{\prime}}} g_i E_i exp(-\frac{E_i}{KT})}{\sum_{i \in G_{i^{\prime}}} g_i exp(-\frac{E_i}{KT})}
\end{equation}
where the subscripts with a prime such as $i^{\prime}$ represent the grouped levels, while the subscripts without the prime such as $i$ denote the actual levels. This convention is followed throughout this paper. $g_i$ and $E_i$ are the degeneracy and the energy of the $i^{th}$ level. 

The individual (\emph{i.e.}, ungrouped) levels within a group follow as Boltzmann distribution and can be computed as:
\begin{equation}
\frac{n_i}{n_{i^{\prime}}}=\frac{g_i}{Q_{i^{\prime}}} \exp \left(-\frac{E_{i}}{K T}\right)
\end{equation}

where the partition function of a grouped level is given as:
\begin{equation}
Q_{i^{\prime}}=\sum_{i \in G_{i^{\prime}}} g_i \exp \left[- \frac{E_i}{KT}\right]
\end{equation}

Definition of the grouped rates for a general excitation and de-excitation process represented by $A^i+B^j \Leftrightarrow A^k+B^l$ is given as\cite{liu2015general}:
\begin{subequations}
\begin{align}
K_{i^{'} j^{'}, k^{'} l^{'}} = & \frac{1}{Q_{i^{'}} Q_{j^{'}}} \sum_{i \in i^{'}} \sum_{j \in j^{'}} \sum_{k \in k^{'}} \sum_{l \in l^{'}} \kappa_{i j, k l}\times g_i \exp \left[- \frac{E_i}{KT}\right]  g_j \exp \left[- \frac{E_j}{KT}\right] \\
K_{k^{'} l^{'}, i^{'} j^{'}} = & \frac{1}{Q_{k^{'}} Q_{l^{'}}} \sum_{i \in i^{'}} \sum_{j \in j^{'}} \sum_{k \in k^{'}} \sum_{l \in l^{'}} \kappa_{k l, i j}\times g_k \exp \left[- \frac{E_k}{KT}\right]  g_l \exp \left[- \frac{E_l}{KT}\right]
\end{align}
\end{subequations}

Grouped rates for general ionization, dissociation, and recombination processes represented by $A^i+B^j \Leftrightarrow C^p+D^q+B^l$ are defined as\cite{liu2015general}:
\begin{subequations}
\begin{align}
K_{i^{'} j^{'}, p^{'} q^{'} l^{'}} = & \frac{1}{Q_{i^{'}} Q_{j^{'}}} \sum_{i \in i^{'}} \sum_{j \in j^{'}} \sum_{p \in p^{'}} \sum_{q \in q^{'}} \sum_{l \in l^{'}} \kappa_{i j, p q l}\times g_i \exp \left[- \frac{E_i}{KT}\right]  g_j \exp \left[- \frac{E_j}{KT}\right] \\
K_{p^{'} q^{'} l^{'},i^{'} j^{'}} = & \frac{1}{Q_{p^{'}} Q_{q^{'}} Q_{l^{'}}} \sum_{i \in i^{'}} \sum_{j \in j^{'}} \sum_{p \in p^{'}} \sum_{q \in q^{'}} \sum_{l \in l^{'}} \kappa_{p q l, i j}\times g_p \exp \left[- \frac{E_p}{KT}\right]  g_q \exp \left[- \frac{E_q}{KT}\right] g_l \exp \left[- \frac{E_l}{KT}\right]
\end{align}
\end{subequations}
Grouped rates of any type of process in the vibronic StS model can be computed by simplification/modification of the above expressions. The backward rates can then be computed in terms of forward rates and equilibrium constants by imposing micro-reversibility. Following the grouping strategy as discussed above, the reduced StS model consisted of 120 species and around \SI{12000}{} reactions which is computationally much more feasible. More details on the grouped levels mapping can be found in \cref{appendix:grouping}.

\section{Numerical Method}\label{sec:numerical_method}
\subsection{Plasma solver}\label{sec:fluid_solver}
The plasma model discussed in \cref{sec:plasma Field} has been implemented in a block-structured finite volume solver \textsc{hegel} (High fidElity tool for maGnEtogas-dynamic appLications) which has already been applied to perform simulations relevant to hypersonic aerothermodynamics, laser-plasma interactions\cite{Munafo_JCP_2020,alberti2022non,alberti2021self,alberti2020collinear,alberti2019laser} and plasma discharges\cite{munafoRGD2022,Kumar_RGD32,kumar2022high,kumar2022self,kumar2023state,oruganti2023modeling}. Inviscid fluxes are evaluated using an all-speed AUSM+up scheme\cite{liou2006sequel} to handle the low mach stiffness associated with ICPs. To achieve second-order accuracy in space, the left and right states provided to the flux function are linearly reconstructed using the MUSCL approach\cite{hirsch1990numerical,van1979towards}. The reconstruction is performed on the set of primitive variables (partial densities, velocity components, heavy-species temperature, and gas pressure) with Van Albada's slope limiter\cite{van1982comparative}. Diffusive fluxes are computed using Green-Gauss’ theorem to evaluate the gradients. The integration in time is done using a fully-implicit Backward-Euler method using a Newton linearization with local CFL-based time stepping to accelerate convergence. Evaluation of thermodynamic and transport properties as well as chemical kinetics source terms are evaluated using \textsc{plato} (PLAsma in Thermodynamic nOn-equilibrium)\cite{munafo2023plato} library. For NLTE simulations, the evaluations are done dynamically where the library is called dynamically by the fluid solver. For LTE computations, \textsc{plato} is used to generate look-up tables for thermodynamic and transport properties as a function of ($\rho$, $\mathrm{T}$) which is then used for CFD calculations for faster calculations.

\subsection{Electromagnetic solver} \label{sec:em_solver}
The electromagnetic equations are solved in a mixed finite element solver \textsc{flux} (Finite-element soLver for Unsteady electromagnetiX)\cite{kumar2022self}. In the context of Galerkin approximations, the choice of the finite element space plays an important role in the stability and convergence of the discretization\cite{rieben2005verification}. For the 2D axi-symmetric module of \textsc{flux} used in this work, where the electric field is a scalar with only a toroidal component, H(Grad) finite element space is used which contains 0-form continuous scalar basis functions that have well-defined gradients. The gradient of a 0-form basis function can be exactly represented by a combination of 1-form basis functions and hence H(Curl) finite element space is chosen for the magnetic field vector which is computed by taking the gradient of the electric field. The weak formulation of the electromagnetic governing equations is based on the formulation presented by Rieben \textit{et. al.} \cite{rieben2005verification} assuming a conforming unstructured finite element mesh composed of tetrahedrons, hexahedrons, wedges, or prisms. The solver uses \textsc{MFEM}\cite{mfem}, a modular finite element library for the finite element capabilities.

\subsection{Coupled multi-physics computational framework}\label{sec:multi_physics_framework}
\begin{figure}[!htb]
\centering
\subfloat[][]{\includegraphics[scale=0.3]{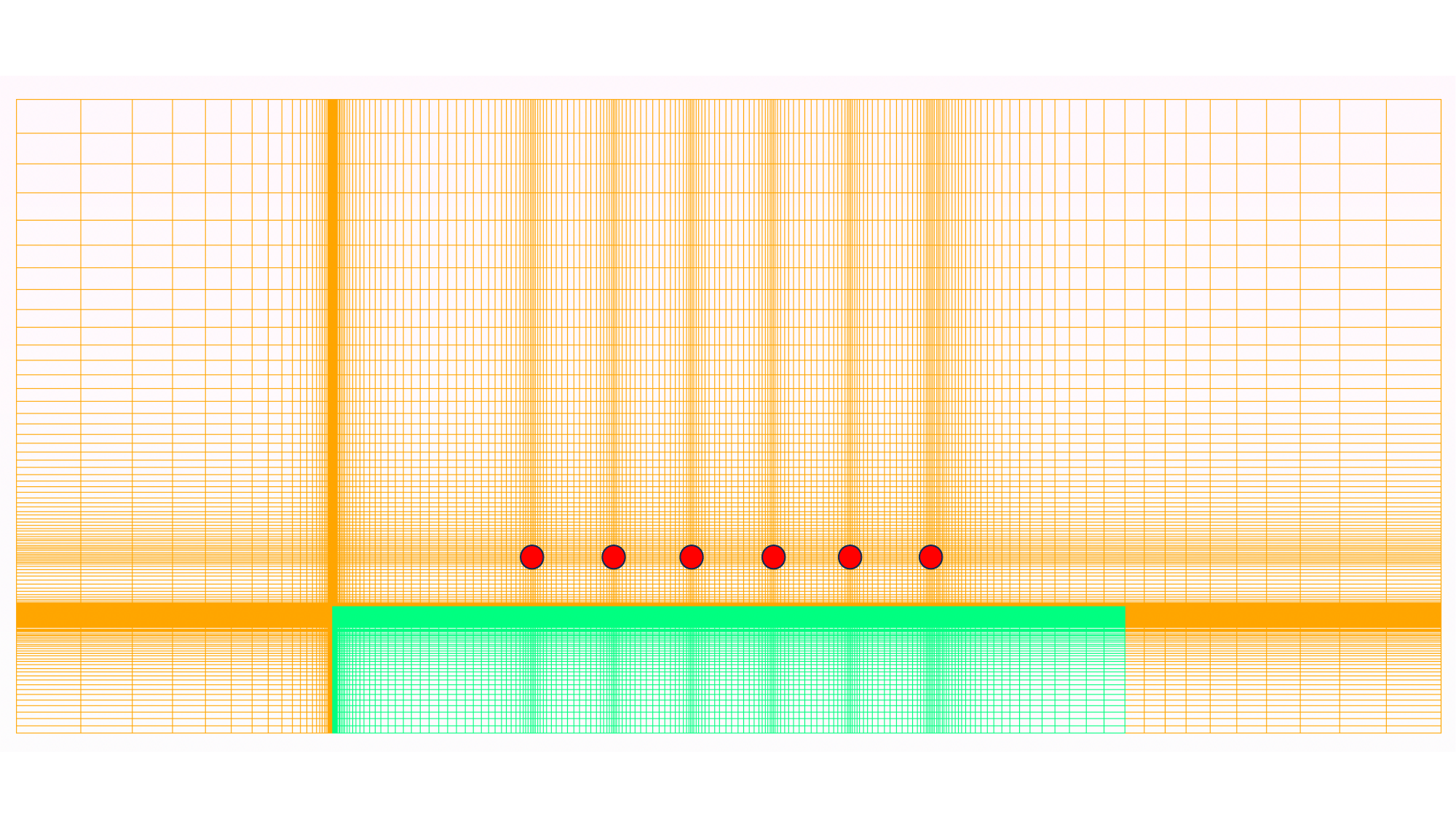}}
\subfloat[][]{\includegraphics[scale=0.225]{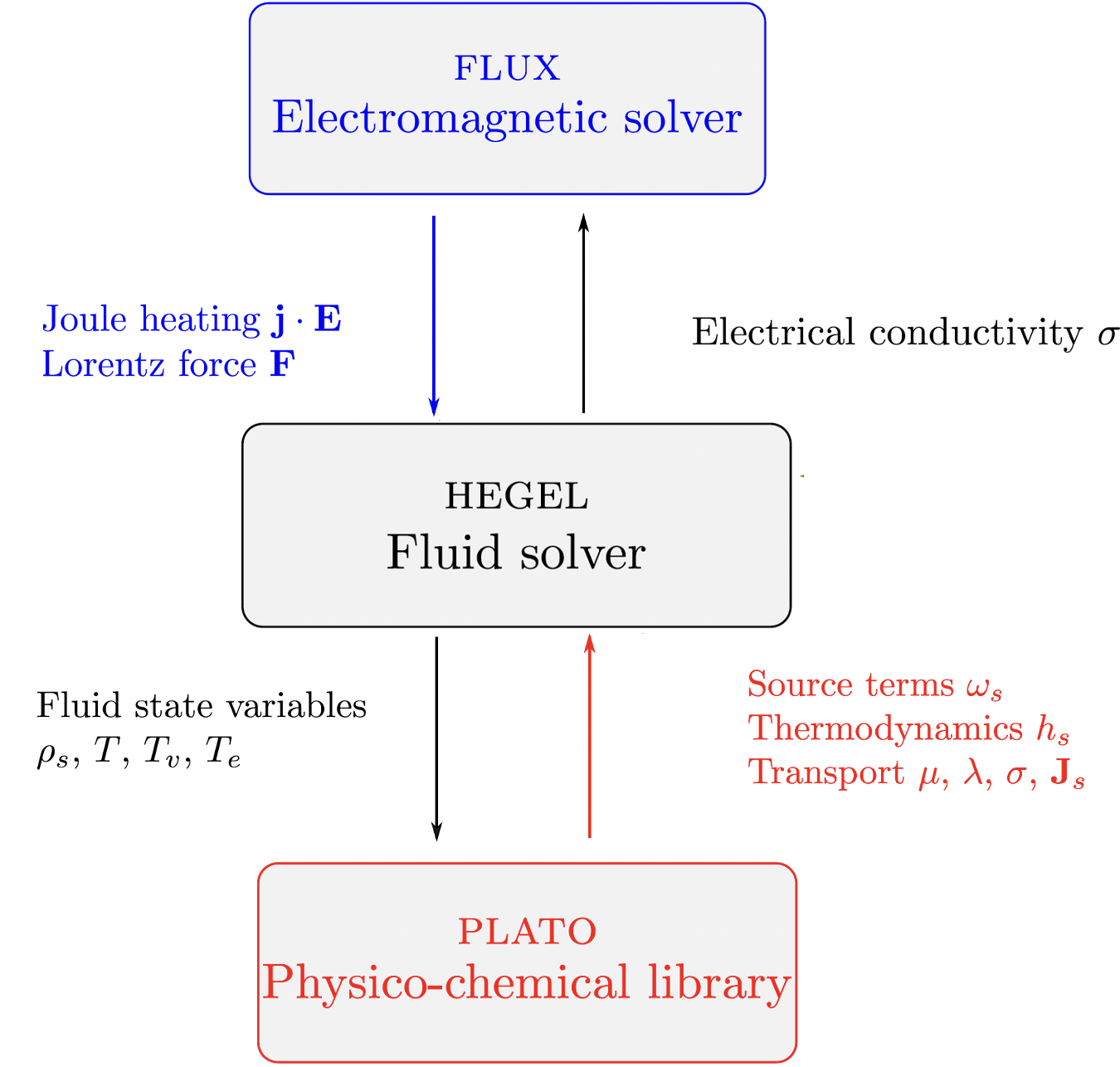}}
\caption{(a) Far-field mesh : green domain represents the fluid solver mesh, while the EM solver mesh contains both the green and the orange domain, (b) flowchart of the coupling framework} 
\label{fig:coupling_framework}
\end{figure}

The plasma and the electromagnetic field are weakly coupled through the source terms (\textit{i.e.} Lorentz forces and Joule heating) in the momentum and energy equations and the electrical conductivity in the governing equations for the electromagnetic field. \textsc{hegel} provides the plasma electrical conductivity distribution to \textsc{flux}, while \textsc{flux} gives back the Joule heating and Lorentz forces to \textsc{hegel}. This communication is accomplished using \textsc{preCICE}\cite{bungartz2016precice}, an open-source coupling library for partitioned multi-physics simulations. The electromagnetic equations are solved on a far-field mesh coinciding with the fluid solver mesh as shown in \cref{fig:coupling_framework}(a). The coupling framework has been depicted in \cref{fig:coupling_framework}(b). The time window for communication is chosen such that \textsc{hegel} calls \textsc{flux} every 10 iterations to avoid slowing down the computations due to communication. Between the coupling time window, the exchanged source terms are frozen.

\section{Results}\label{sec:results}

 \subsection{Verification of the grouping strategy for the vibronic StS model}\label{sec:box_verification}
\begin{figure}[!htb]
    \centering
    \subfloat[][]{\includegraphics[scale=0.25]{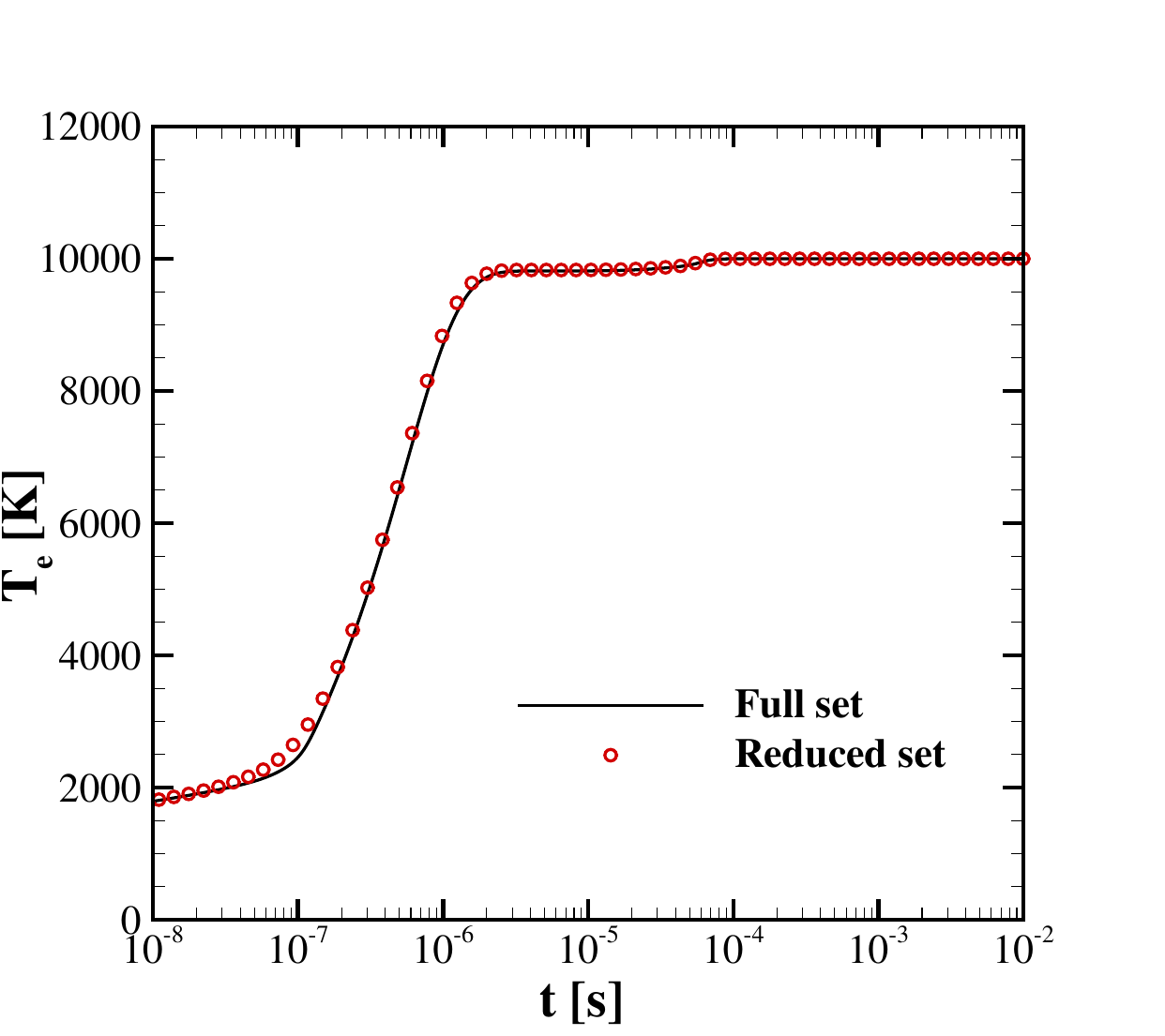}}
    \subfloat[][]{\includegraphics[scale=0.25]{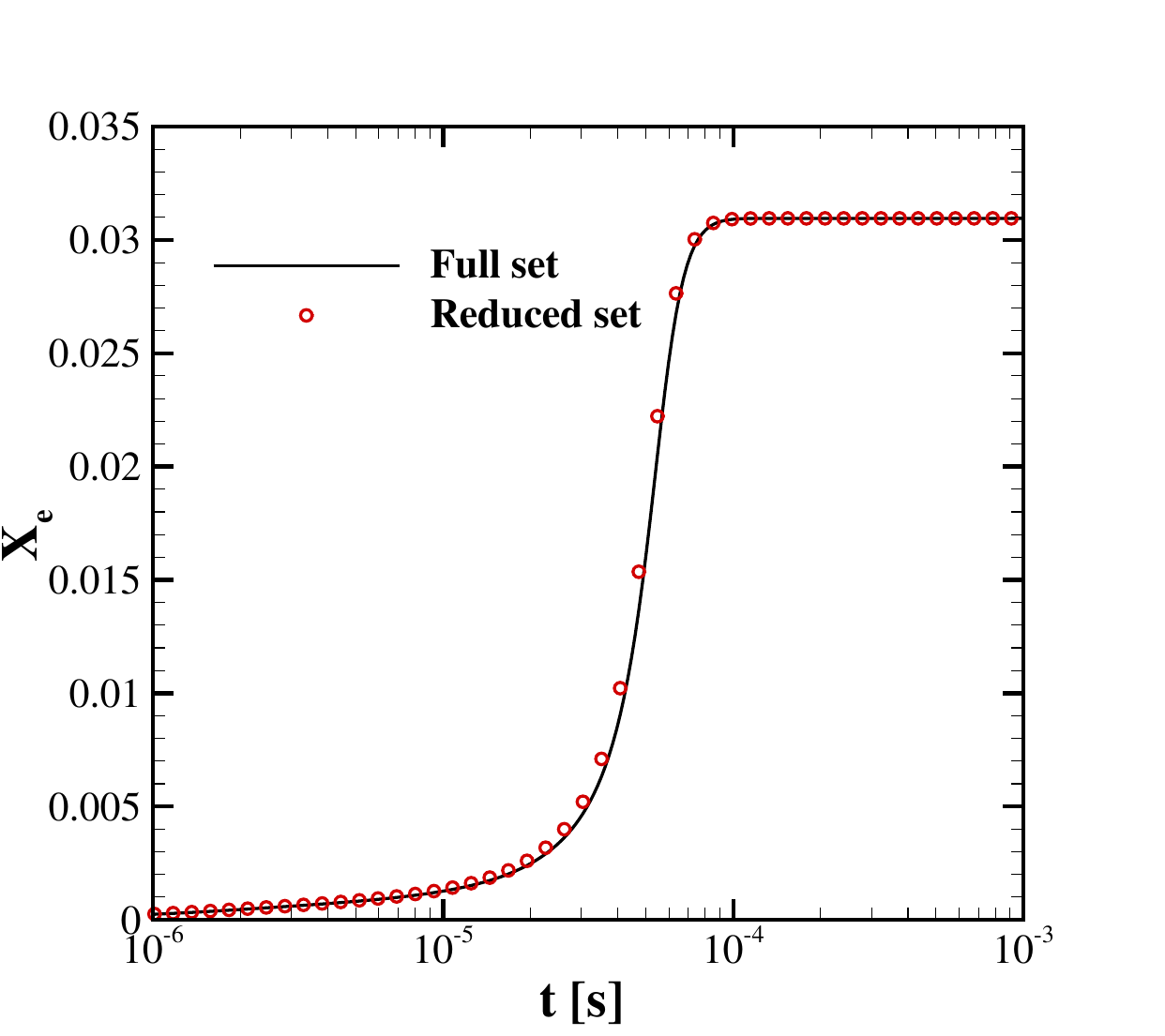}}
    \subfloat[][]{\includegraphics[scale=0.25]{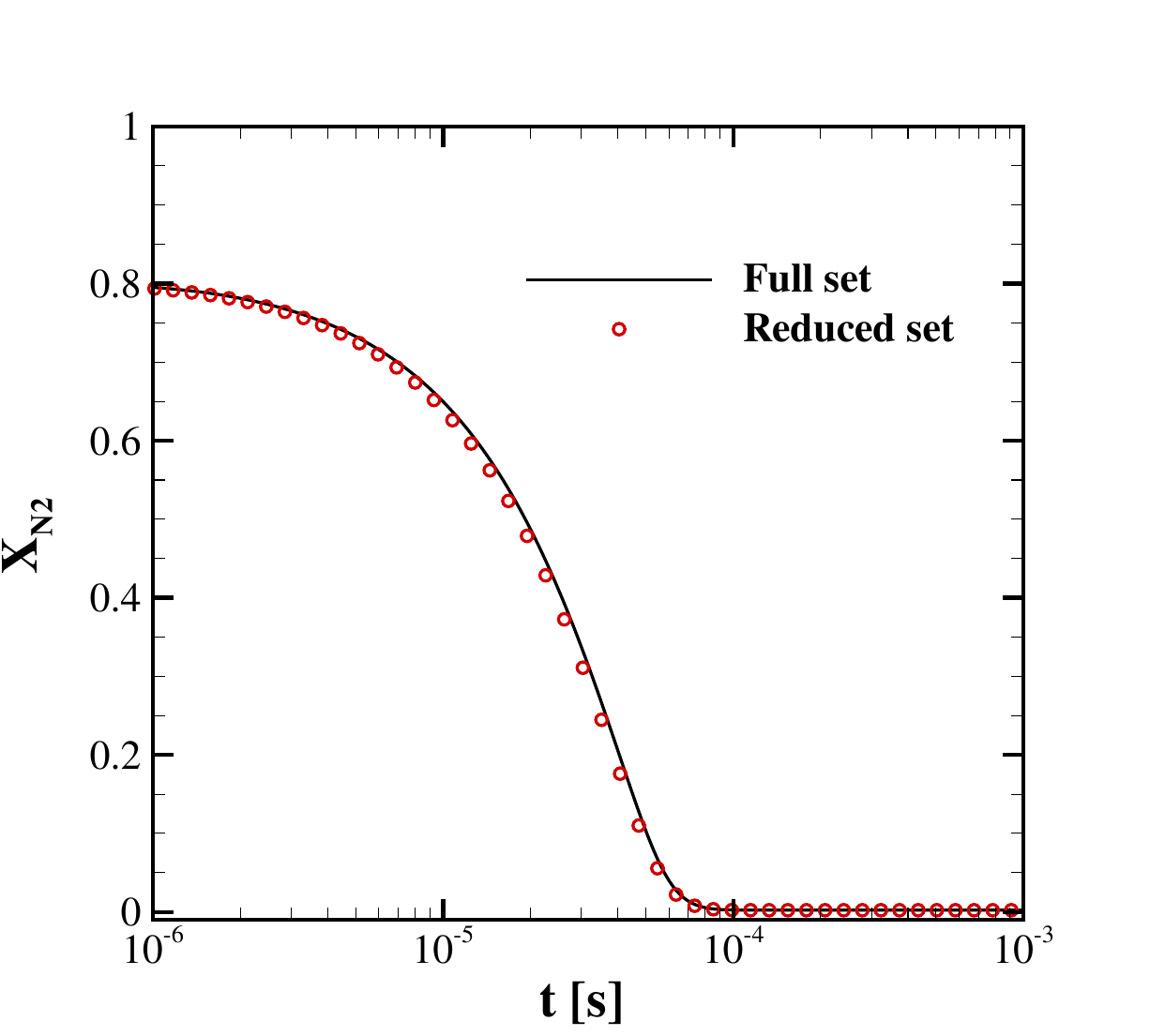}}\\ 
    \subfloat[][]{\includegraphics[scale=0.25]{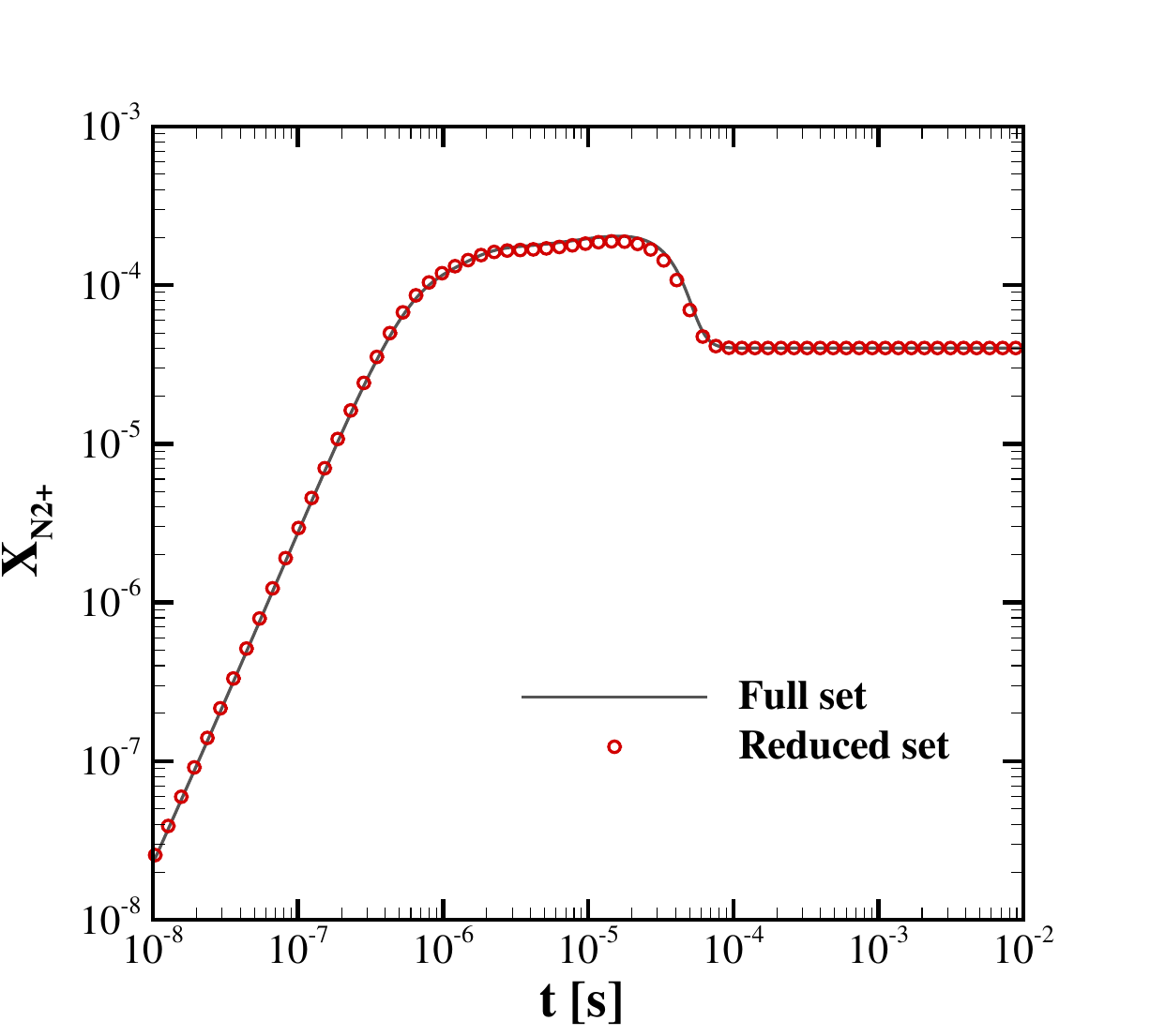}}
    \subfloat[][]{\includegraphics[scale=0.25]{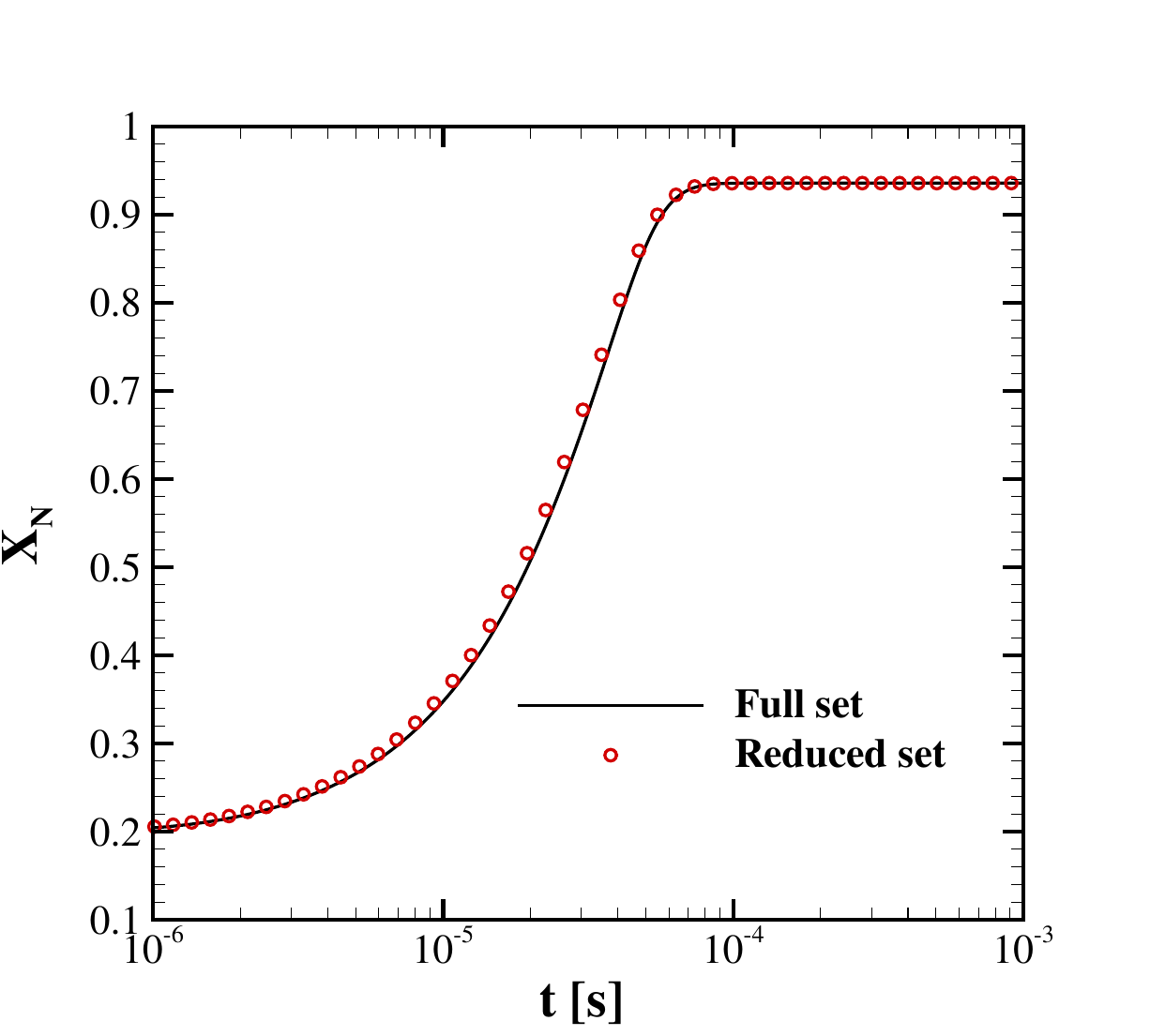}}
    \subfloat[][]{\includegraphics[scale=0.25]{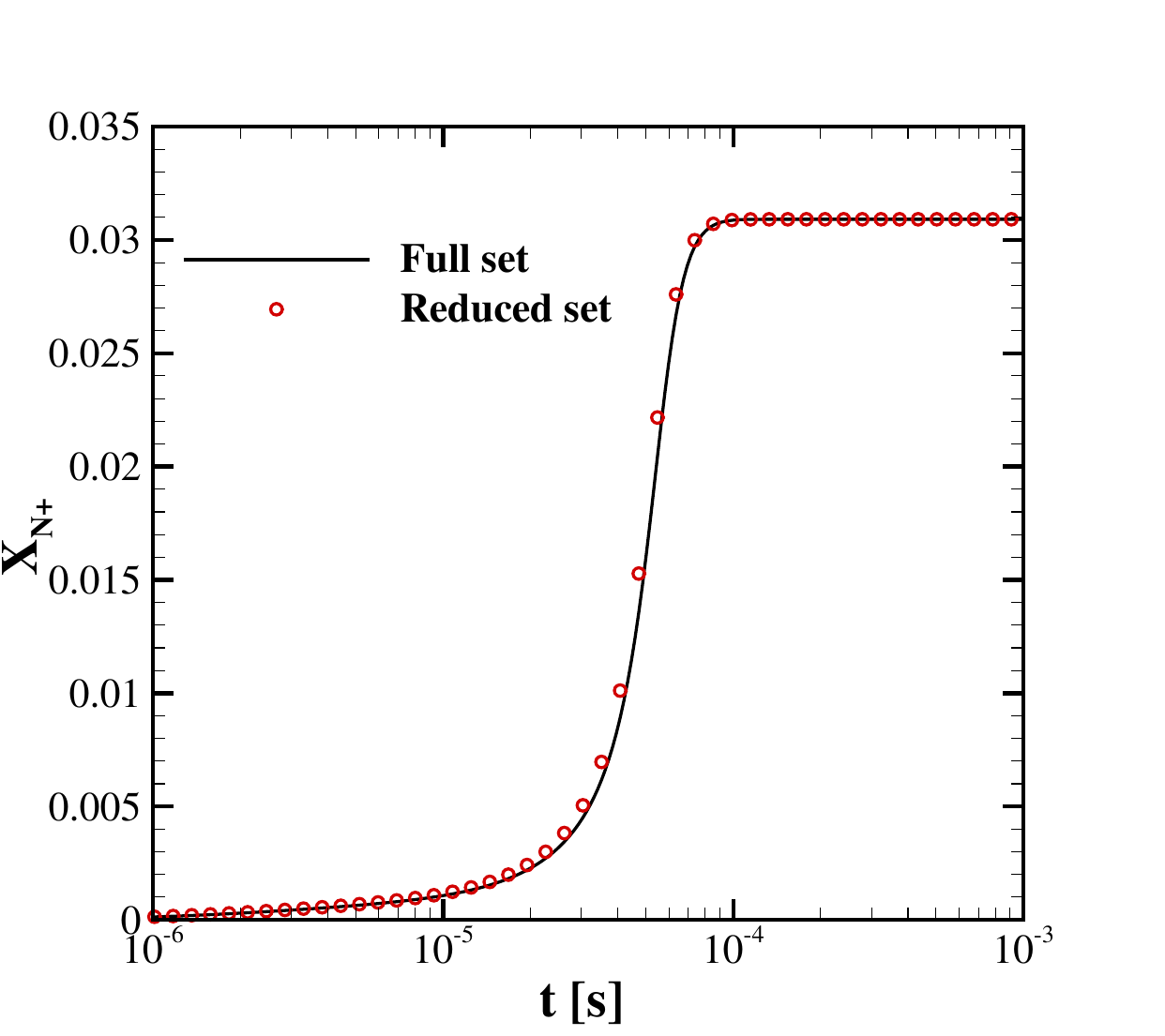}}
    
    \caption{Time evolution of various quantities for 0D isochoric reactor under compression: (a) electron temperature, (b) electron mole-fraction, (c) N\textsubscript{2} mole-fraction, (d) $\mathrm{N}_2^+$ mole-fraction, (e) N mole-fraction and (f) N\textsuperscript{+} mole-fraction}
    \label{fig:full_vs_reduced_macro_heating}
    \end{figure}

 To verify that the reduced StS model can capture the dynamics of the full StS model with reasonable accuracy, 0D isochoric-isothermal reactor simulations were performed for two cases: compression and expansion. For the compression case, the nitrogen mixture initially at \SI{300}{K} was suddenly heated to a bath temperature of \SI{10000}{K} allowing the system to evolve in time and relax to its final equilibrium state. The initial conditions are as followed: T\textsubscript{h} = T\textsubscript{e} = \SI{300}{K}, P = \SI{1000}{Pa}, X\textsubscript{N} = 0.2 and X\textsubscript{N\textsubscript{2}} = 0.8. \cref{fig:full_vs_reduced_macro_heating} shows the evolution of various quantities with time obtained with the full set as well as the reduced set, showing that the reduced StS model can accurately represent the full StS kinetics. 
 
\begin{figure}[!htb]
    \centering
    \subfloat[][]{\includegraphics[scale=0.25]{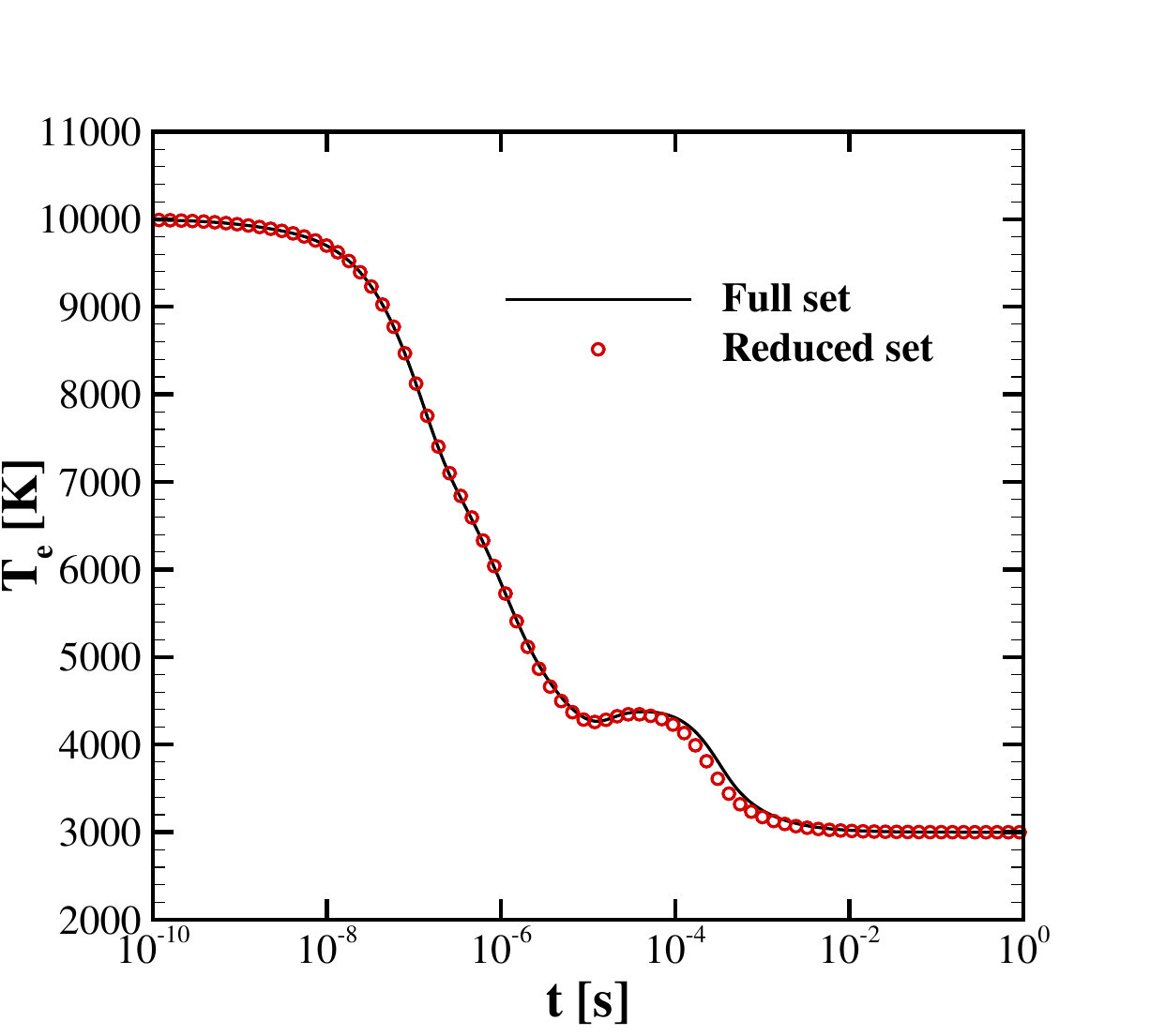}}
    \subfloat[][]{\includegraphics[scale=0.25]{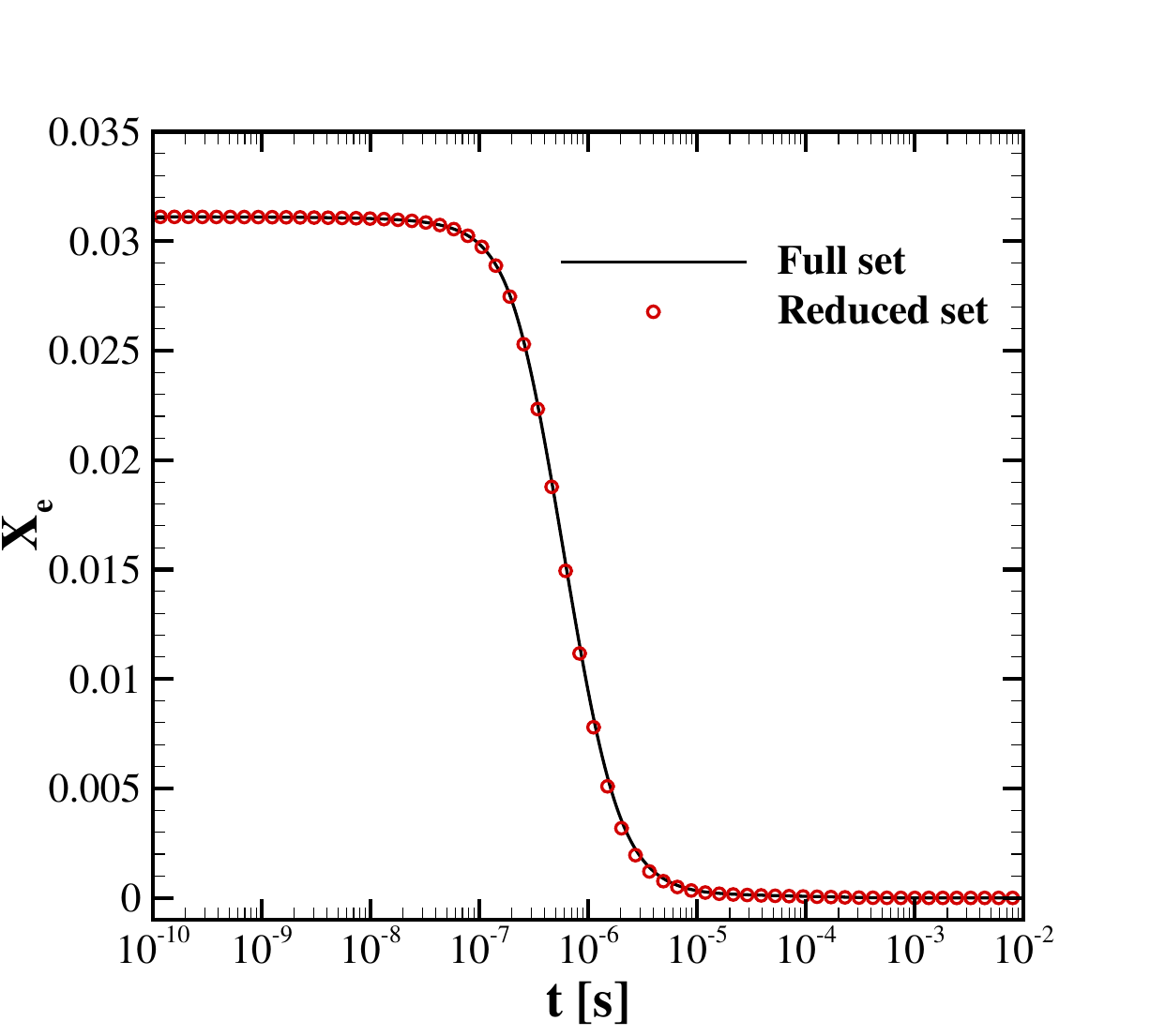}}
    \subfloat[][]{\includegraphics[scale=0.25]{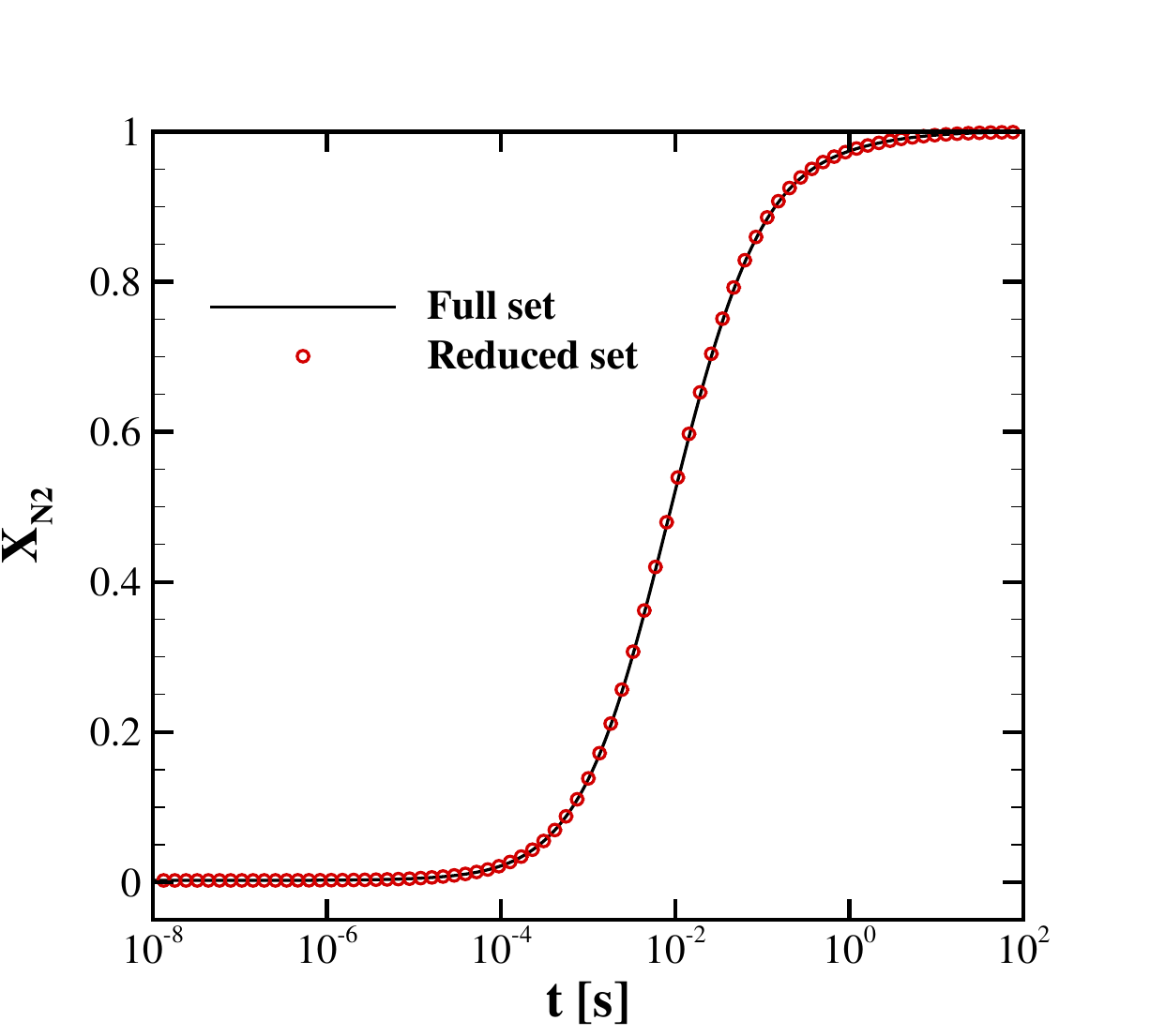}}\\ 
    \subfloat[][]{\includegraphics[scale=0.25]{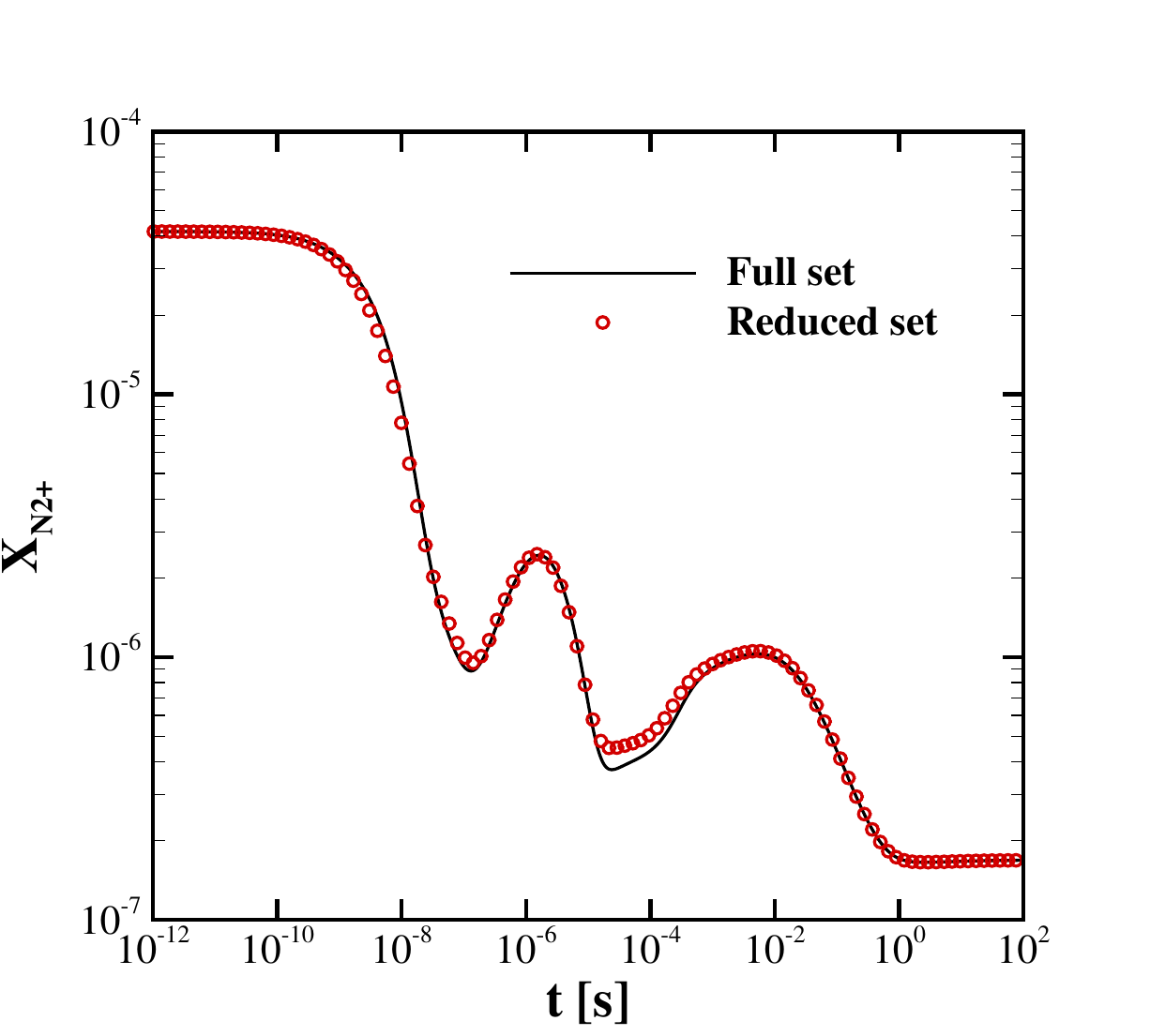}}
    \subfloat[][]{\includegraphics[scale=0.25]{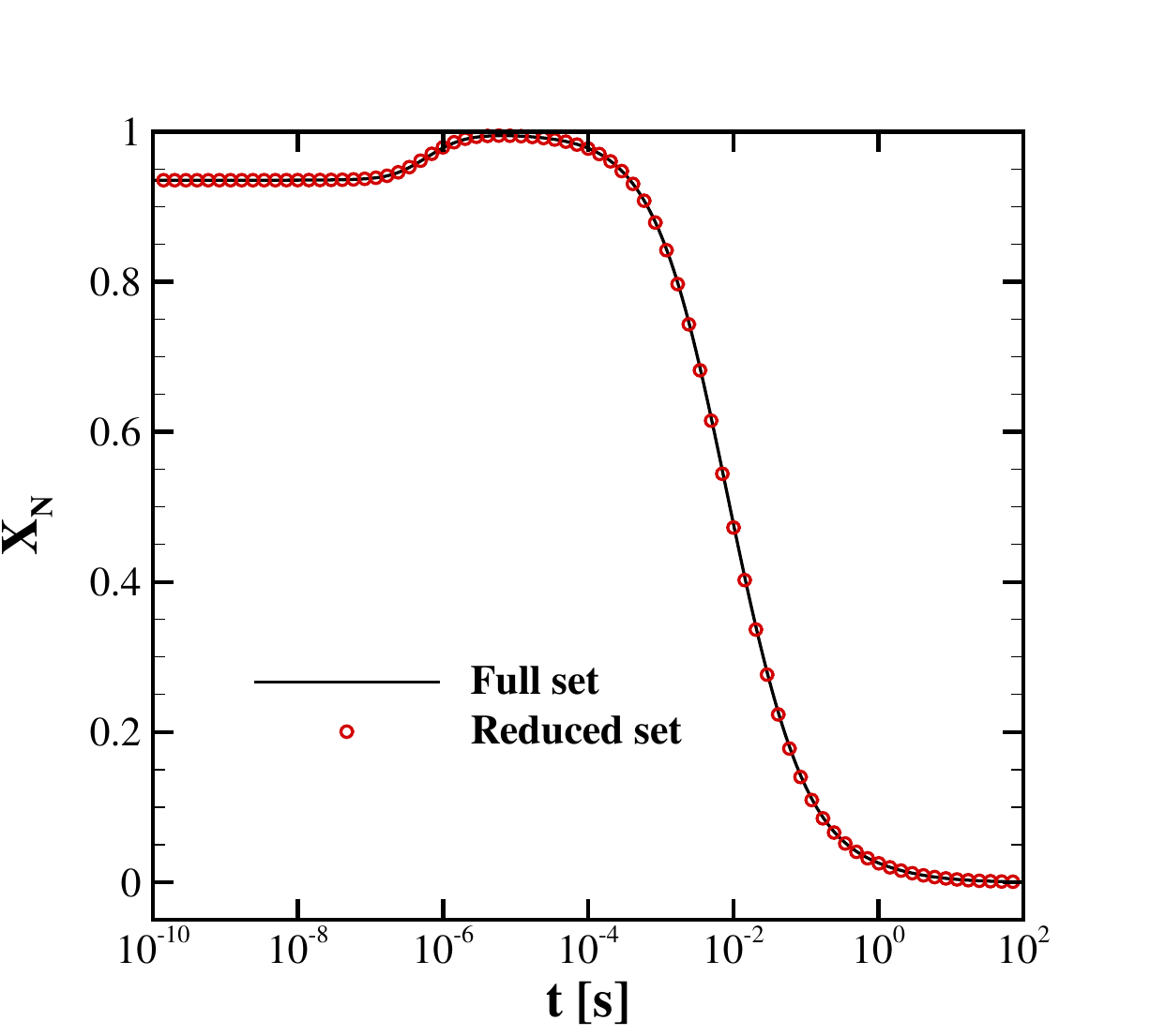}}
    \subfloat[][]{\includegraphics[scale=0.25]{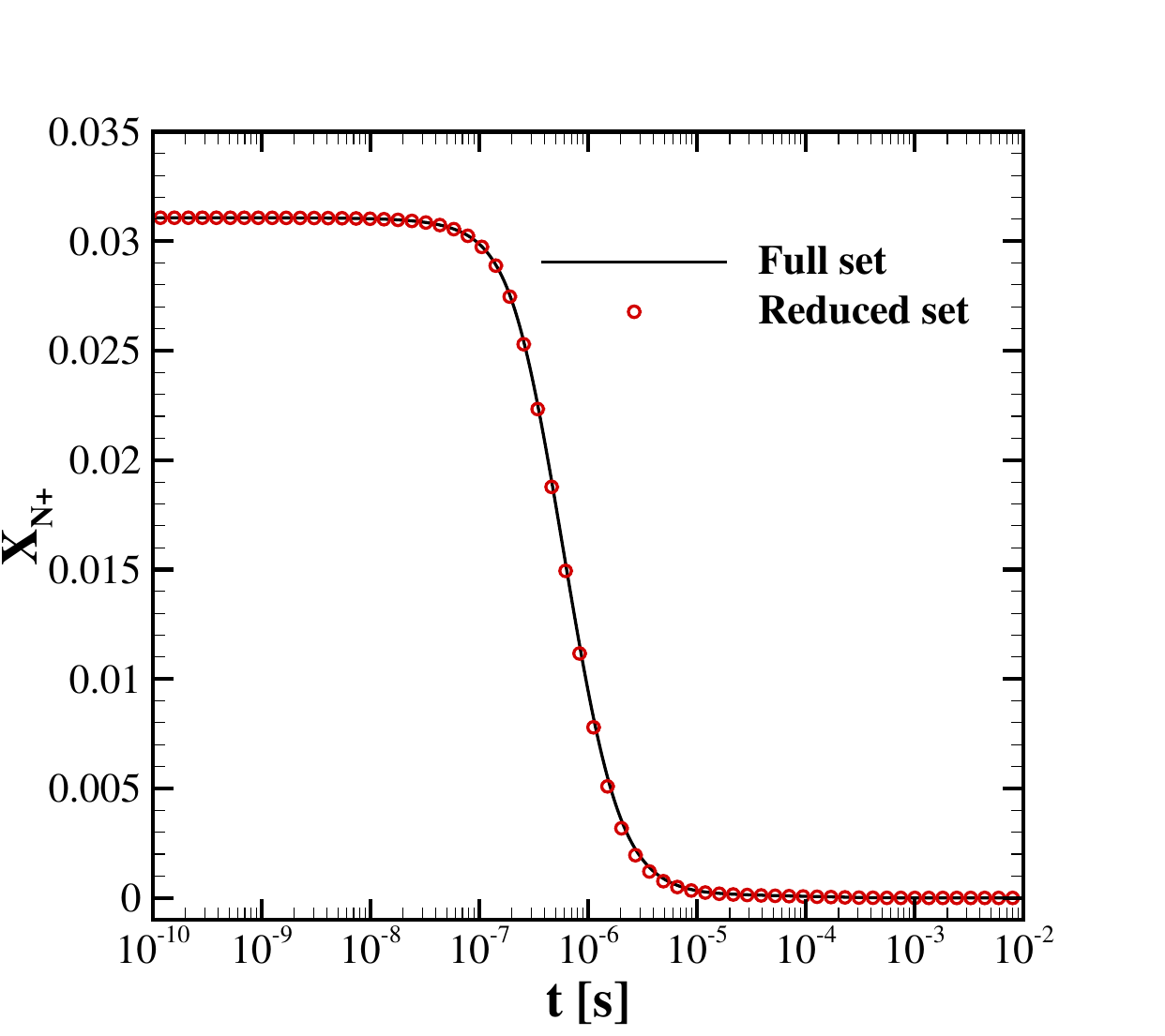}}
    
    \caption{Time evolution of various quantities for 0D isochoric reactor under expansion: (a) electron temperature, (b) electron mole-fraction, (c) N\textsubscript{2} mole-fraction, (d) $\mathrm{N}_2^+$ mole-fraction, (e) N mole-fraction and (f) N\textsuperscript{+} mole-fraction}
    \label{fig:full_vs_reduced_macro_cooling}
    \end{figure}

 For the expansion case, the nitrogen mixture initially at \SI{10000}{K} was suddenly cooled to a bath temperature of \SI{3000}{K}. The initial conditions are as followed: T\textsubscript{h} = T\textsubscript{e} = \SI{10000}{K}, P = \SI{61106.32}{Pa}, X\textsubscript{e} = \SI{0.0311099}{}, X\textsubscript{N} = \SI{0.935280}{}, X\textsubscript{N\textsubscript{2}} = \SI{0.00249988}{}, $\mathrm{X}_{\mathrm{N}_2^+} = 0.41682\times10^{-4}$ and X\textsubscript{N\textsuperscript{+}} = \SI{0.0310683}{}, which is the LTE composition at the given pressure and temperature. \cref{fig:full_vs_reduced_macro_cooling} shows the evolution of various quantities with time for the expansion case showing very good agreement between the reduced and the full StS model. Hence, the results from the 0D isochoric reactor simulations confirm that the reduced StS model captures the dynamics of the full StS model with excellent accuracy.

\begin{figure}[!htb]
    \hspace*{-1cm}
    \subfloat[][]{\includegraphics[scale=0.3]{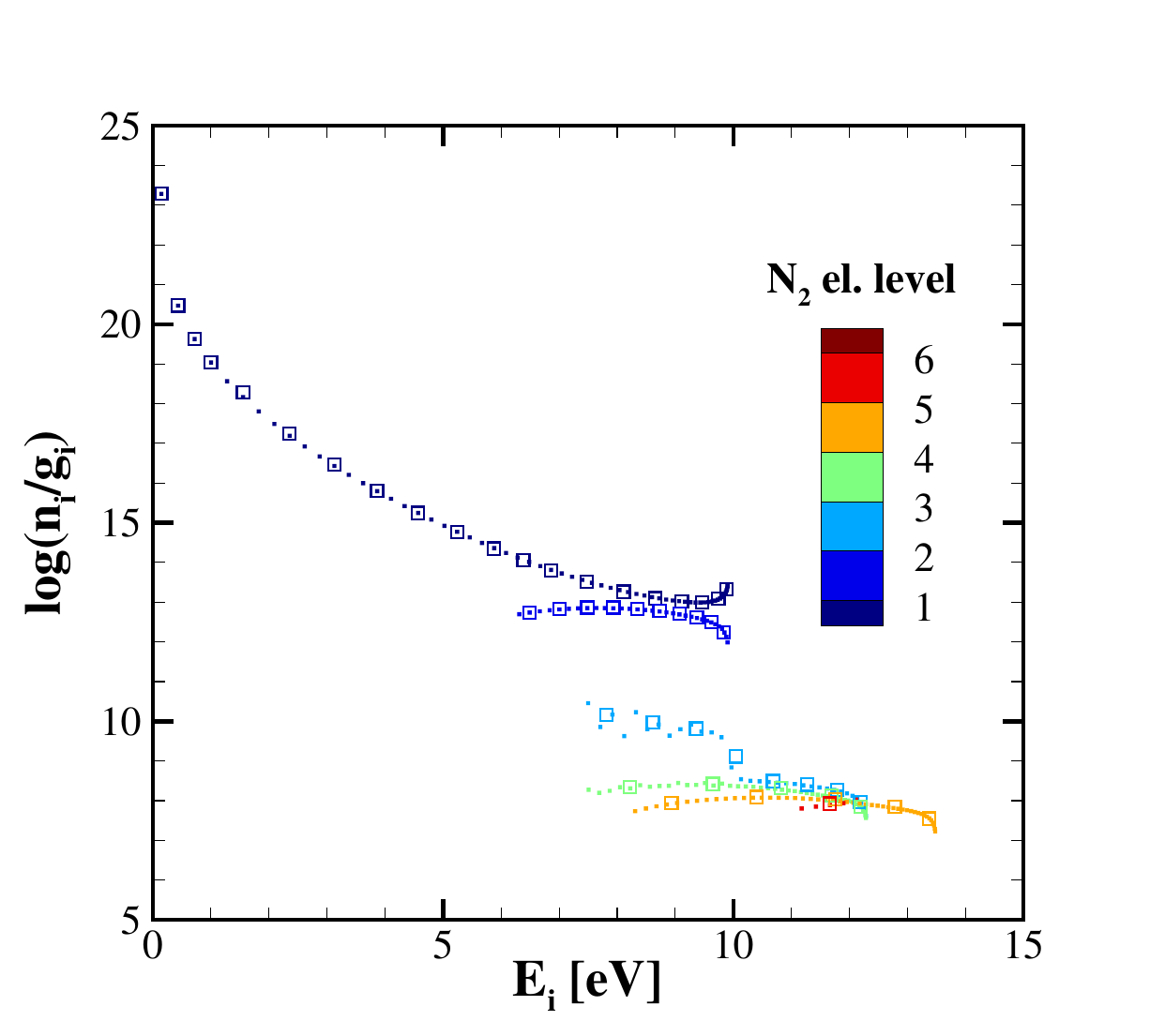}}
    \subfloat[][]{\includegraphics[scale=0.3]{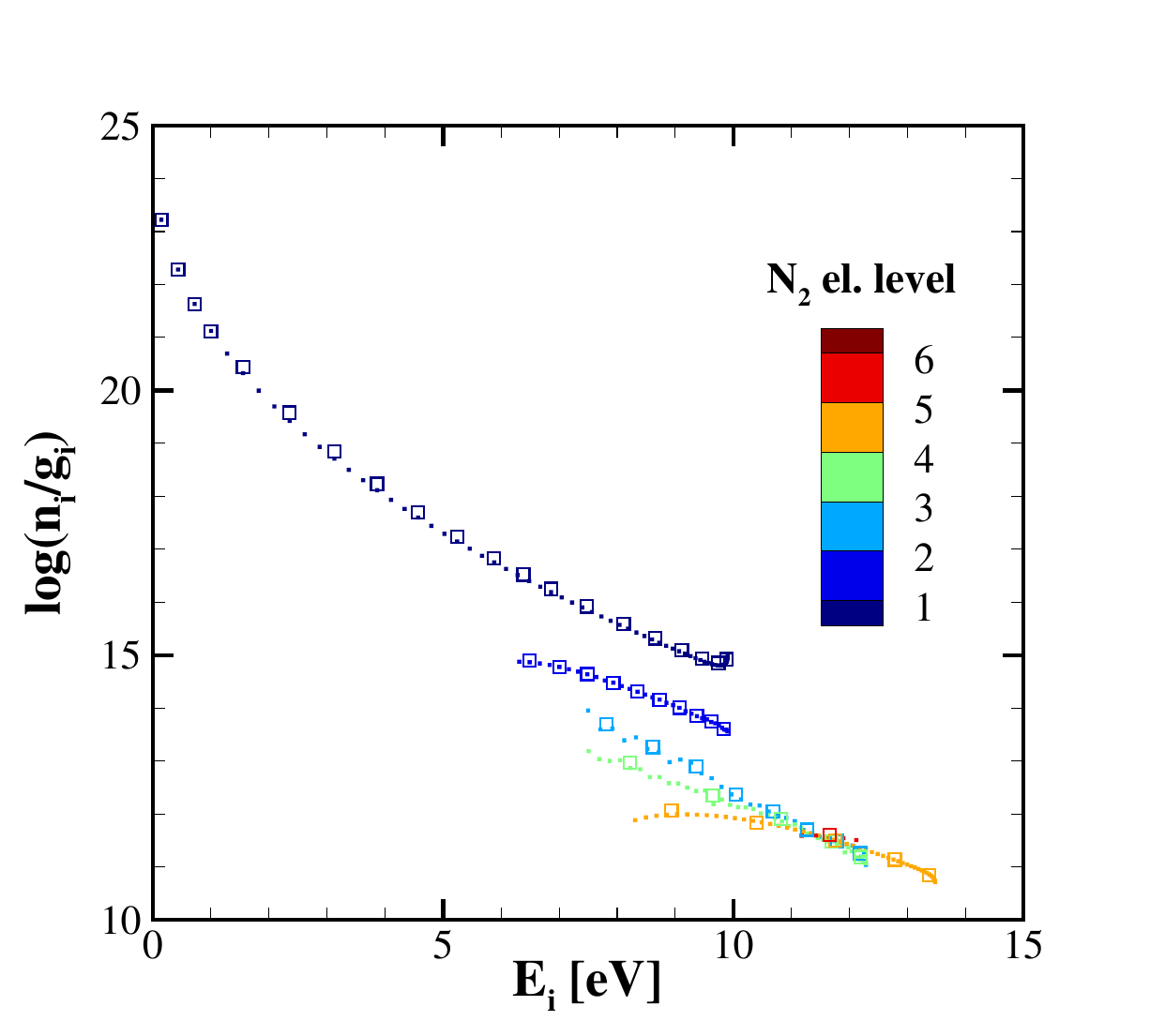}}
    \subfloat[][]{\includegraphics[scale=0.3]{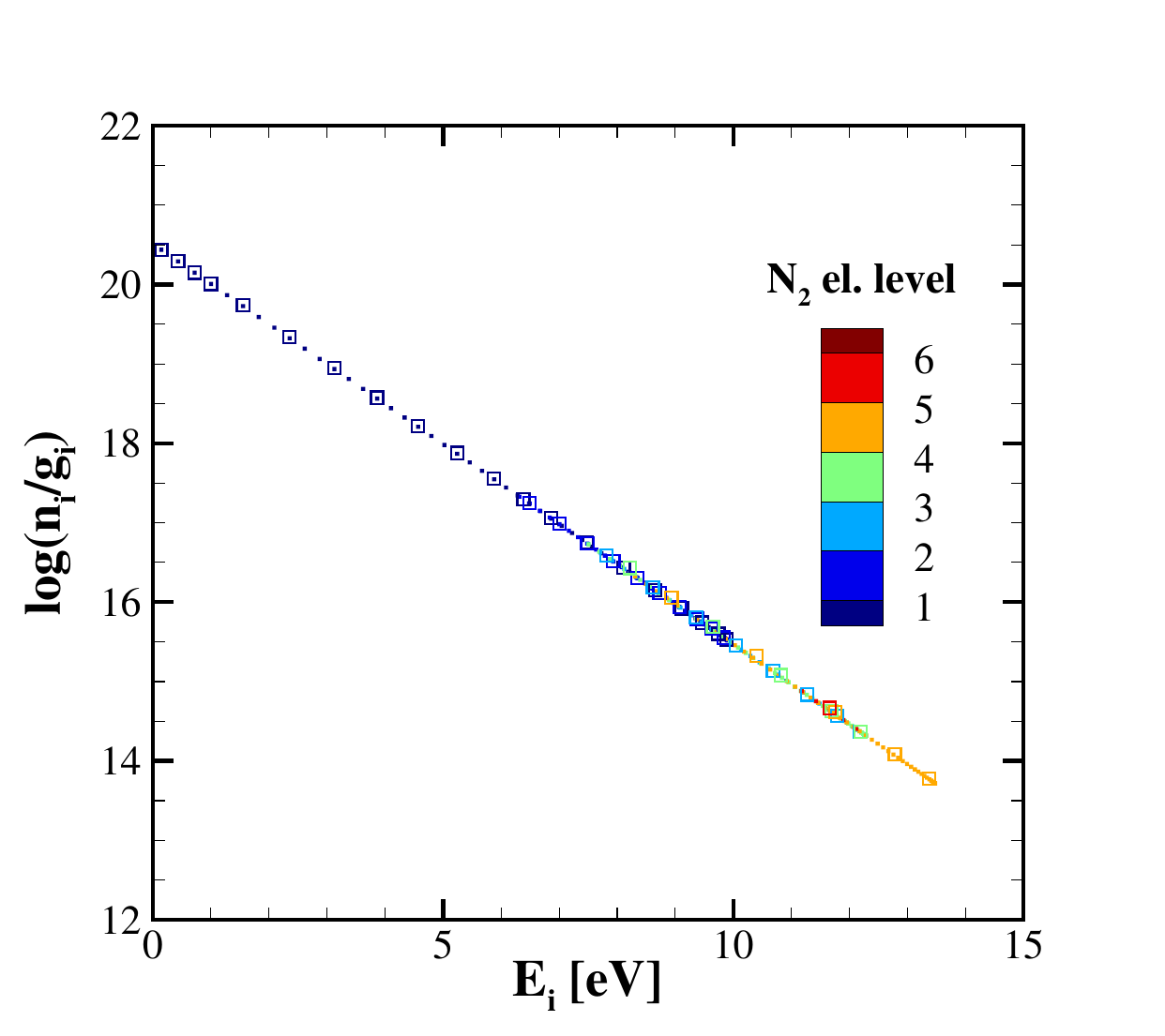}}
    \caption{Population distribution of N\textsubscript{2} for 0D isochoric reactor under compression at: (a) t = $5.23 \times 10^{-10}$s, (b) t = $4.4 \times 10^{-8}$s and (c) t = t\textsubscript{final}. Small dots represent the actual states while the big squares represent the grouped states. Vibrational states within different electronic levels are shown by different colors. }
    \label{fig:full_vs_reduced_pop_N2_heating}
    \end{figure}

\begin{figure}[!htb]
    \hspace*{-1cm}
    \subfloat[][]{\includegraphics[scale=0.3]{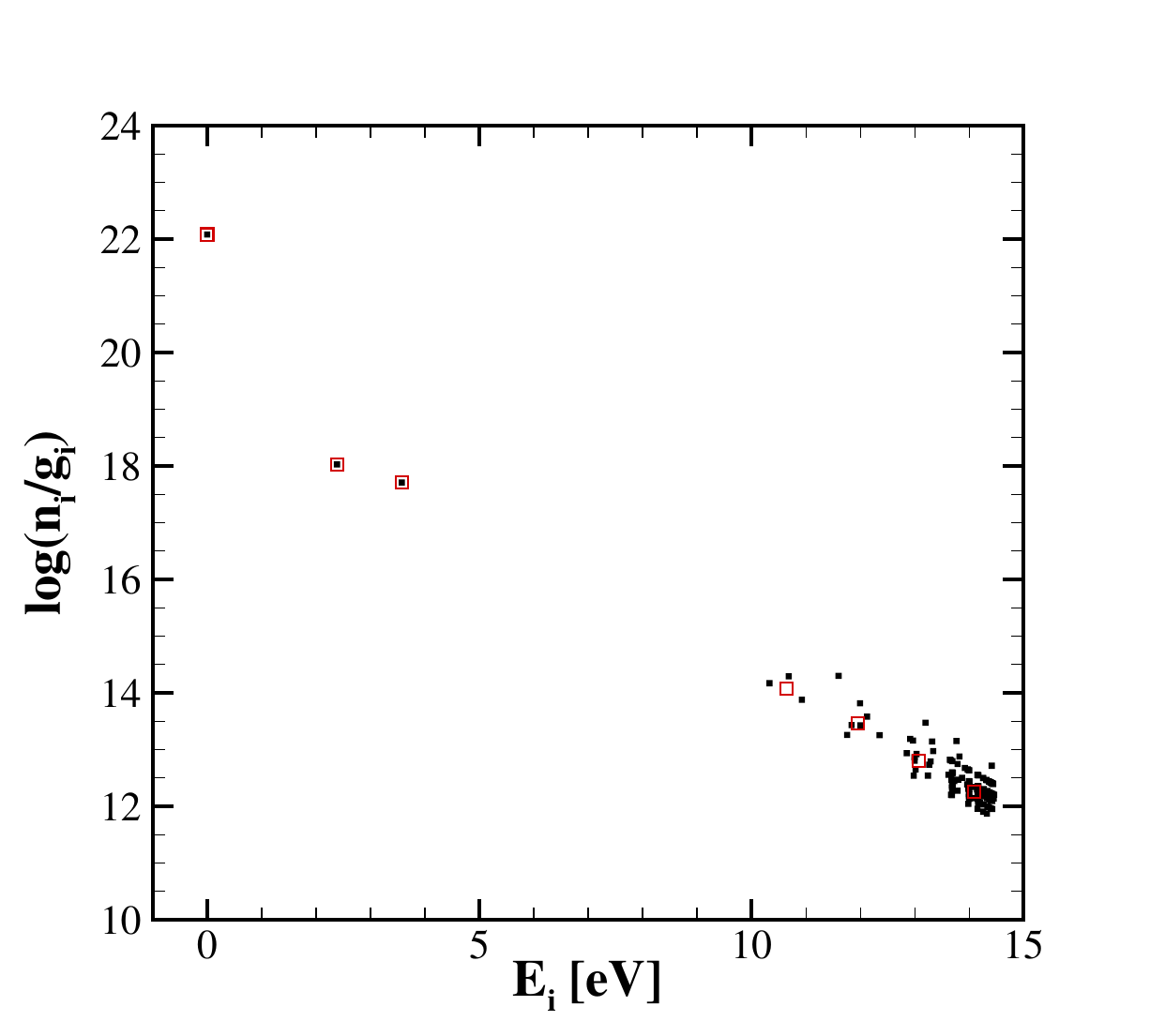}}
    \subfloat[][]{\includegraphics[scale=0.3]{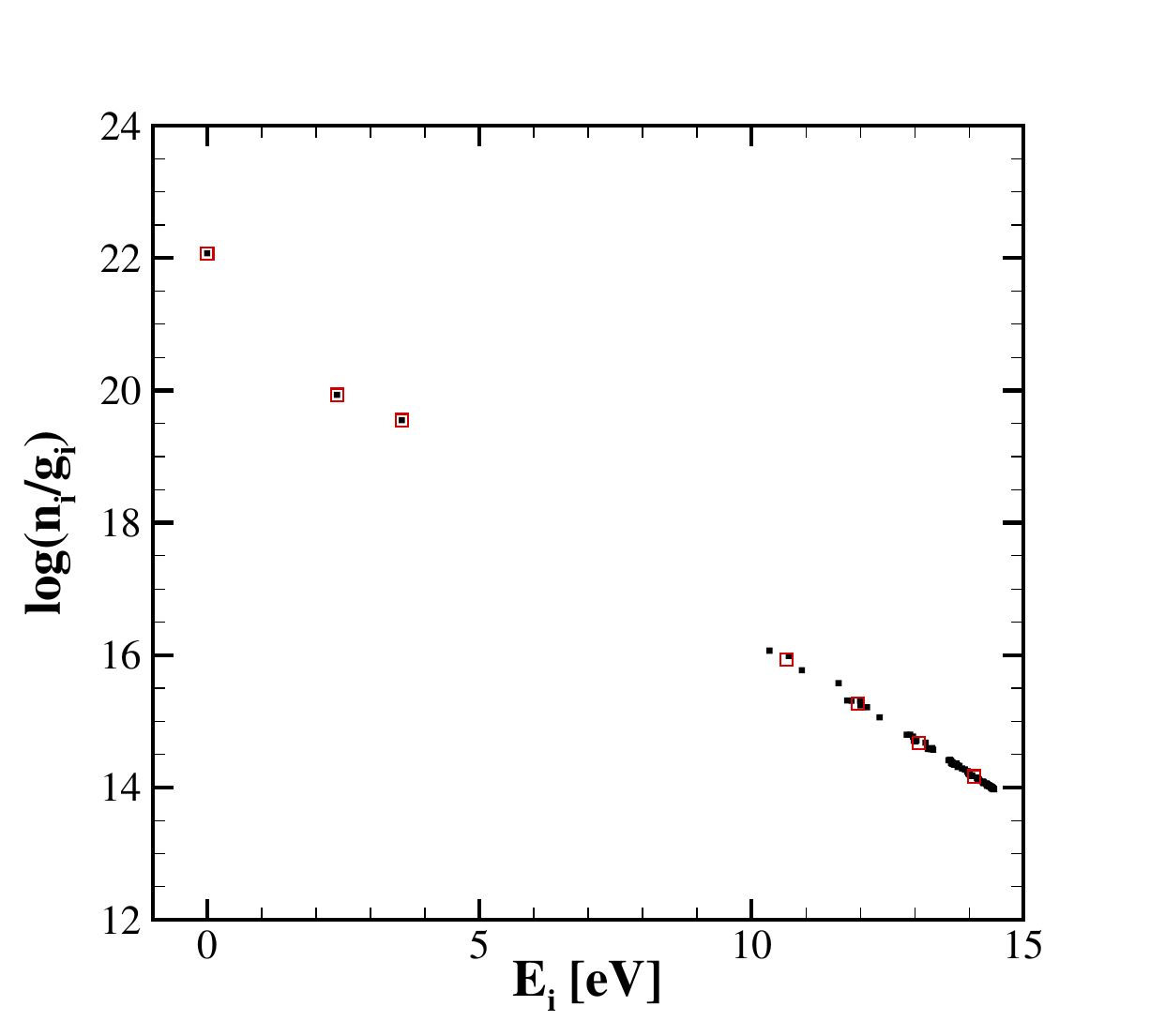}} 
    \subfloat[][]{\includegraphics[scale=0.3]{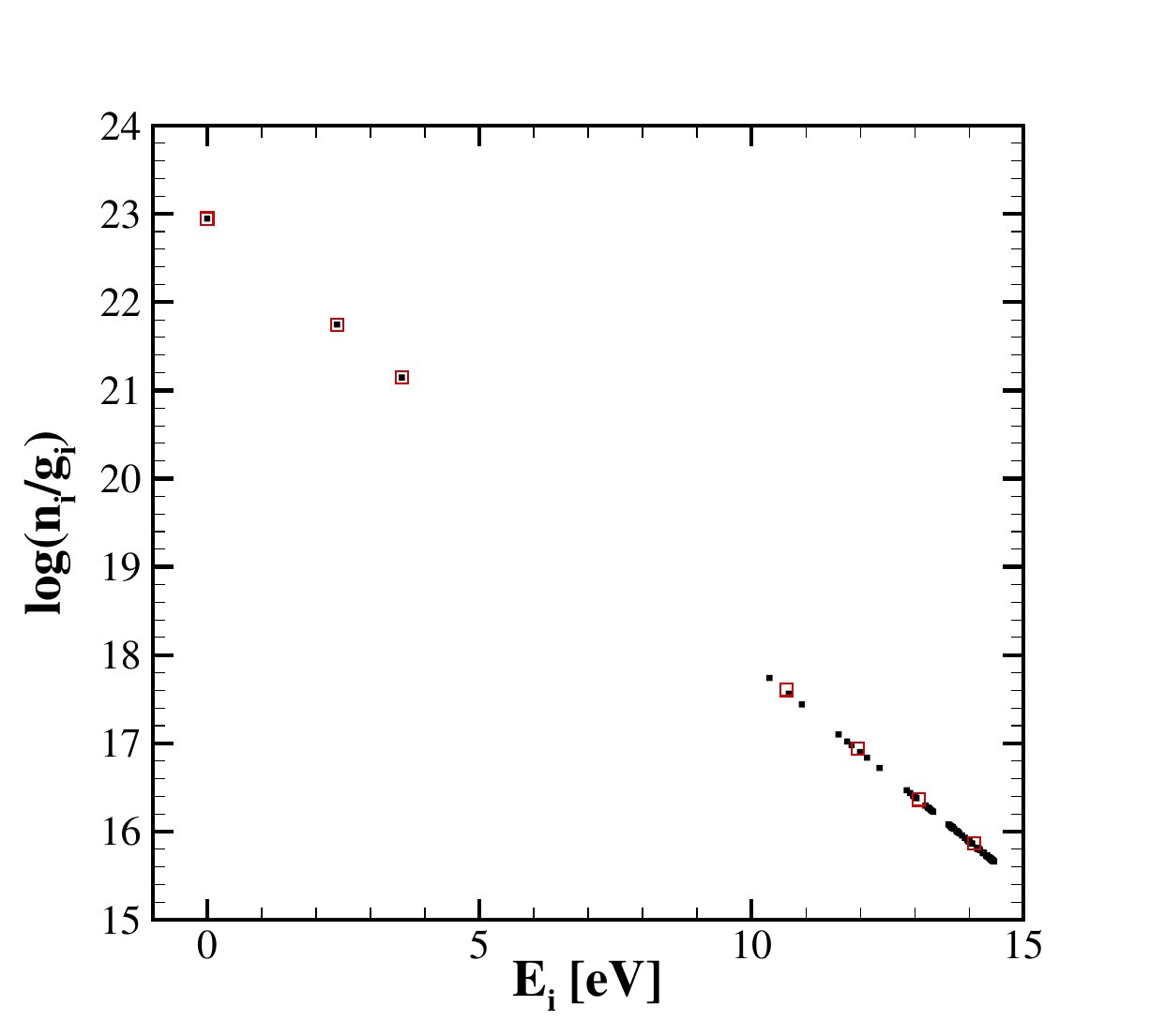}}
    \caption{Population distribution of N atom for 0D isochoric reactor under compression at: (a) t = $5.23 \times 10^{-10}$s, (b) t = $4.4 \times 10^{-8}$s and (c) t = t\textsubscript{final}. Small dots represent the actual electronic states while the big squares represent the grouped states.}
    \label{fig:full_vs_reduced_pop_N_heating}
    \end{figure}

\begin{figure}[!htb]
    \hspace*{-1cm}
    \subfloat[][]{\includegraphics[scale=0.3]{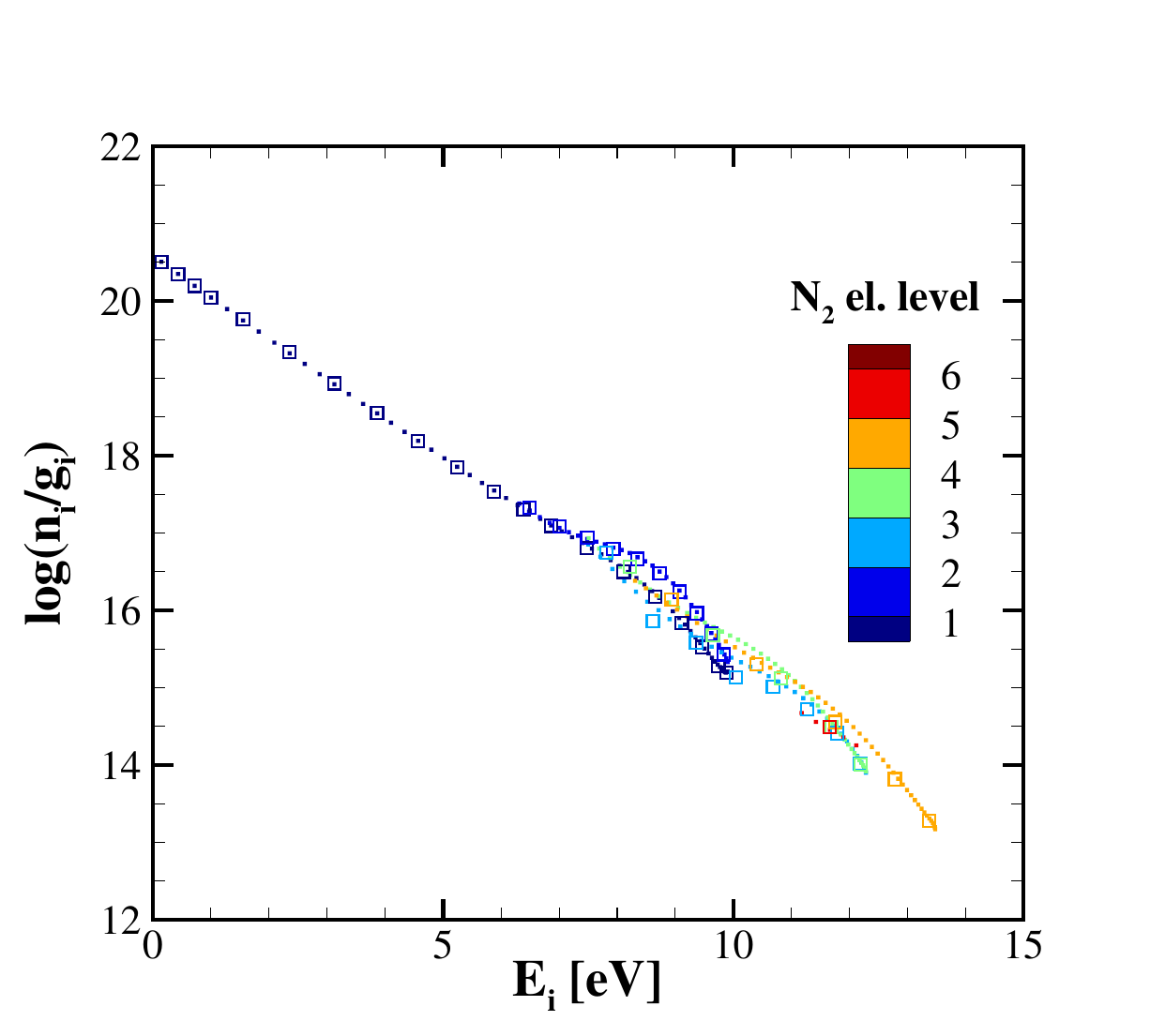}}
    \subfloat[][]{\includegraphics[scale=0.3]{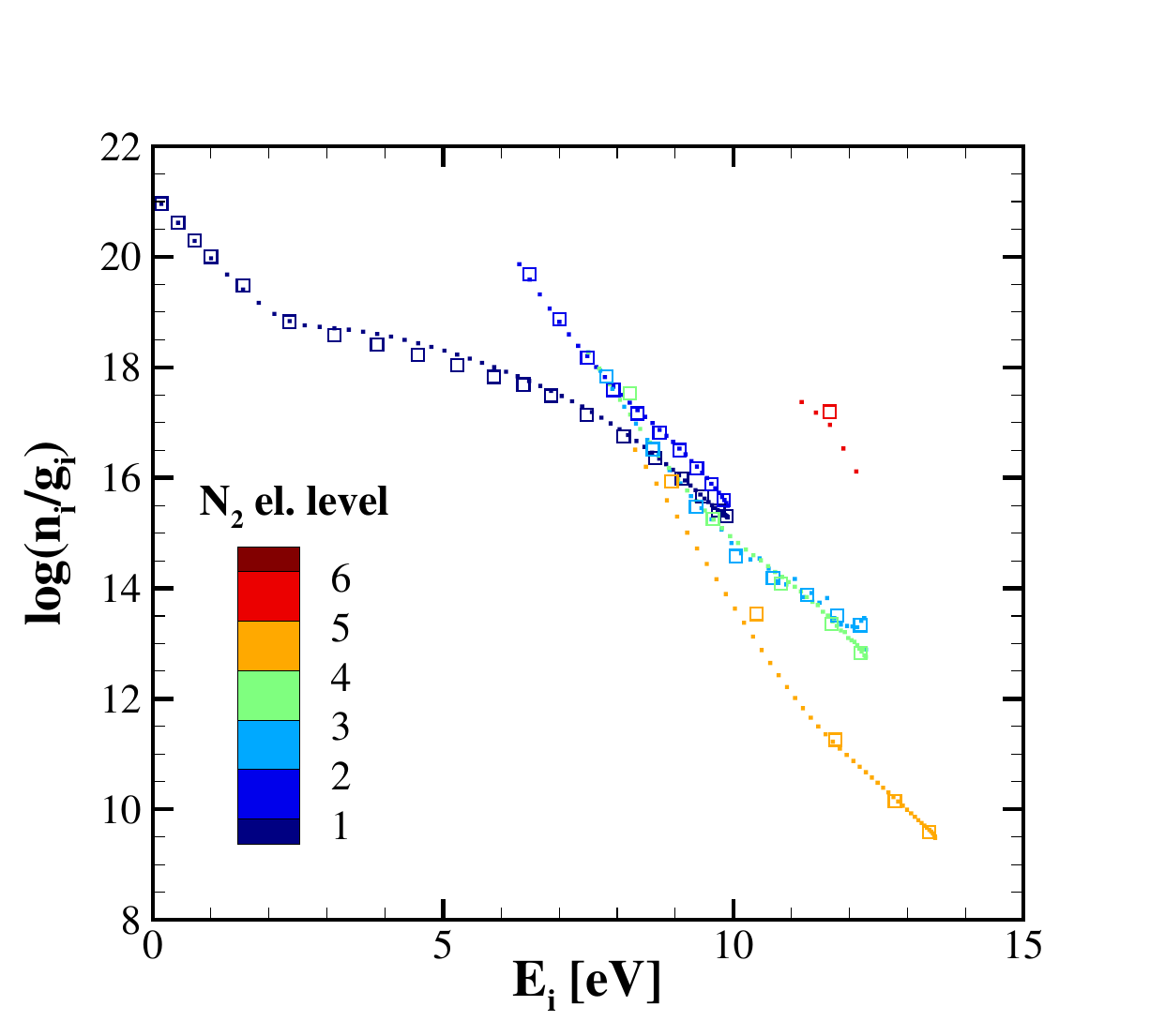}}
    \subfloat[][]{\includegraphics[scale=0.3]{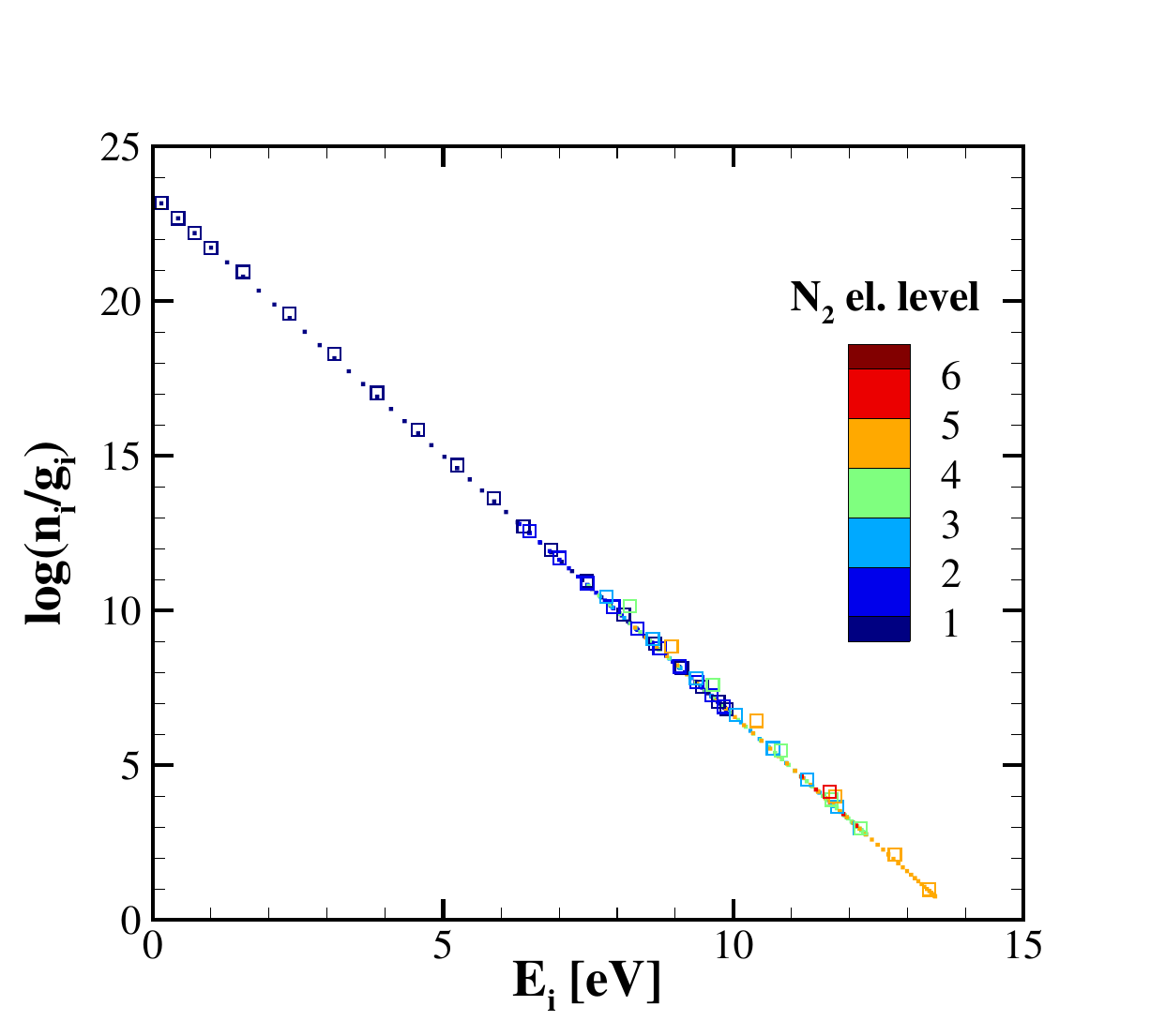}}
    \caption{Population distribution of N\textsubscript{2} for 0D isochoric reactor under expansion at: (a) t = $3.13 \times 10^{-8}$s, (b) t = $1.15 \times 10^{-4}$s and (c) t = t\textsubscript{final}. Small dots represent the actual states while the big squares represent the grouped states. Vibrational states within different electronic levels are shown by different colors. }
    \label{fig:full_vs_reduced_pop_N2_cooling}
    \end{figure}

\begin{figure}[!htb]
    \hspace*{-1cm}
    \subfloat[][]{\includegraphics[scale=0.3]{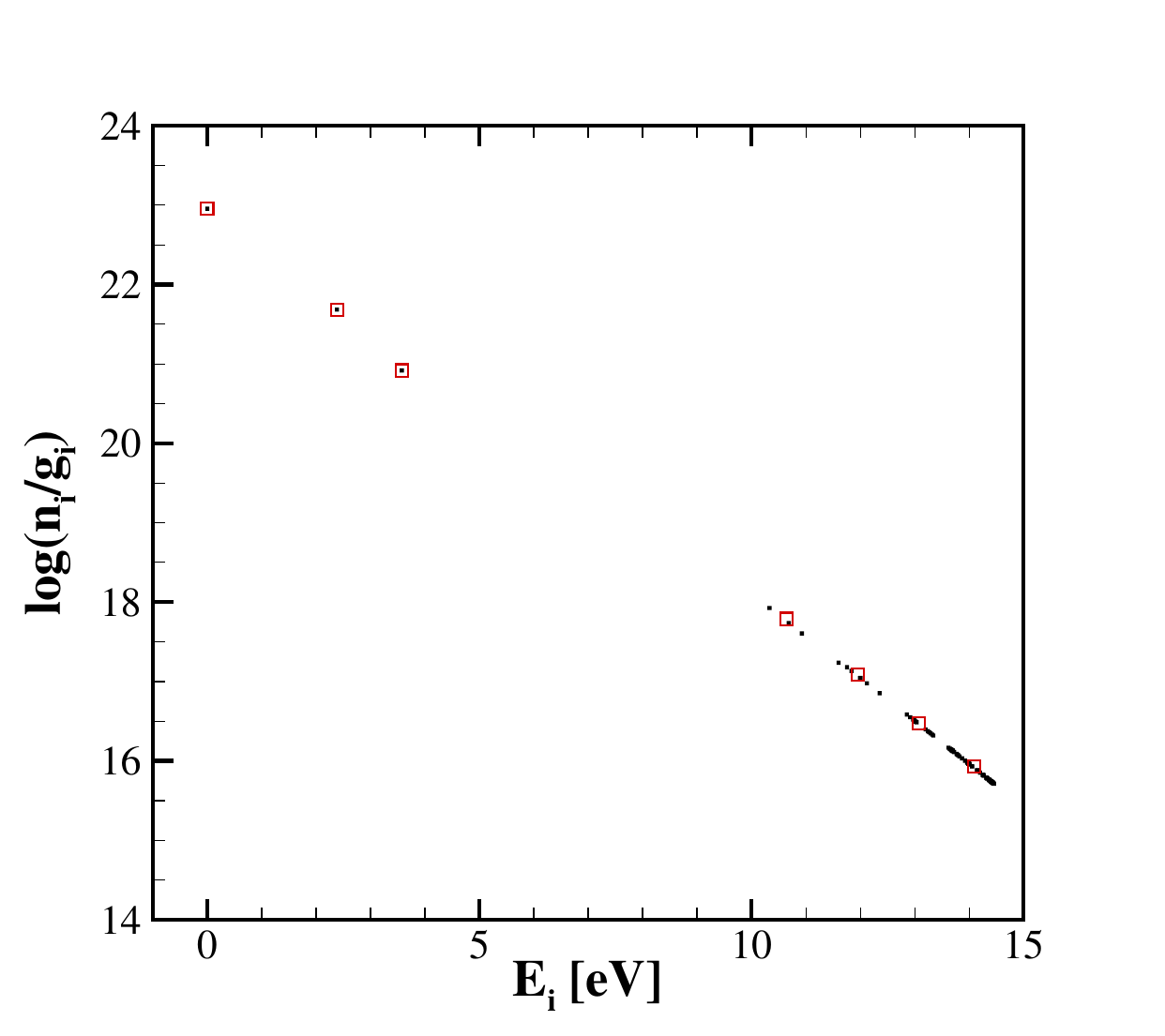}}
    \subfloat[][]{\includegraphics[scale=0.3]{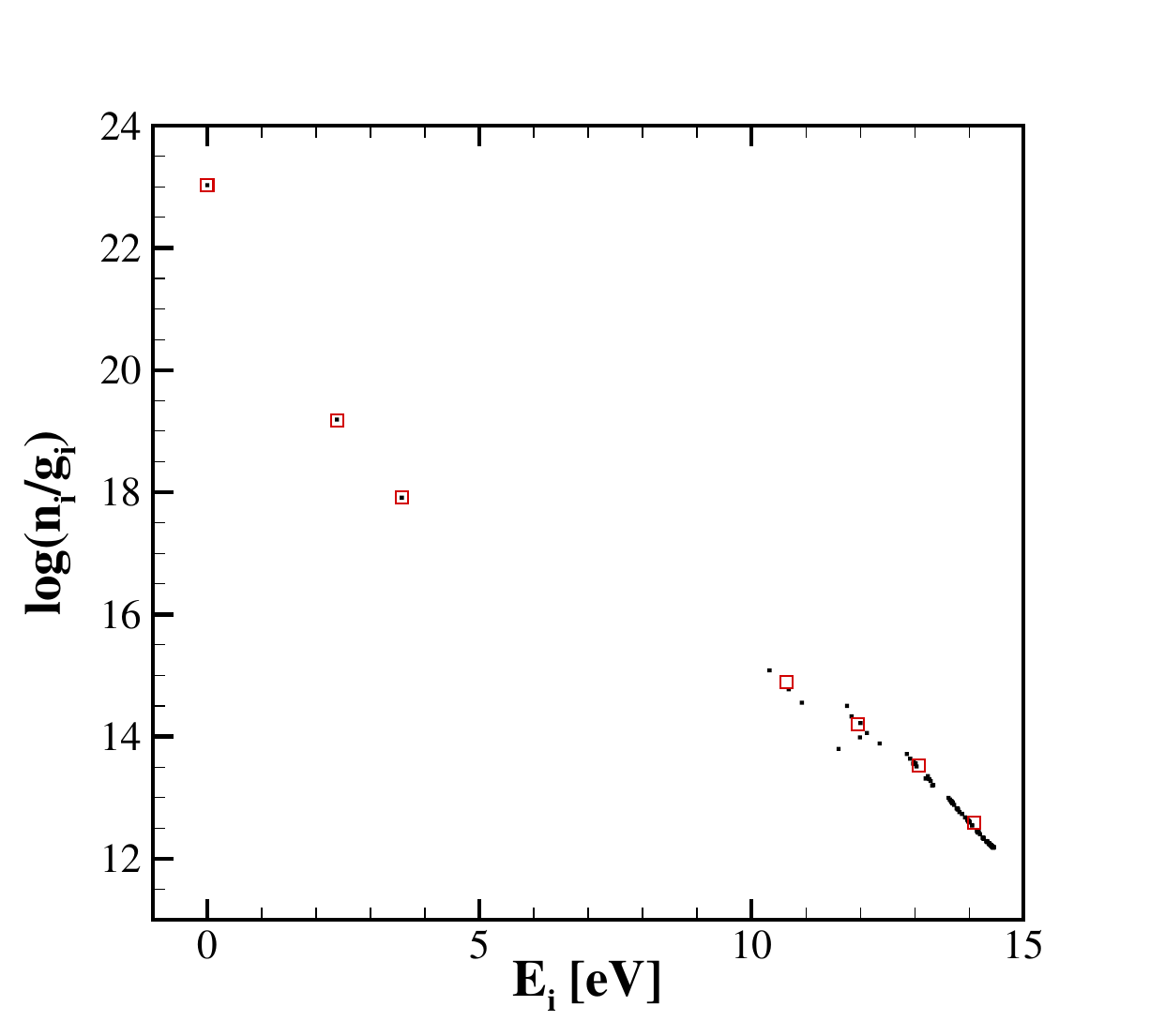}}   
    \subfloat[][]{\includegraphics[scale=0.3]{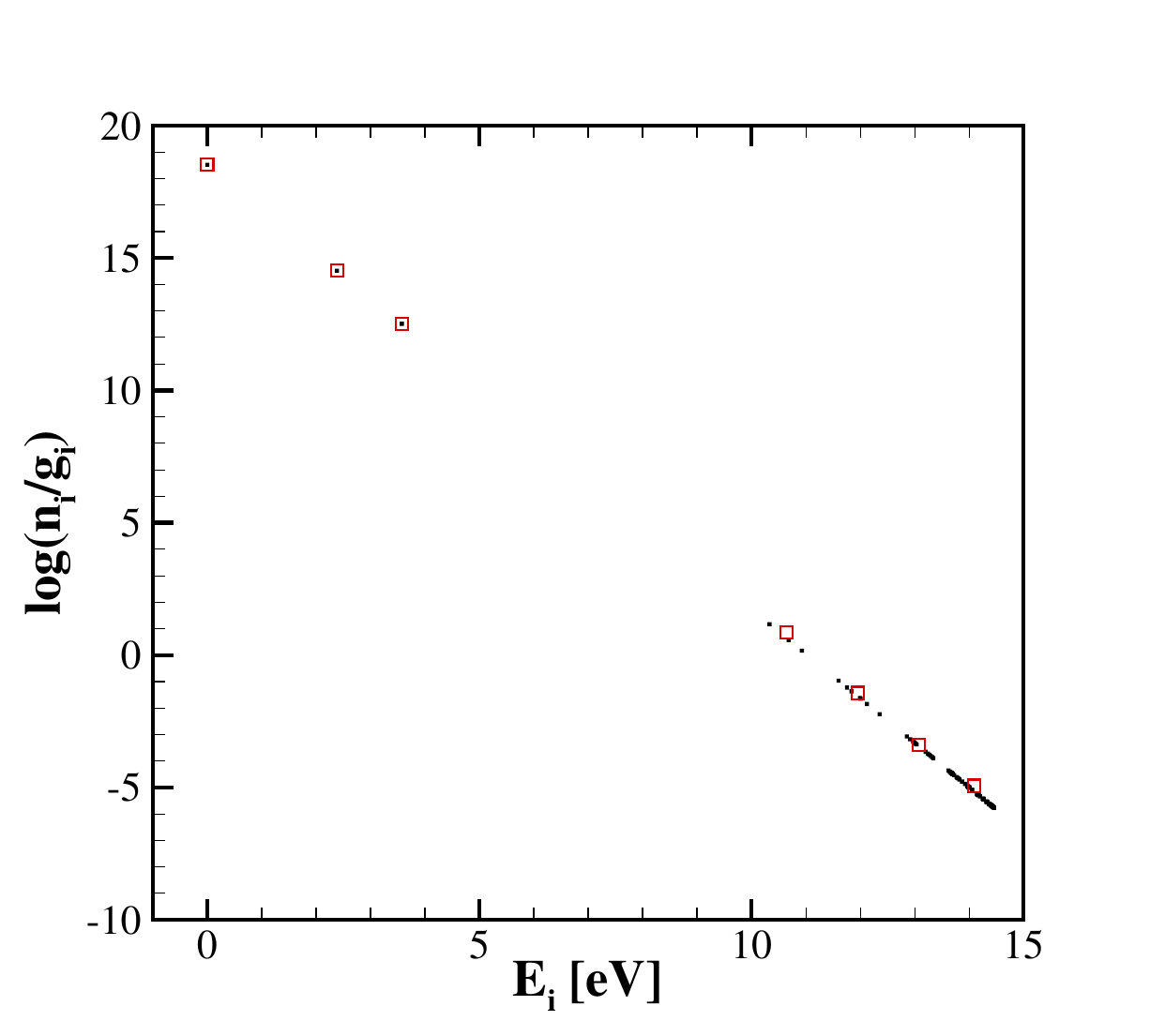}}
    \caption{Population distribution of N atom for 0D isochoric reactor under expansion at: (a) t = $3.13 \times 10^{-8}$s, (b) t = $1.15 \times 10^{-4}$s and (c) t = t\textsubscript{final}. Small dots represent the actual electronic states while the big squares represent the grouped states.}
    \label{fig:full_vs_reduced_pop_N_cooling}
    \end{figure}

  \cref{fig:full_vs_reduced_pop_N2_heating,fig:full_vs_reduced_pop_N_heating,fig:full_vs_reduced_pop_N2_cooling,fig:full_vs_reduced_pop_N_cooling} further show a comparison of the population distributions of N\textsubscript{2} and N at various times obtained using the full and the reduced StS model for both the compression and expansion cases re-affirming the ability of the grouped model to capture the dynamics of the full model.

    \subsection{ICP torch simulation}\label{sec:ICP_torch}
    This subsection investigates the non-equilibrium effects inside the ICP torch via simulations using the reduced vibronic StS model discussed in previous sections. The StS flowfield is compared against the LTE flowfield to highlight the extent of discrepancies due to the difference in the physico-chemical model. 
    
     \subsubsection{Problem description}\label{sec:problem_description}
\begin{figure}[!ht]
\centering
\subfloat[][]{\includegraphics[scale=0.2,valign=c]{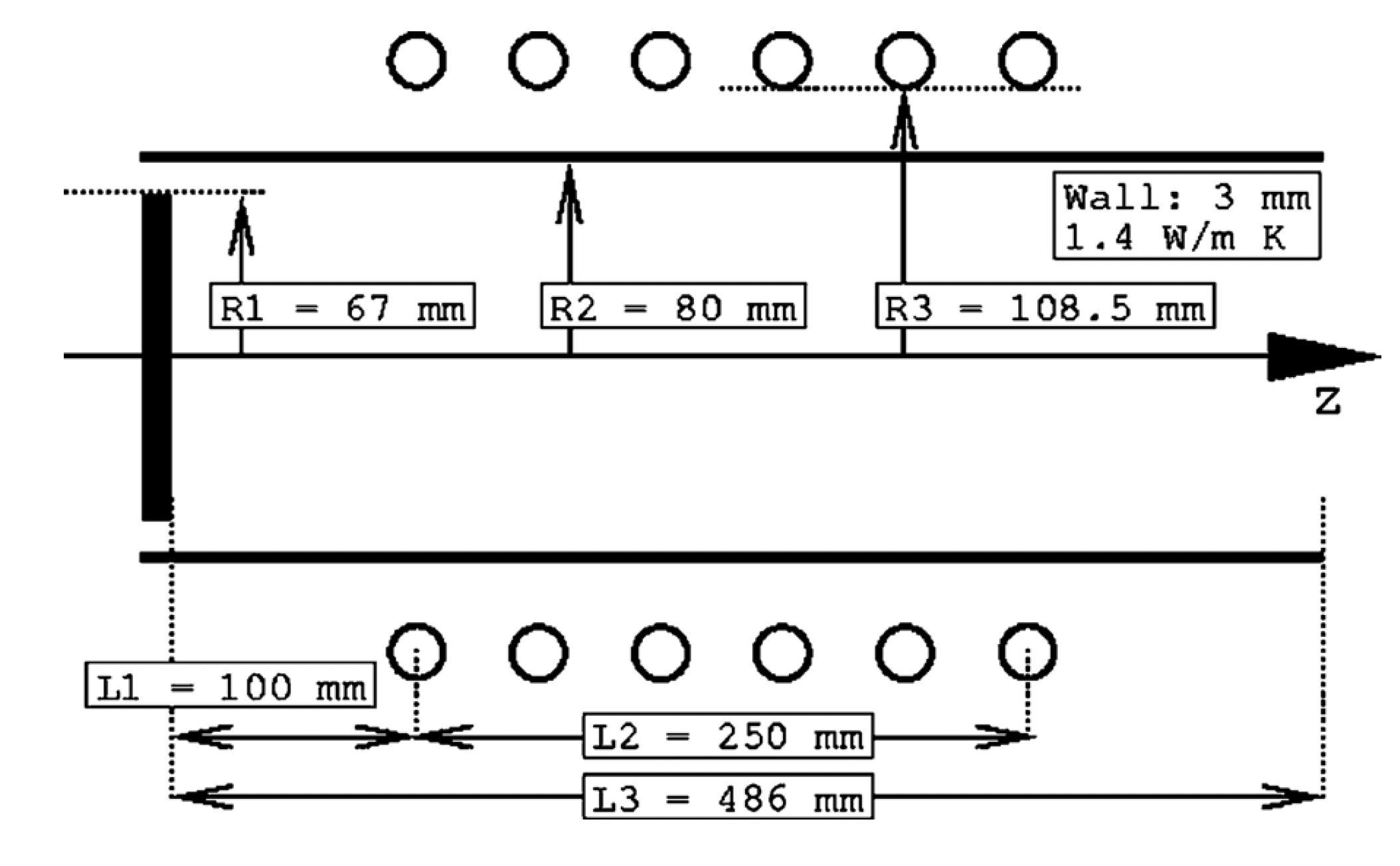}
\vphantom{\includegraphics[scale=0.175,valign=c]{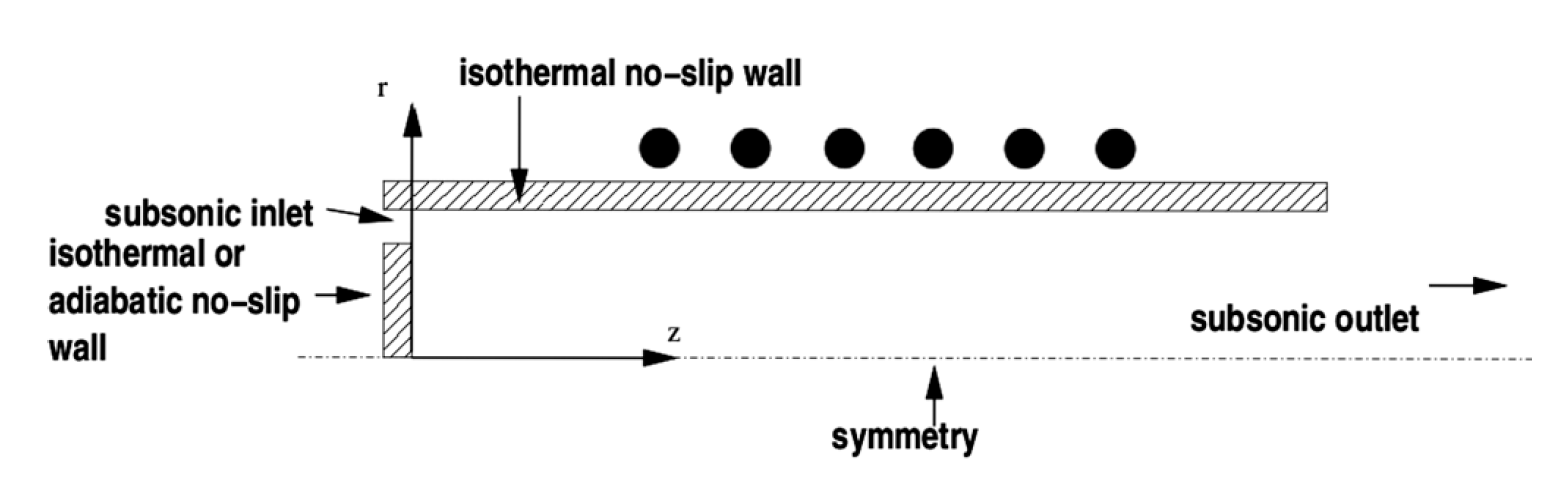}}}
\subfloat[][]{\includegraphics[scale=0.175,valign=c]{torch_bc.png}}
\caption{(a) Schematic of the ICP torch used for simulations (credits: Von Karman Institute for Fluid Dynamics\cite{abeele2000efficient}), (b) boundary conditions.}
\label{fig:torch}
\end{figure}

    For 2D axi-symmetric simulations of the ICP torch, a simplified geometry of the VKI torch\cite{abeele2000efficient} has been used in the current work as shown in \cref{fig:torch} (a). Corresponding torch boundary conditions are shown in \cref{fig:torch} (b). Cold gas is injected via a thin annular injector which gets heated by radio-frequency inductor coils. The coils are assumed to be parallel to enforce symmetry about the torch axis. The rings are assumed to be infinitely thin and are located at the innermost part of the coil cross-section which is a good approximation since most of the current is concentrated there due to skin-effect\cite{abeele2000efficient}. Current from the coils are modeled as point current sources as:
    \begin{equation}
    \mathbf{\tilde{J}}_{\mathrm{s}}= \mathbf{J}_0 \exp\left({i \omega t}\right) \sum_{i=1}^{N_{\mathrm{c}}}\delta(\mathbf{r} - \mathbf{r}_i) \, \mathbf{e}_{\theta},
    \end{equation}
    where $\omega$ denotes the angular frequency of the current running through the $N_{\mathrm{c}}$ inductor coils, whereas $\mathbf{r}_i$ is the location of the center of the $i$-th coil. The symbols $\imath$ and $\delta$ stand for the imaginary unit and Dirac's delta function, respectively. The frequency of the coils is \SI{0.45}{MHz}. The operating conditions for the simulation are as follows: mass flow \SI{6}{g/s}, pressure \SI{1000}{Pa}, and power \SI{50}{kW}.
    \\
    
    A structured grid consisting of 100$\times$50 cells has been used for all the simulations which has been found to be sufficient to provide grid-converged solutions. The implementation of the boundary conditions is as follows:
    \begin{itemize}
    \item inlet (subsonic):
$$
\rho u=\frac{\dot{m}}{A}, \quad  y_s = y_{a,s}, \quad \frac{\partial p}{\partial x}=0, \quad T_h=T_{a} \quad \text{and} \quad T_{e}=T_{a},
$$
where $A$ denotes the area of the annular injector, $y_s$ denotes the mass fraction of species $s$ and the subscript $a$ denotes the ambient conditions.
\item centerline (symmetry):
$$
\frac{\partial \rho_s}{\partial r}=\frac{\partial u}{\partial r}=\frac{\partial p}{\partial r}=0 \quad \text{and} \quad v=0.
$$
\item walls (isothermal):
$$
u=v=0, \quad T_h=T_w \quad \text{and} \quad T_{e}=T_w.
$$
\item outlet (subsonic):
$$
p=p_{\infty} .
$$
\end{itemize} 

where, $T_w$ and $T_{a}$ are taken to be \SI{350}{K} for all the simulations.

    \begin{figure}[!htb]
    \hspace*{-0.75cm}
    \subfloat[][]{\includegraphics[scale=0.16]{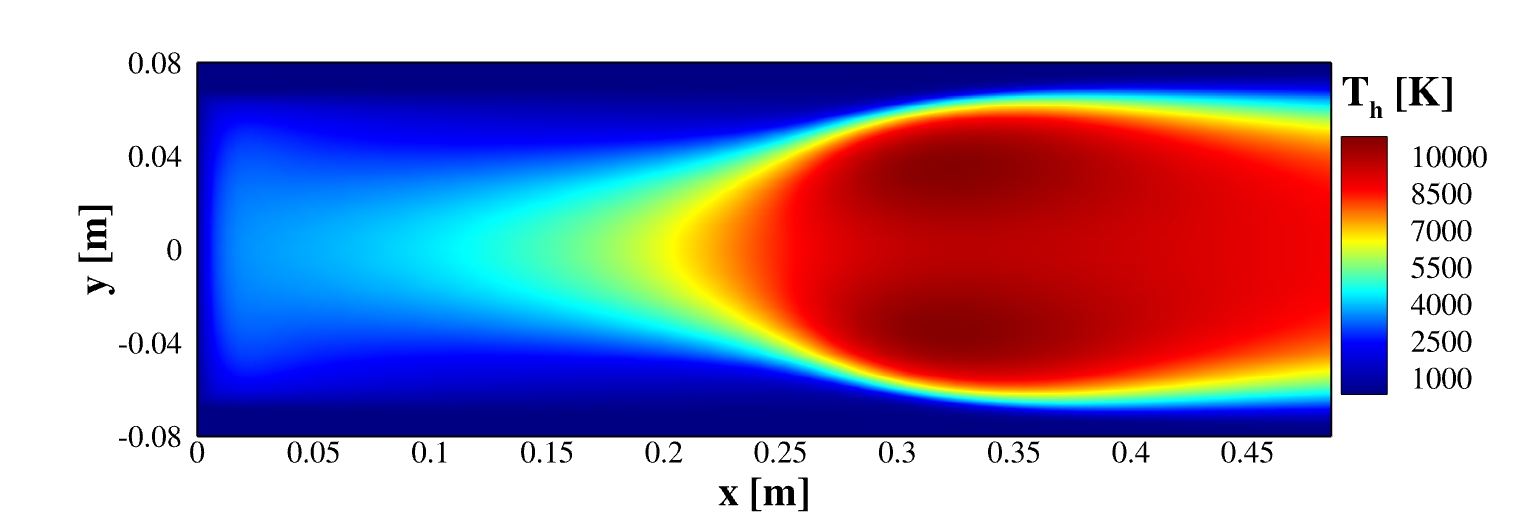}}
    \subfloat[][]{\includegraphics[scale=0.16]{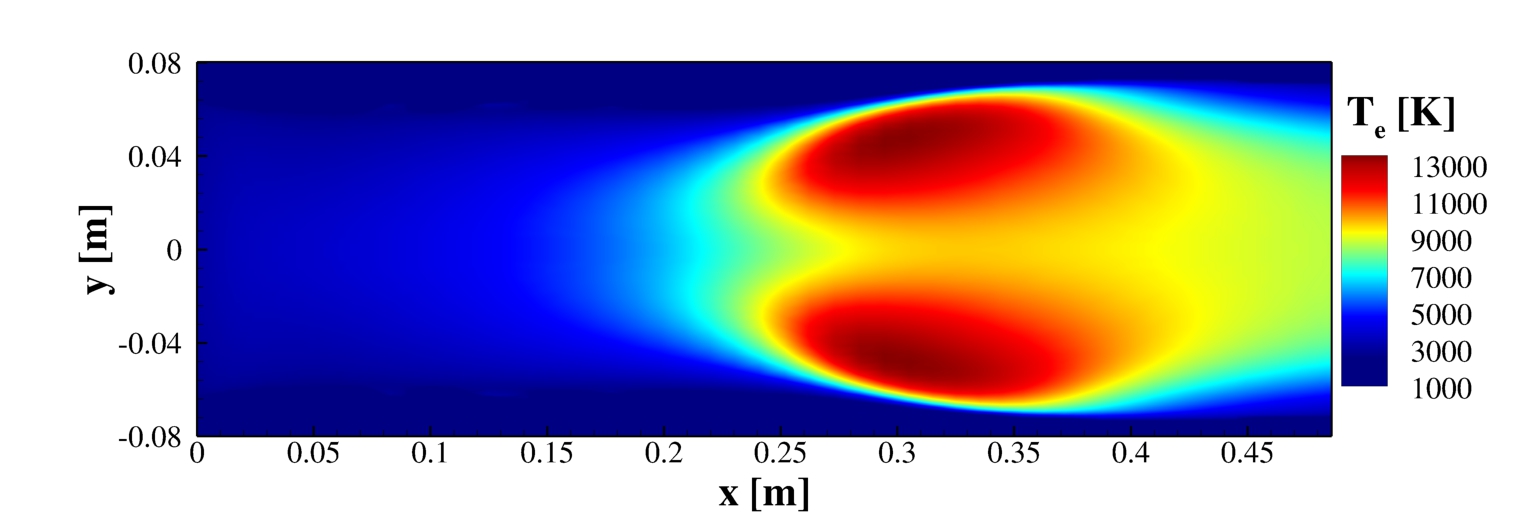}}
    \\
    \hspace*{-0.75cm}
    \subfloat[][]{\includegraphics[scale=0.16]{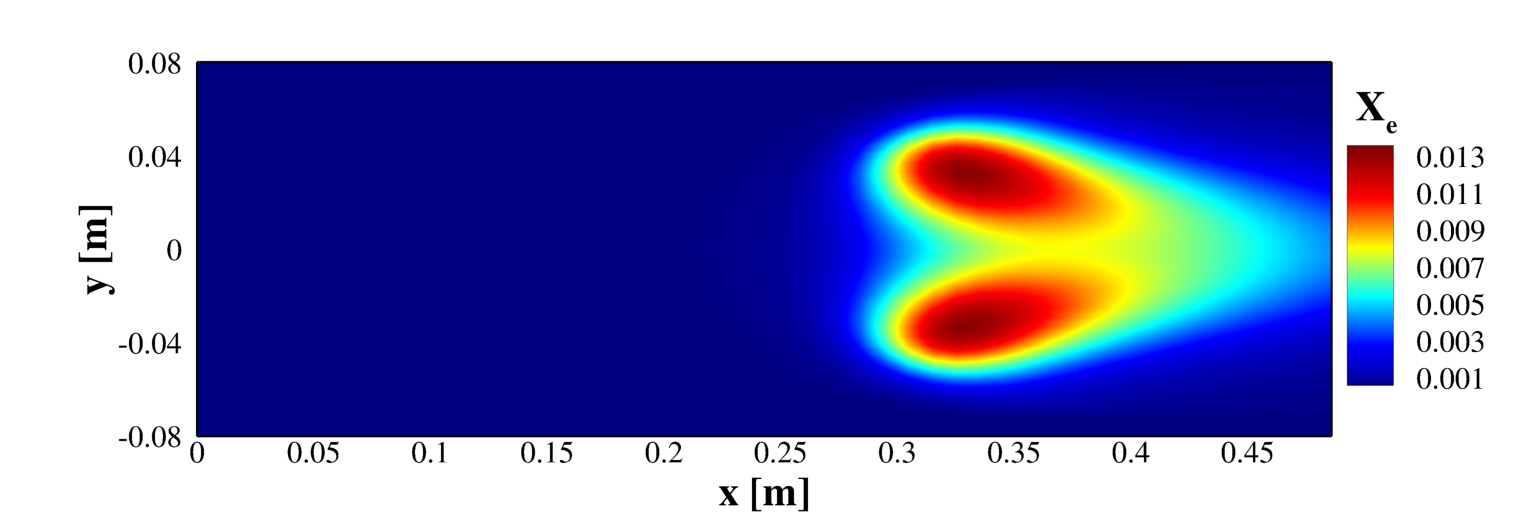}}
    \subfloat[][]{\includegraphics[scale=0.16]{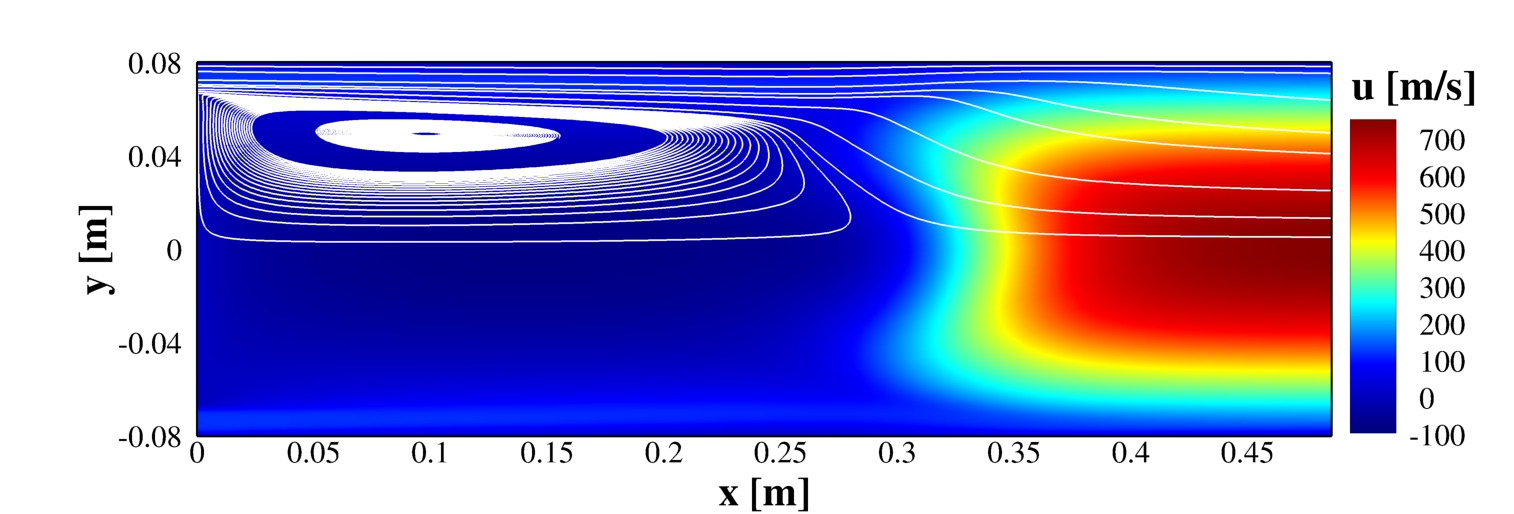}}
    \caption{Plasma flowfield inside the ICP torch obtained from vibronic StS simulation: (a) heavy-species temperature, (b) electron-temperature, (c) electron mole-fraction, and (d) velocity. Operating conditions: \SI{1000}{Pa}, \SI{50}{kW} and \SI{6}{g/s}. }
    \label{fig:sts_flow_contours}
    \end{figure}

        \begin{figure}[!htb]
    \hspace*{-0.75cm}
    \subfloat[][]{\includegraphics[scale=0.16]{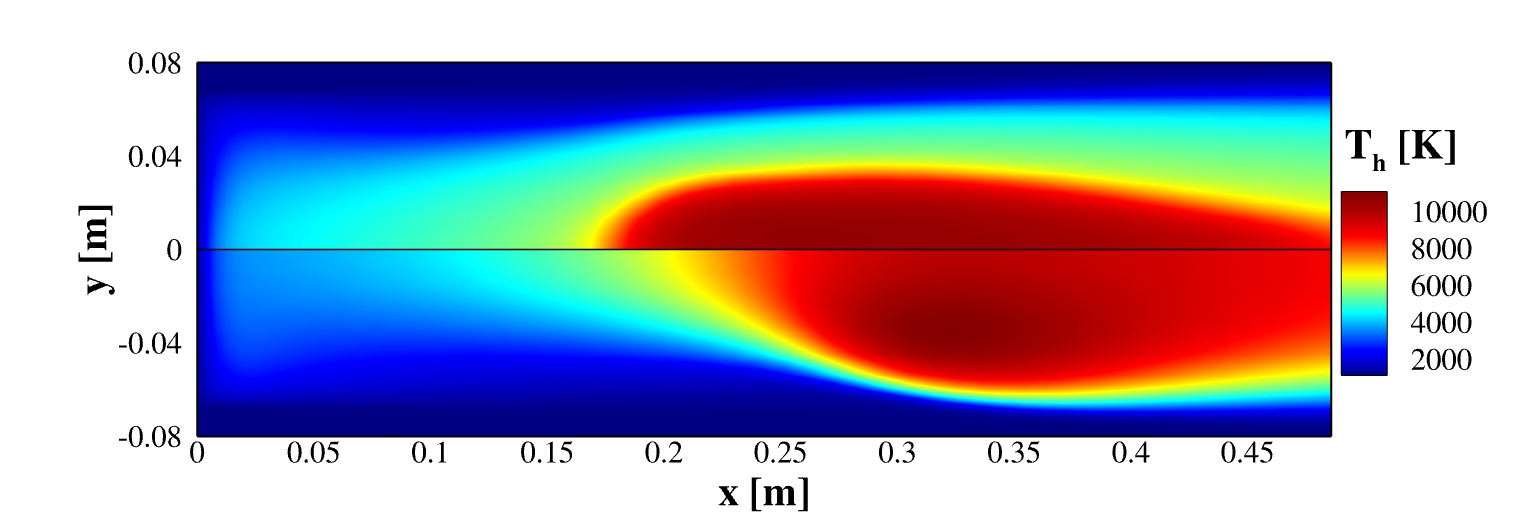}}
    \subfloat[][]{\includegraphics[scale=0.16]{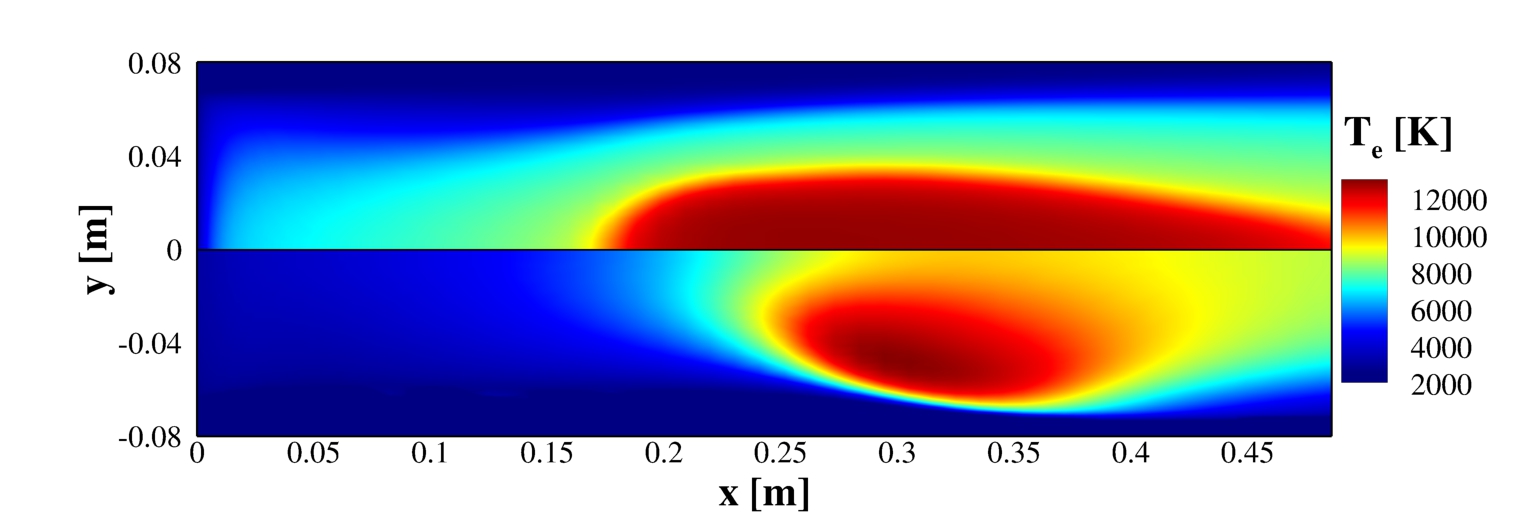}}
    \caption{Comparison of temperature fields obtained using LTE and vibronic StS simulations: (a) heavy-species temperature and (b) electron temperature. Top: LTE, bottom: vibronic StS. Operating conditions: \SI{1000}{Pa}, \SI{50}{kW} and \SI{6}{g/s}.}
    \label{fig:lte_vs_vibronic_contours}
    \end{figure}

    \subsubsection{Vibronic StS versus LTE flowfield}\label{sec:sts_vs_lte_ICP}
    
\begin{figure}[!htb]
    \centering
    \subfloat[][]{\includegraphics[scale=0.5]{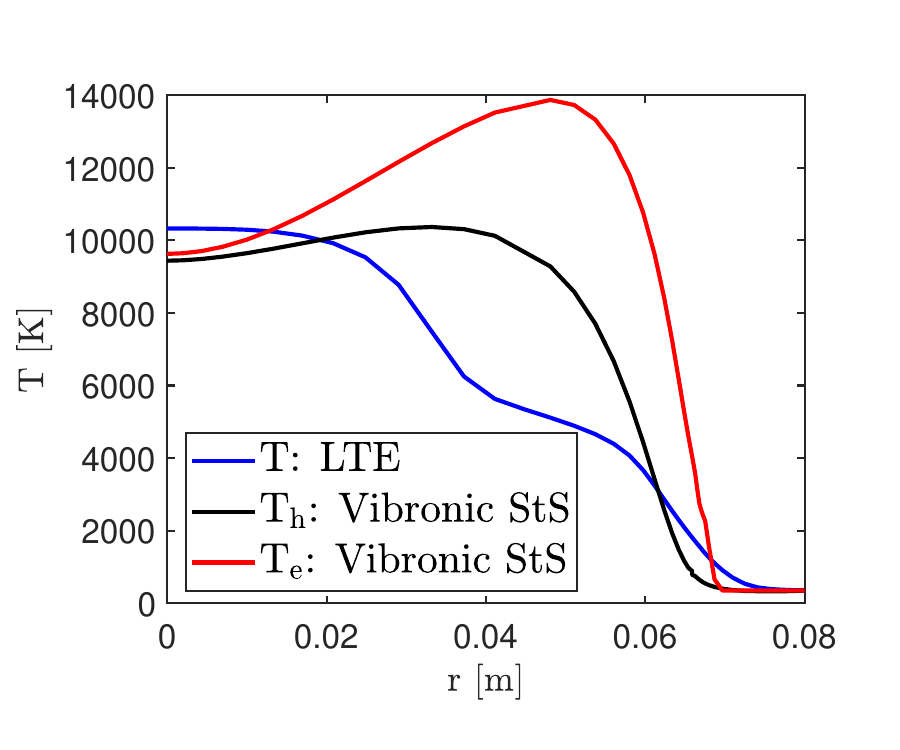}}
    \subfloat[][]{\includegraphics[scale=0.5]{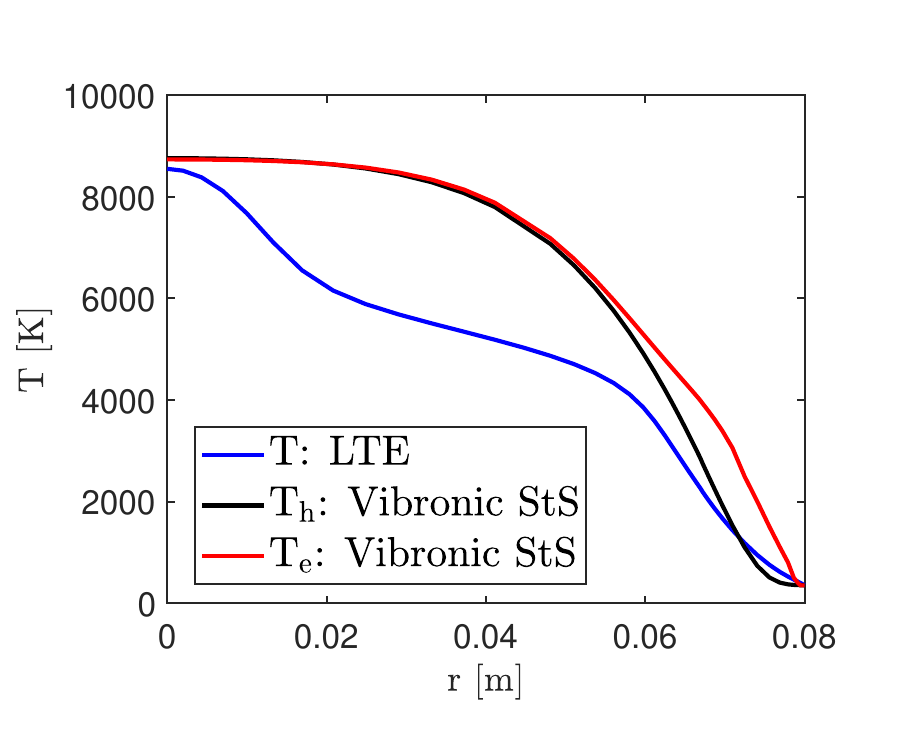}}
    \caption{Radial temperature profiles at: (a) x = \SI{0.3}{m} (mid-torch location) and (b) x = \SI{0.485}{m} (torch outlet). Operating conditions: \SI{1000}{Pa}, \SI{50}{kW} and \SI{6}{g/s}. }
    \label{fig:T_profiles_lte_vs_vibronic}
    \end{figure}

\begin{figure}[!htb]
    \hspace*{-0.75cm}
    \subfloat[][]{\includegraphics[scale=0.27,clip, trim=0.25in 1in 0.5in 1in]{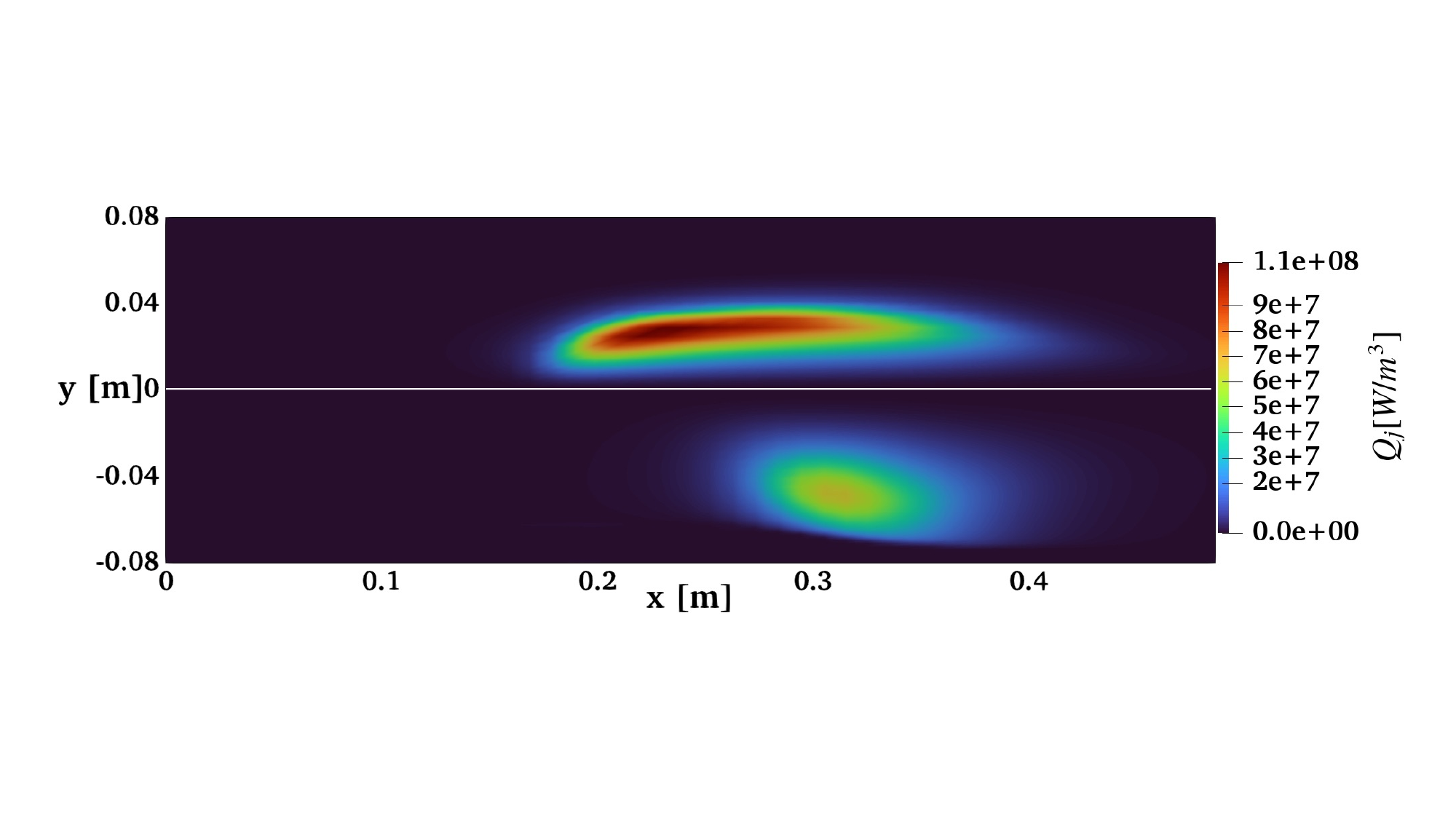}}
    \subfloat[][]{\includegraphics[scale=0.27,clip, trim=0.25in 1in 0.5in 1in]{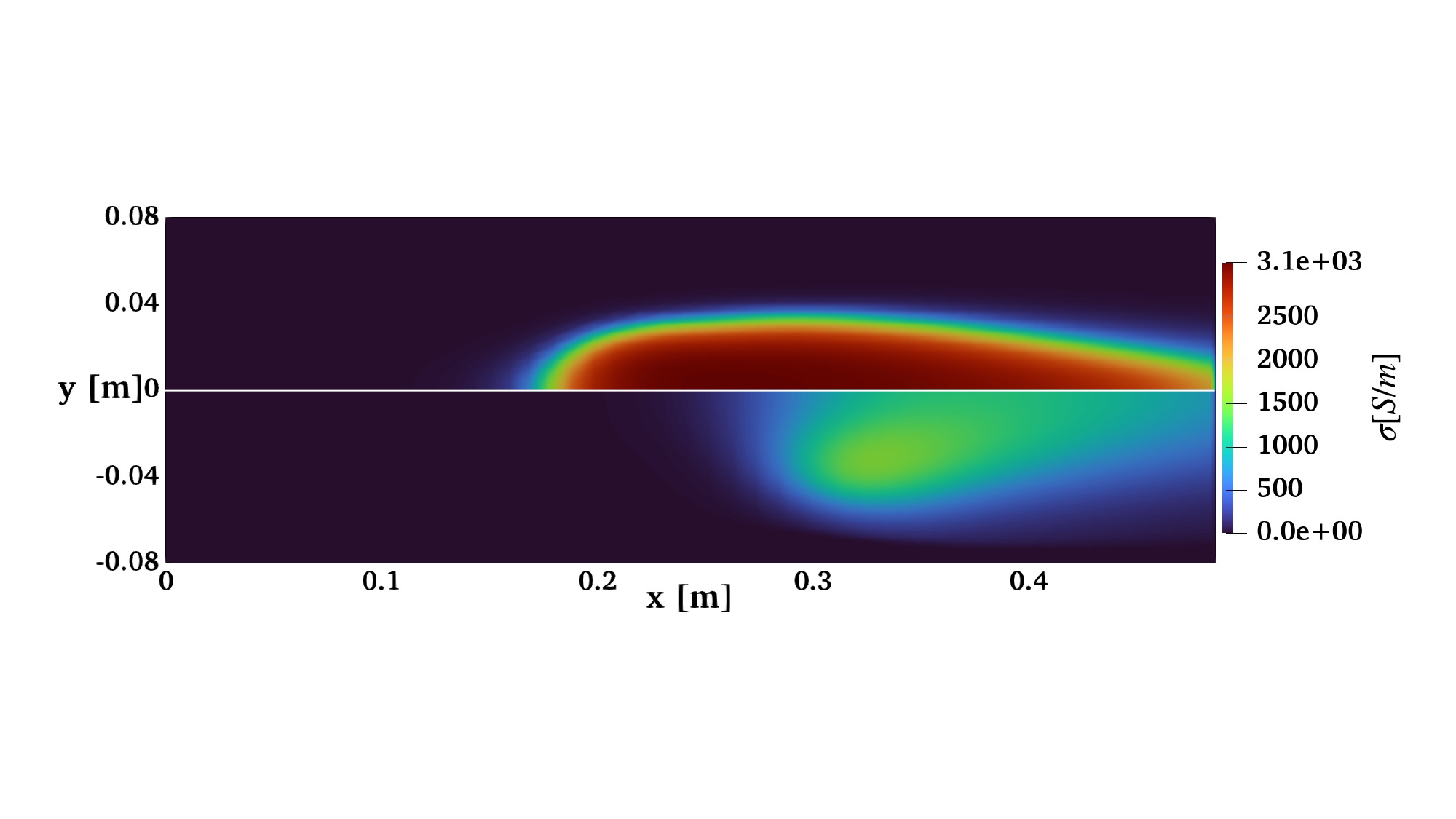}}
    \caption{(a) Joule heating [$W/m^3$] and (b) electrical conductivity [$S/m$] distribution inside the ICP torch obtained using LTE and vibronic StS simulation. Top: LTE, bottom: vibronic StS. Operating conditions: \SI{1000}{Pa}, \SI{50}{kW} and \SI{6}{g/s}.}
    \label{fig:lte_vs_sts_EM_contours}
    \end{figure}

\begin{figure}[!htb]
    \centering
    \subfloat[][]{\includegraphics[scale=0.5]{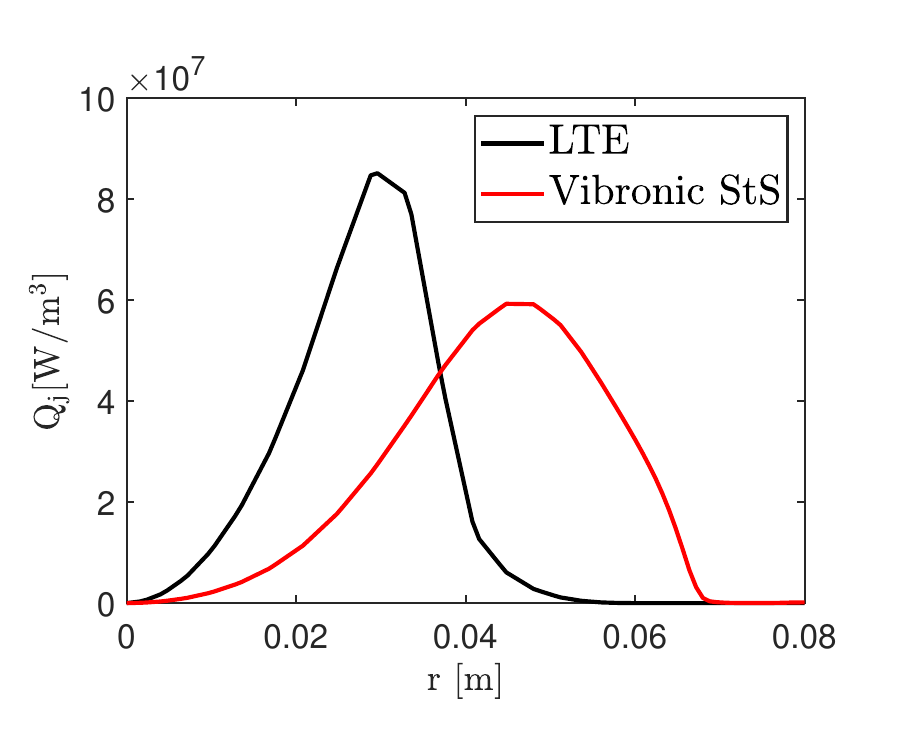}}
    \subfloat[][]{\includegraphics[scale=0.5]{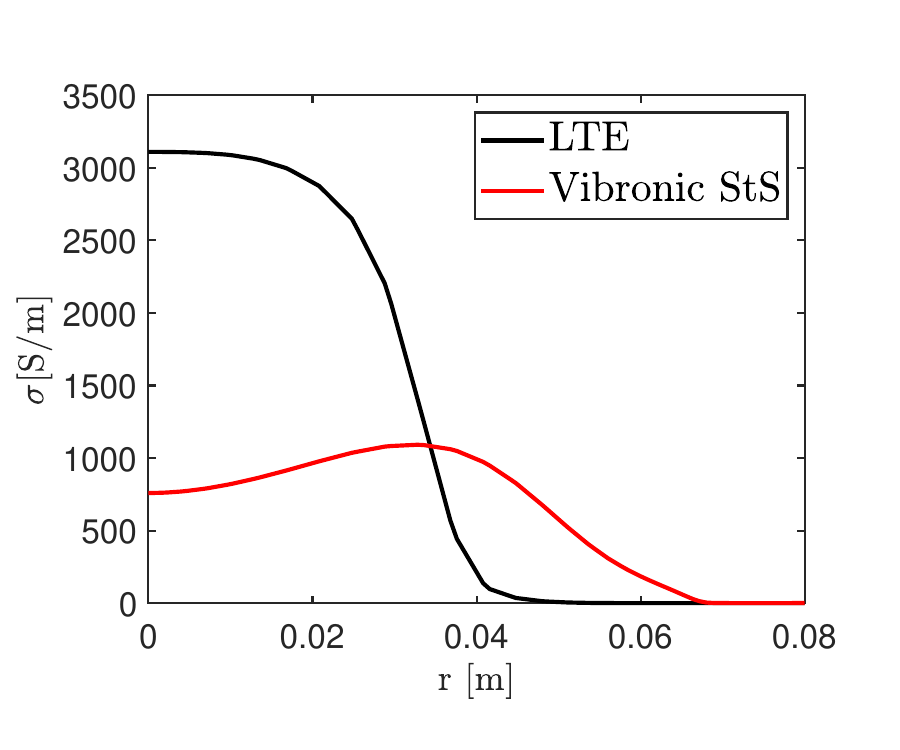}}
    \caption{(a) Joule heating and (b) electrical conductivity profiles at x = \SI{0.3}{m} (mid-torch location). Operating conditions: \SI{1000}{Pa}, \SI{50}{kW} and \SI{6}{g/s}. }
    \label{fig:EM_profiles_lte_vs_vibronic}
    \end{figure}

     Non-local thermodynamic equilibrium (NLTE) simulation was performed using the vibronic StS model for a nitrogen ICP torch for the operating conditions mentioned above. Also, LTE simulation was conducted for the same conditions to compare against the NLTE results. \cref{fig:sts_flow_contours} shows the plasma flow field inside the torch obtained from vibronic StS simulation. The streamlines show a large re-circulation region in the torch which is a typical flow feature of the ICP torches\cite{abeele2000efficient,zhang2016analysis,munafoRGD2022}. The re-circulation helps in sustaining the plasma by pulling the hot plasma core towards the inlet and thus preventing it from getting advected away by the flow. The temperature contours show a significant difference between the heavy-species temperature (T\textsubscript{h}) and the free-electron temperature (T\textsubscript{e}) showcasing significant non-equilibrium conditions in the torch. \cref{fig:lte_vs_vibronic_contours} compares the thermal field obtained from vibronic StS simulation against the one obtained using an LTE simulation. The vibronic StS simulation gives a significantly different thermal field in the torch as compared to the LTE simulation which is physically consistent. The energy from the coils is dissipated into the plasma through the interaction of the electromagnetic field with the electrons. At low pressures such as those used in the current investigation, the electrons in the coil region are not able to equilibrate with the heavy-species due to very low collisional frequency. As a result, a large difference between heavy-species and electron temperatures is observed at the mid-torch location where the effect of the coil is maximum, as shown in \cref{fig:T_profiles_lte_vs_vibronic}(a). Also, the extent of non-equilibrium is the highest at around r = \SI{0.05}{m} which happens to be the radial location where peak Joule heating occurs as shown in \cref{fig:lte_vs_sts_EM_contours}(a) and \cref{fig:EM_profiles_lte_vs_vibronic}(a). This re-affirms that the source of the non-equilibrium is indeed the heating due to the coils. As we move away from the coils, thermal equilibrium between heavy species and electrons starts to prevail. However, the effect of NLTE propagates towards the outlet, and even though the thermal equilibrium exists, the temperature profile is still very different from the one given by the LTE simulation which is evident from \cref{fig:T_profiles_lte_vs_vibronic} (b). The difference between the StS and LTE temperature profiles at the torch outlet impacts the prediction of the plasma state for TPS material testing. Also, the peak plasma temperature in the case of LTE simulation occurs at the axis whereas it occurs somewhere in between the axis and the top wall for the vibronic StS simulation. This can be explained from \cref{fig:lte_vs_sts_EM_contours} (a) which shows that the peak of the Joule heating distribution lies closer to the axis in the case of LTE simulation whereas it lies much further away from the axis in the case of vibronic StS simulation. These results show that the LTE model is unable to capture the excitation of the internal energy modes for ICPs at such low-pressure conditions and needs accurate NLTE models for a correct description of the plasma. 

    \begin{figure}[!htb]
    \centering
    \subfloat[][]{\includegraphics[scale=0.5]{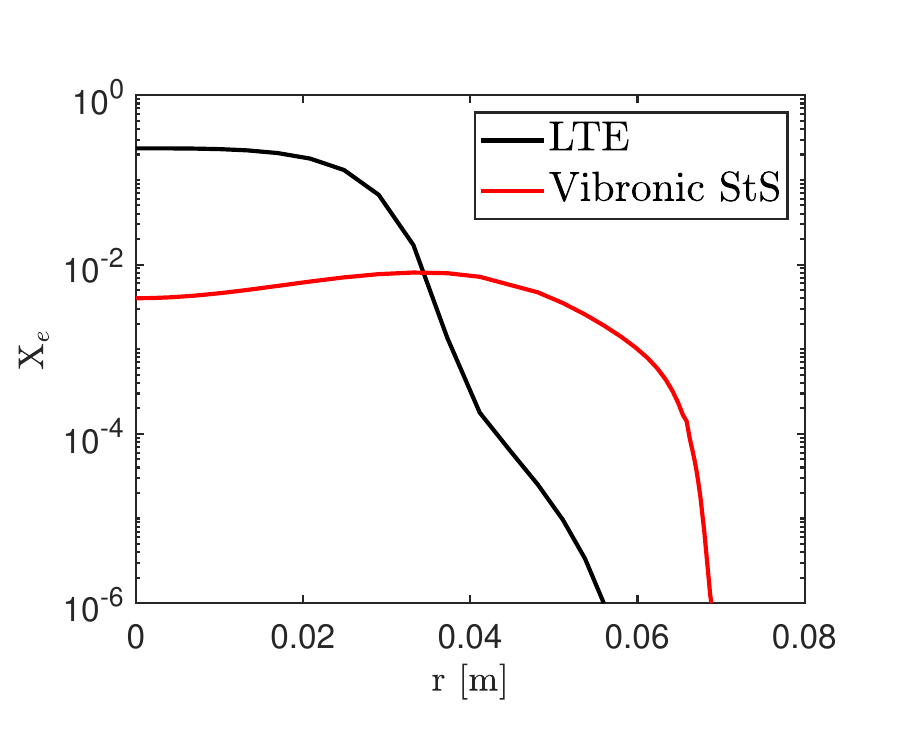}}
    \subfloat[][]{\includegraphics[scale=0.5]{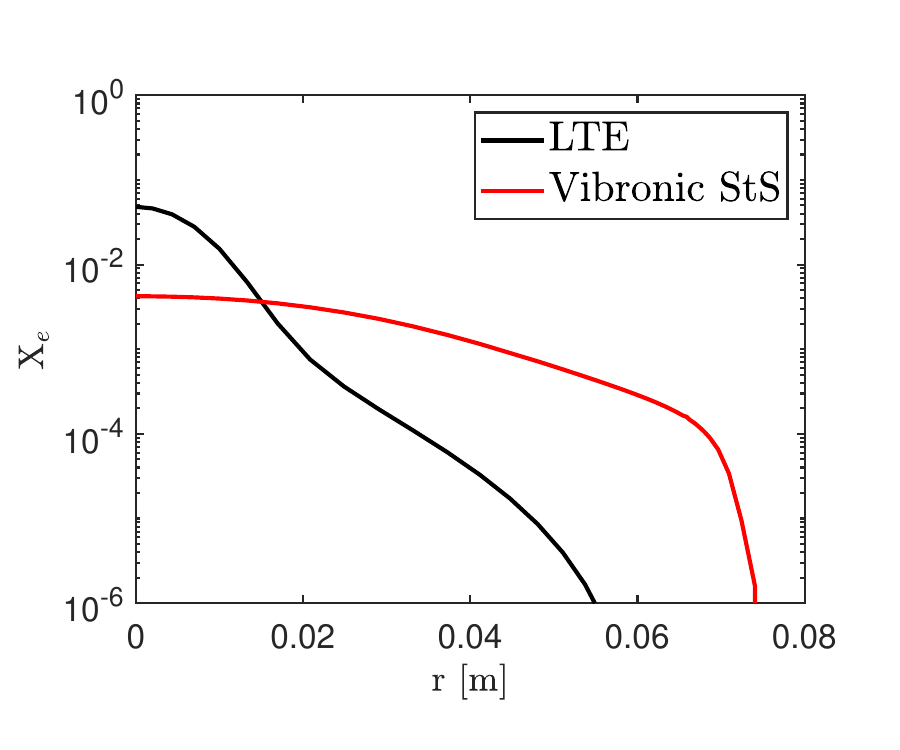}}
    \caption{Radial electron mole-fraction profiles at: (a) x = \SI{0.3}{m} (mid-torch location) and (b) x = \SI{0.485}{m} (torch outlet). Operating conditions: \SI{1000}{Pa}, \SI{50}{kW} and \SI{6}{g/s}. }
    \label{fig:Xe_profiles_lte_vs_vibronic}
    \end{figure}

    \begin{figure}[!htb]
        \hspace*{-5cm}
        \includegraphics[scale=0.75]{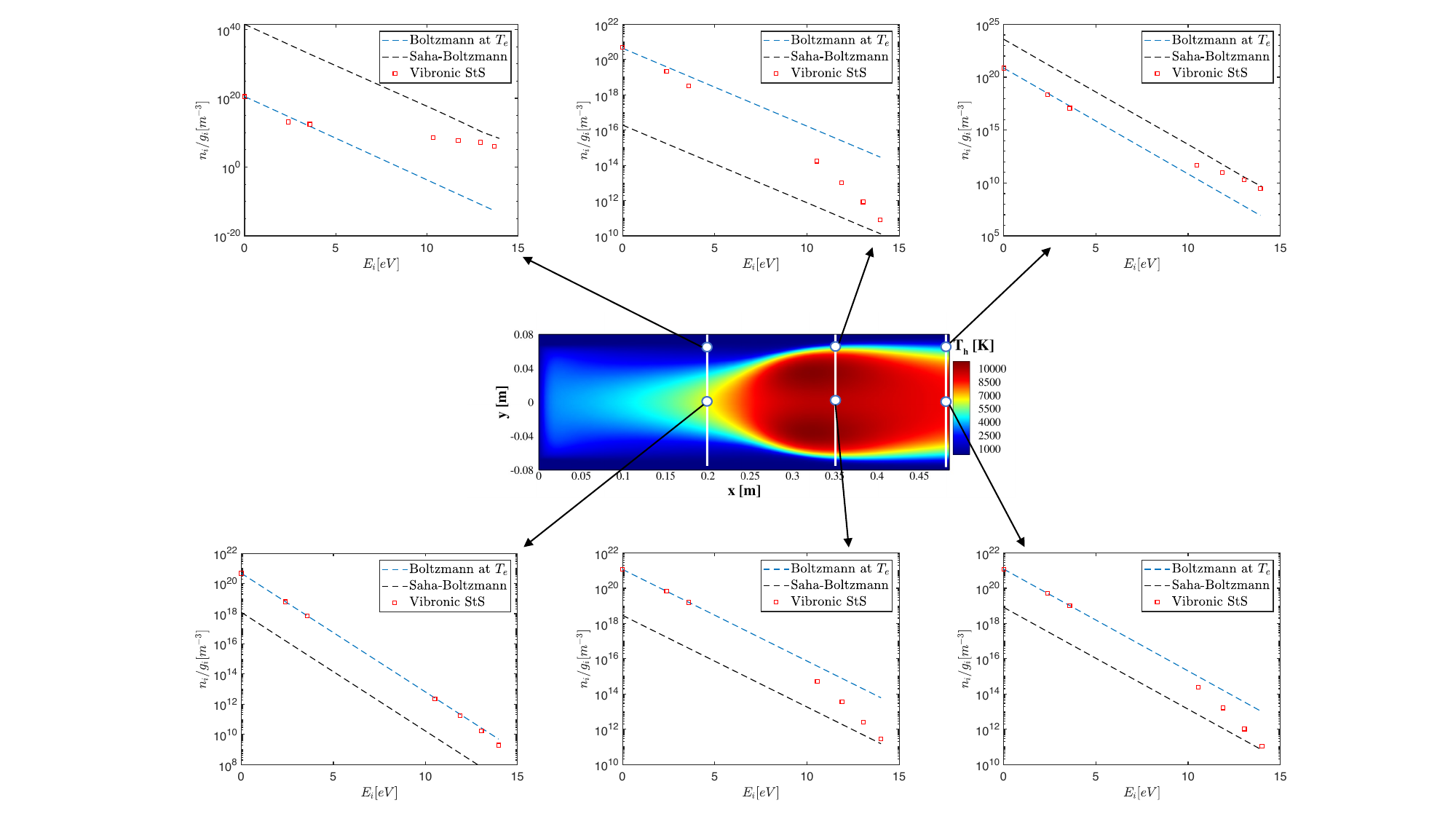}
        \caption{Population distribution of N atom at various locations in the torch: along the axis and $r = \SI{0.06}{m}$ (vicinity of the cold wall). Operating conditions: \SI{1000}{Pa}, \SI{50}{kW} and \SI{6}{g/s}.}
        \label{fig:pop_N}
    \end{figure}
    
    \begin{figure}[!htb]
        \hspace*{-5cm}
        \includegraphics[scale=0.75]{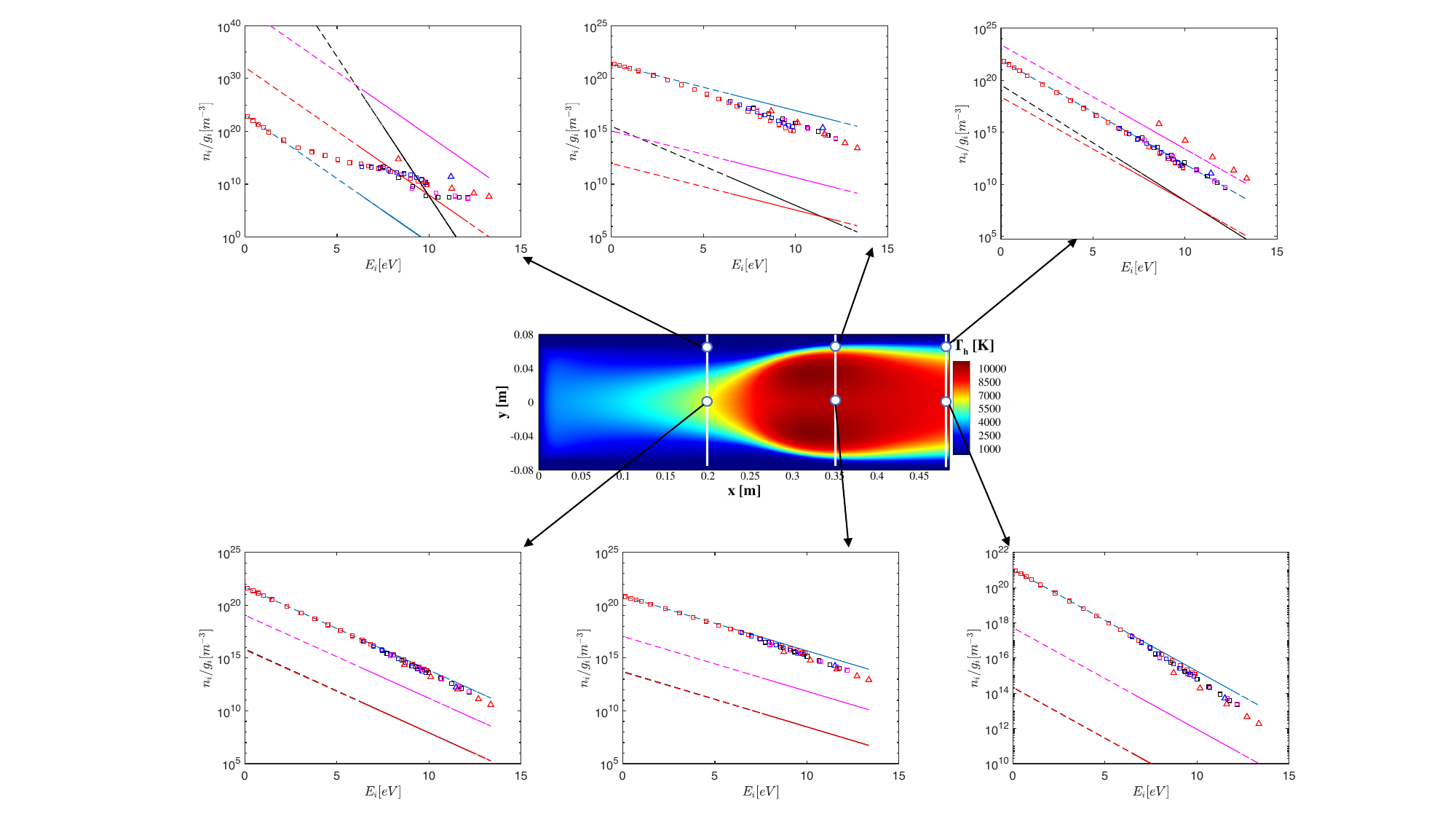}
        \caption{Population distribution of N\textsubscript{2} at various locations in the torch: along the axis and $r = \SI{0.06}{m}$ (vicinity of the cold wall). Operating conditions: \SI{1000}{Pa}, \SI{50}{kW} and \SI{6}{g/s}. Dashed blue line: Boltzmann at $\mathrm{T}_\mathrm{e}$, dashed red line: dissociation equilibrium at $\mathrm{T}_\mathrm{e}$, dashed black line: dissociation equilibrium at $\mathrm{T}_\mathrm{h}$, dashed magenta line: Saha-Boltzmann, red squares: StS ($\mathrm{N}_2$(X)), blue squares: StS ($\mathrm{N}_2$(A)), black squares: StS ($\mathrm{N}_2$(B)), magenta squares: StS ($\mathrm{N}_2$(W)), red triangles: StS ($\mathrm{N}_2$(B')), and blue triangle: StS ($\mathrm{N}_2$(C))  }
        \label{fig:pop_N2}
    \end{figure}

        \begin{figure}[!htb]
        \hspace*{-5cm}
        \includegraphics[scale=0.75]{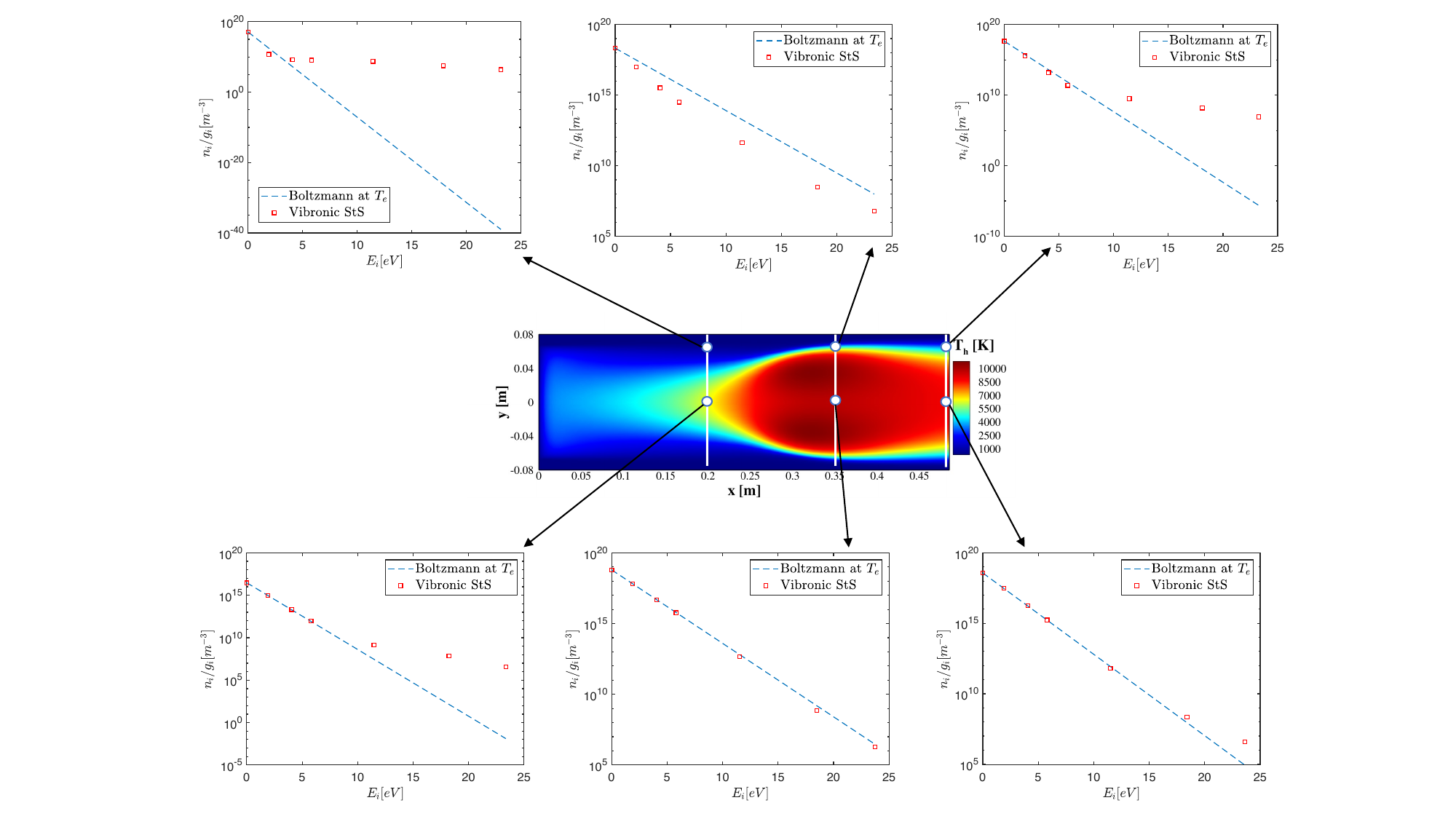}
        \caption{Population distribution of N\textsuperscript{+} at various locations in the torch: along the axis and $r = \SI{0.06}{m}$ (vicinity of the cold wall). Operating conditions: \SI{1000}{Pa}, \SI{50}{kW} and \SI{6}{g/s}.}
        \label{fig:pop_Np}
    \end{figure}

        \begin{figure}[!htb]
        \hspace*{-5cm}
        \includegraphics[scale=0.75]{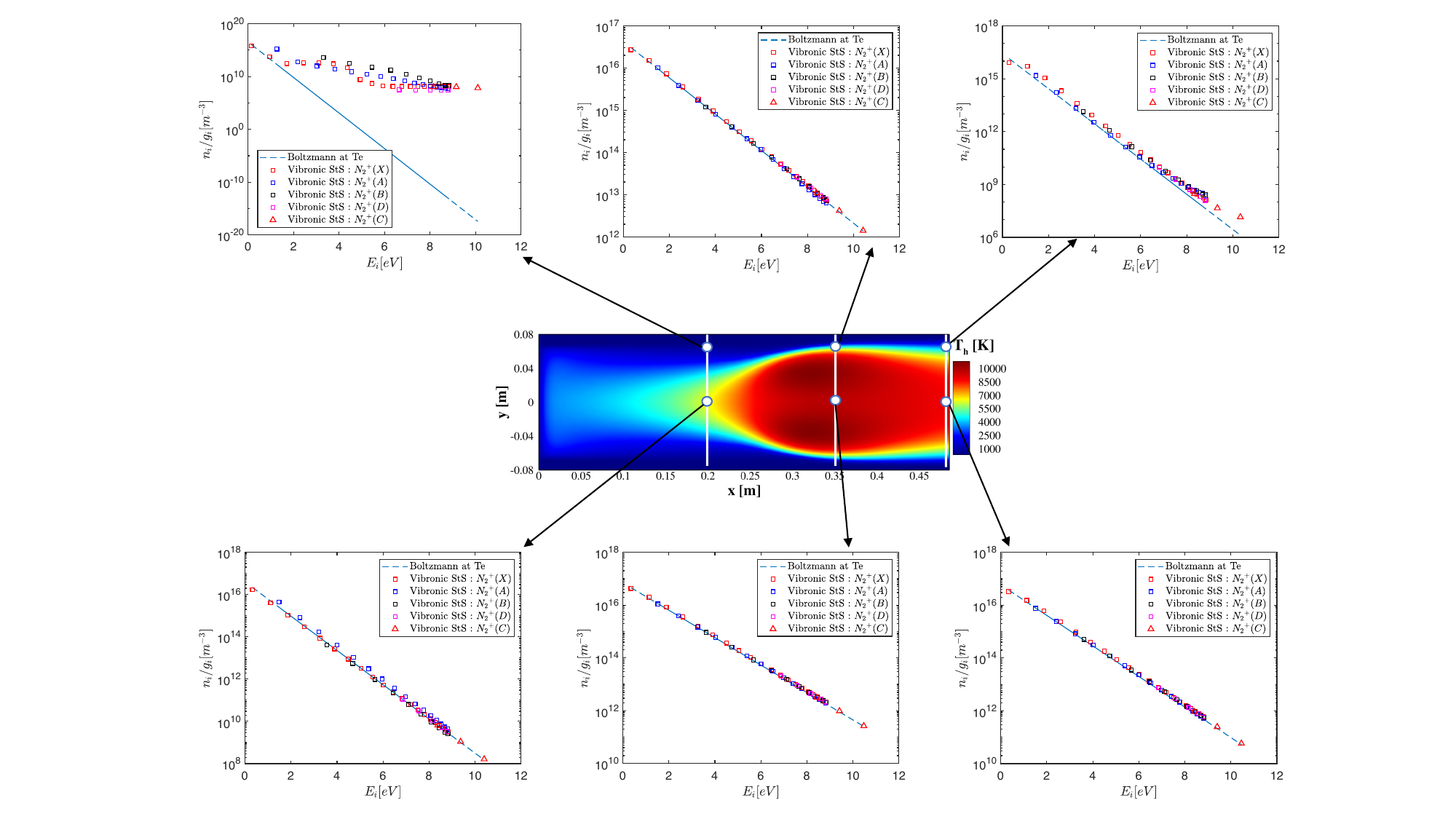}
        \caption{Population distribution of $\mathrm{N}_2^+$ at various locations in the torch: along the axis and $r = \SI{0.06}{m}$ (vicinity of the cold wall)). Operating conditions: \SI{1000}{Pa}, \SI{50}{kW} and \SI{6}{g/s}.}
        \label{fig:pop_N2p}
    \end{figure}

    \subsubsection{Dynamics of the internal state populations}
    \cref{fig:pop_N2,fig:pop_N2p,fig:pop_N,fig:pop_Np} present the population distribution of the vibronic states of molecules and electronic states of atoms at various locations in the torch obtained from the state-to-state simulation for the same operating conditions as mentioned above. \cref{fig:pop_N} indicates that the electronic population distributions of N atom show a significant distortion from the Boltzmann distribution at electron temperature (T\textsubscript{e}) at most of the locations. The excited electronic levels at the axial locations in the hot plasma core (x = \SI{0.35}{m} and torch outlet) are severely under-populated with respect to the Boltzmann distribution due to ionization non-equilibrium. In the colder regions near the cold wall, the high-lying states are over-populated with respect to Boltzmann distribution due to recombination non-equilibrium. Ionization non-equilibrium leads to strong depletion of the excited states, while the recombining condition induces an overpopulation of the excited states. Due to the high efficiency of ionization/recombination processes involving the high-lying states as compared to the excitation/de-excitation processes and ionization/recombination processes involving the low-lying states, the high-lying states closer to the ionization limit tend to reach Saha-equilibrium with the free electrons before reaching Maxwell-Boltzmann equilibrium with lower states. In this case, however, it is observed that even the high-lying states are not in complete equilibrium with the Saha-Boltzmann line, indicating a very strong ionization and recombination non-equilibrium. For the axis location at x = \SI{0.2}{m}, the plasma is not hot enough to initiate ionization and hence it shows a complete Boltzmann distribution.

    \cref{fig:pop_N2} shows the population distributions of the vibronic states of $\mathrm{N}_2$ at various locations in the torch along with several other limits: Boltzmann at $\mathrm{T}_e$, dissociation equilibrium at $\mathrm{T}_e$, dissociation equilibrium at $\mathrm{T}_h$ and Saha-equilibrium. Similar to ionization (recombination) non-equilibrium, the high-lying states of molecules tend to get underpopulated due to dissociation non-equilibrium and overpopulated due to recombination non-equilibrium. Again, the high-lying states due to higher efficiency dissociation/recombination tend to reach the dissociation equilibrium limit before reaching the Boltzmann equilibrium with lower states. Along with dissociation non-equilibrium, the molecules further show ionization non-equilibrium similar to the atoms. In this case, at all the axial locations, the population distribution is close to the Boltzmann distribution with vibrational levels of only the excited electronic states showing mild under-population. In the cold region near the wall at x = \SI{0.2}{m}, severe over-population of excited states can be seen due to recombination non-equilibrium. At the torch outlet near the wall, it is interesting to observe that all the states follow Boltzmann distribution except $\mathrm{N}_2 (\mathrm{B}^\prime)$ which shows strong electronic non-equilibrium although having vibrational equilibrium within the electronic state. 

    \cref{fig:pop_Np} presents the population distribution of $\mathrm{N}^+$ showing mostly Boltzmann distribution in the hot plasma core while deviating from it in the colder recombination regions where large over-population of excited states is seen.
    \cref{fig:pop_N2p} shows the population distribution of the vibronic states of $\mathrm{N}_2^+$ which again follow Boltzmann-distribution while showing over-population only in the colder region near the wall at x = \SI{0.2}{m}. At the torch outlet however, the ground state and the first excited state show large deviation from Boltzmann slope which would give a different internal temperature for $\mathrm{N}_2^+$ as compared to $\mathrm{T}_\mathrm{e}$ as will be shown below.   
    
\begin{figure}[!htb]
    \centering
    \subfloat[][]{\includegraphics[scale=0.5]{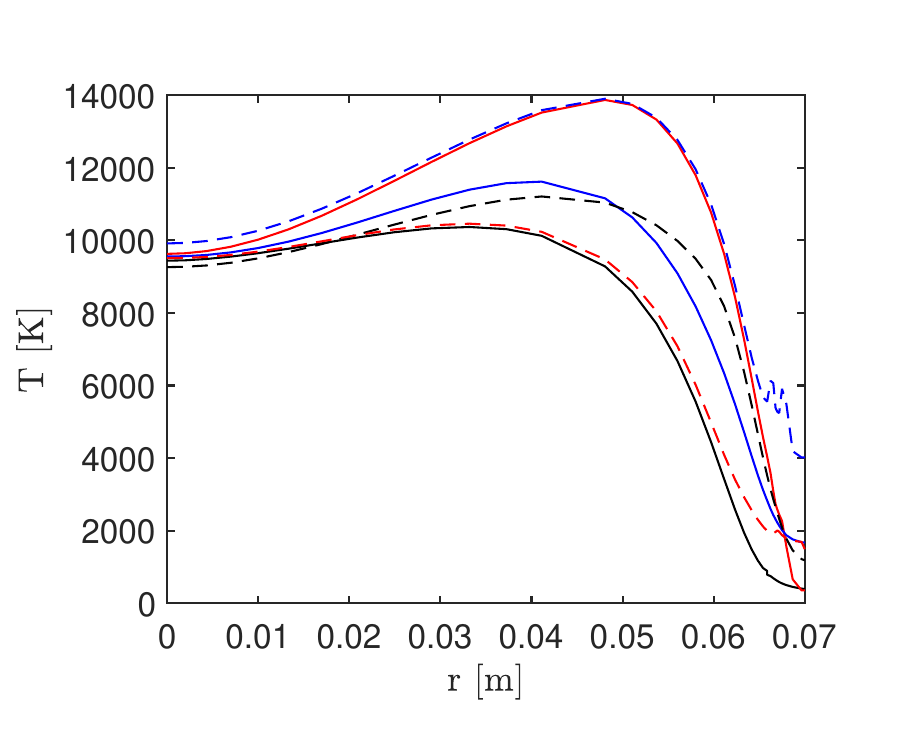}}
    \subfloat[][]{\includegraphics[scale=0.5]{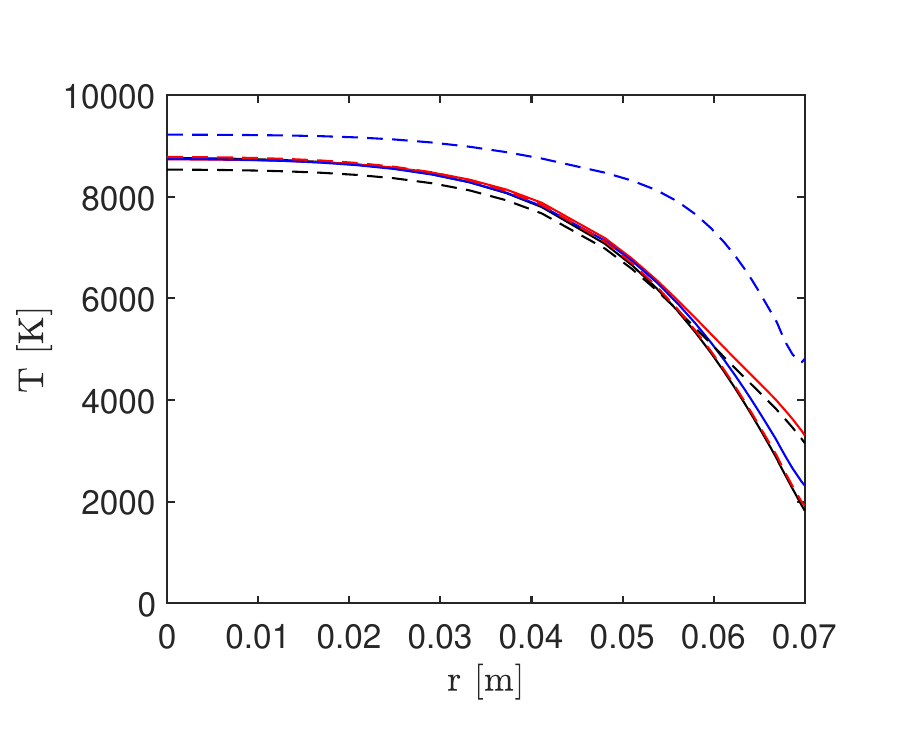}}
    \caption{Radial temperature profiles inside the ICP torch obtained from vibronic StS simulation: (a) x = \SI{0.3}{m} (mid-torch location) and (b) x = \SI{0.485}{m} (torch outlet). Operating conditions: \SI{1000}{Pa}, \SI{50}{kW} and \SI{6}{g/s}. Solid black line: T\textsubscript{h}, solid red line: T\textsubscript{e}, solid blue line: T\textsubscript{N}, dashed black line: T\textsubscript{N\textsubscript{2}}, dashed red line: T\textsubscript{N\textsuperscript{+}}, and dashed blue line: $\mathrm{T}_{\mathrm{N}_2^+}$.  }
    \label{fig:Tvibronic}
    \end{figure}

    Next, to see the extent of the non-Boltzmann effect on the internal temperatures, average vibronic temperatures of molecules ($\mathrm{N}_2$ and $\mathrm{N}_2^+$) and average electronic temperatures of atoms (N and $\mathrm{N}^+$) were reconstructed iteratively from the population distributions of the internal states as:

    \begin{equation}
        \frac{\sum_{i} n_i E_i}{\sum_i n_i}= \frac{\sum_i E_i \exp \left(\frac{-E_i}{k T}\right)}{\mathcal{Q}}
    \end{equation}
    where, the subscript $\mathrm{i}$ refers to the vibronic states in the case of molecules and electronic states in the case atoms, $\mathrm{E}_i$ is the energy of the $\mathrm{i}^{\mathrm{th}}$ energy level, $k$ is the Boltzmann constant, $\mathcal{Q}$ is the partition function and $T$ is the vibronic/electronic temperature that is to be solved for. The vibronic temperatures of $\mathrm{N}_2$ and $\mathrm{N}_2^+$ are denoted as $\mathrm{T}_{\mathrm{N}_2}$ and $\mathrm{T}_{\mathrm{N}_2^+}$, while the electronic temperatures of N and $\mathrm{N}^+$ are denoted as $\mathrm{T}_\mathrm{N}$ and $\mathrm{T}_{\mathrm{N}^+}$.  This definition of the internal temperature tries to find an average slope for the given population distribution which is close to the slope corresponding to the first few energy levels, since their populations are much higher. The populations of the high-lying states have small contributions in determining the internal temperatures unless they show a very large over/under population with respect to the Boltzmann distribution. \cref{fig:Tvibronic} shows the radial temperature profiles inside the torch at x = \SI{0.3}{m} (mid-torch location) and x = \SI{0.485}{m} (torch outlet). At the mid-torch location, which is in the coil region and hence has a large non-equilibrium effect, the internal temperatures of $\mathrm{N}_2$, N and $\mathrm{N}^+$  have large deviations from free-electron temperature $\mathrm{T}_e$ as a result of non-Boltzmann effect. At the axis, all the temperatures are very close due to thermal equilibrium as discussed previously. Away from the axis $\mathrm{T}_{\mathrm{N}_2}$, $\mathrm{T}_\mathrm{N}$ and $\mathrm{T}_{\mathrm{N}^+}$ are lower than $\mathrm{T}_e$ indicating under-population with respect to the Boltzmann distribution, while near the wall, the $\mathrm{T}_{\mathrm{N}_2}$, $\mathrm{T}_\mathrm{N}$ and $\mathrm{T}_{\mathrm{N}^+}$ are higher than $\mathrm{T}_e$ due to over-population. $\mathrm{T}_{\mathrm{N}_2^+}$ however, is very close to $\mathrm{T}_e$ everywhere due to populations being close to Boltzmann distribution at $\mathrm{T}_\mathrm{e}$, except close to the wall where it becomes much higher than $\mathrm{T}_e$ due to large over-population with respect to the Boltzmann distribution. At the torch outlet, as discussed previously, LTE conditions start prevailing and all the modes get into thermal equilibrium which causes all the temperature profiles to collapse together, except very close to the cold wall where still some non-equilibrium effect is observed. However, $\mathrm{T}_{\mathrm{N}_2^+}$ is much higher than other temperatures at all radial locations at the torch outlet as a result of over-population of the first few low-lying states of $\mathrm{N}_2^+$ with respect to Boltzmann distribution as shown in \cref{fig:pop_N2p}. The internal temperatures of various components computed from the state-to-state simulations provide a more accurate way to compare the temperatures reconstructed from the optical emission spectroscopy data during experiments, as compared to the temperatures predicted by a two-temperature or LTE simulation where the internal state populations are assumed to be in Boltzmann.

    \begin{figure}[!htb]
        \hspace*{-1cm}
        \includegraphics[scale=0.75]{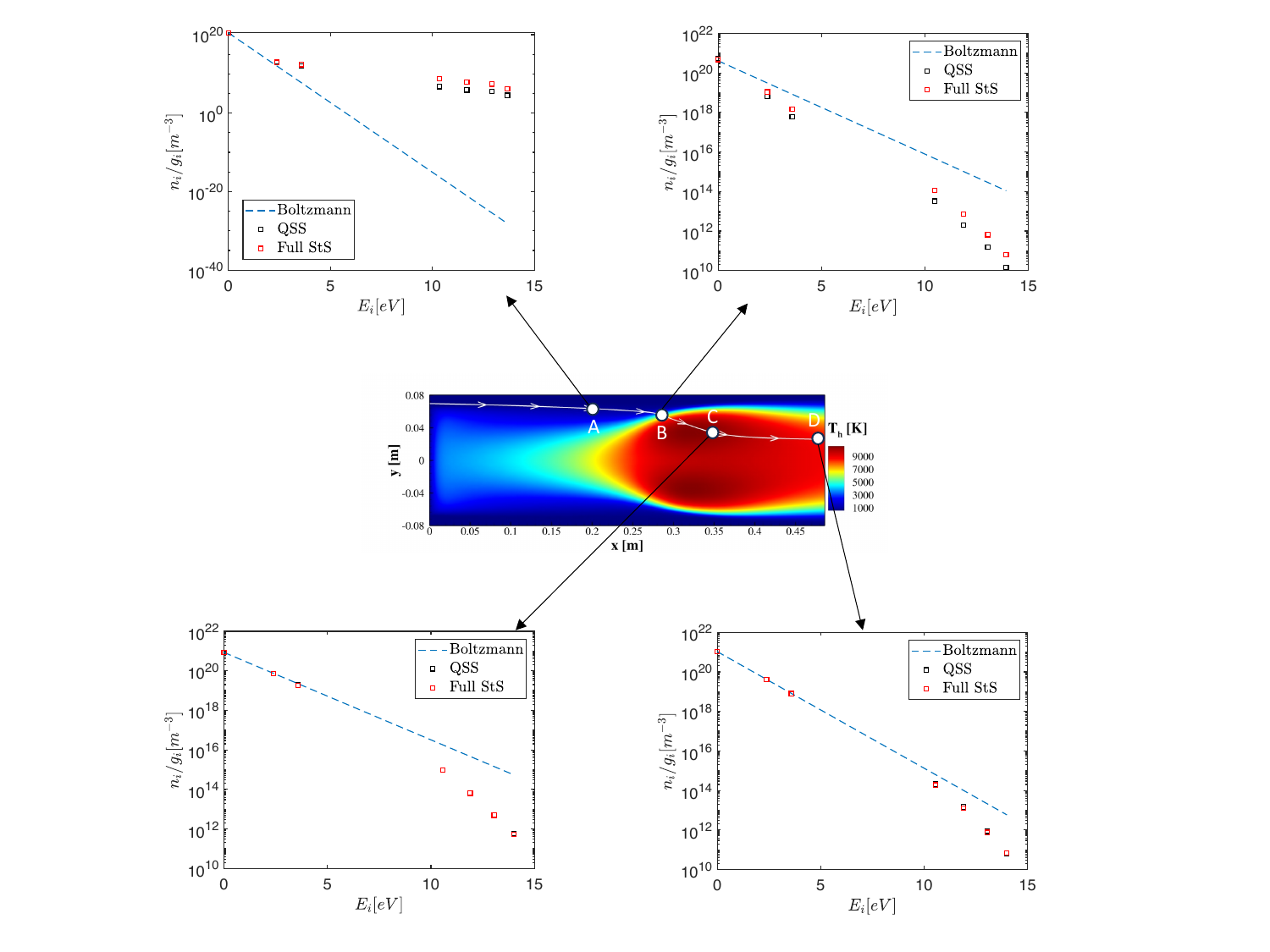}
        \caption{Population distribution of $\mathrm{N}$ atom at various locations along a streamline. Operating conditions: \SI{1000}{Pa}, \SI{50}{kW} and \SI{6}{g/s}.}
        \label{fig:qss_pop_N}
    \end{figure}

    \begin{figure}[!htb]
        \hspace*{-1cm}
        \includegraphics[scale=0.75]{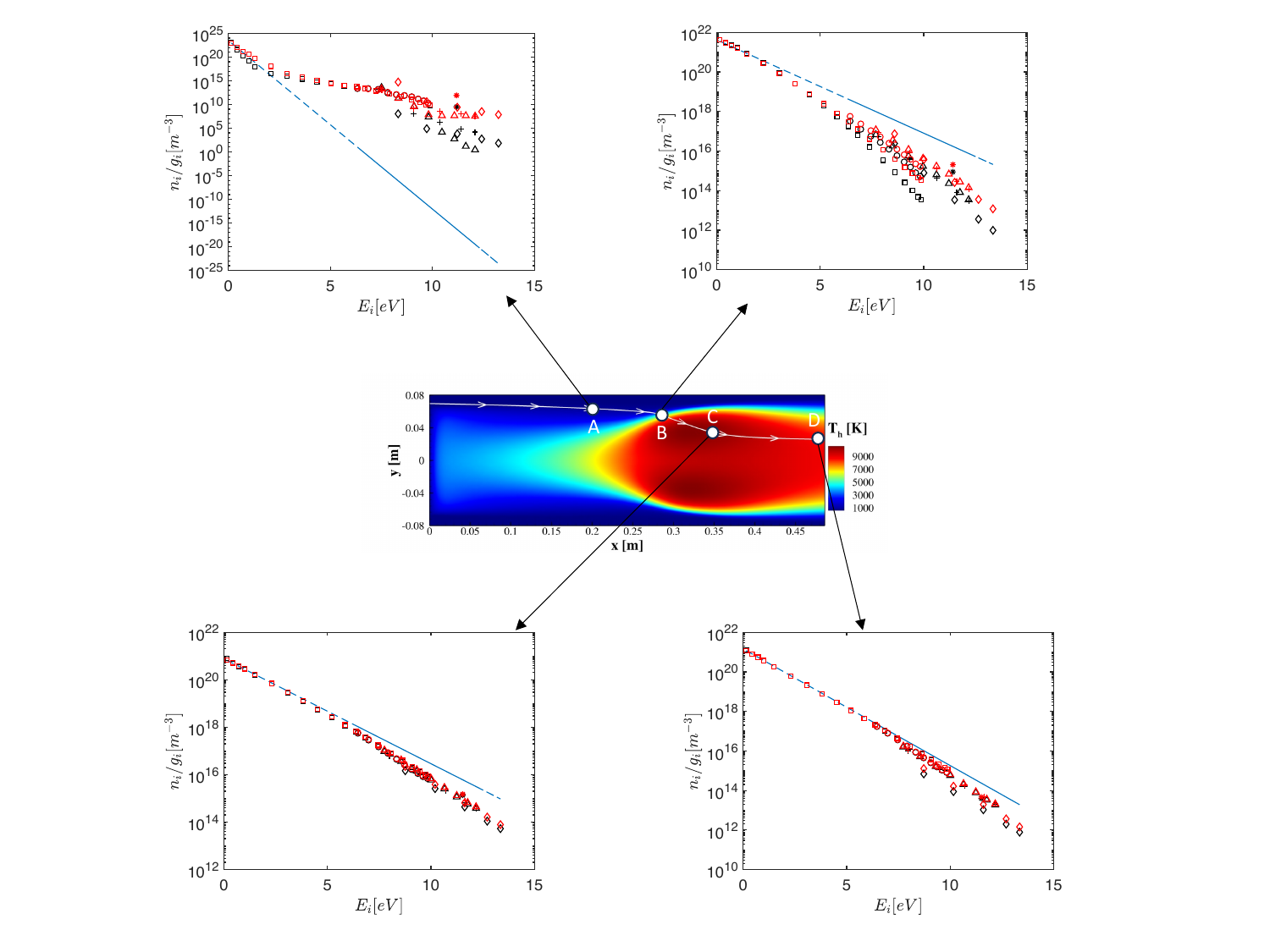}
        \caption{Population distribution of $\mathrm{N}_2$ at various locations along a streamline. Operating conditions: \SI{1000}{Pa}, \SI{50}{kW} and \SI{6}{g/s}. Dashed blue line: Boltzmann, black markers: QSS, and red markers: full StS. Vibrational states within different electronic levels are denoted by different symbols [square: $\mathrm{N}_2$(X), circle: $\mathrm{N}_2$(A), delta: $\mathrm{N}_2$(B), plus sign: $\mathrm{N}_2$(W), diamond: $\mathrm{N}_2$(B'), and asterisk: $\mathrm{N}_2$(C)].  } 
        \label{fig:qss_pop_N2}
    \end{figure}

    \subsection{Validity of the quasi-steady-state (QSS) assumption}\label{sec:qss}
     The quasi-steady-state assumption is often employed in the literature\cite{park1989nonequilibrium,johnston2006nonequilibrium,laux2012state} to study internal state populations in non-equilibrium conditions, where the hypothesis of Boltzmann distribution for the internal states does not hold good. In the case of the QSS condition, the rate of change of population of an excited state is much smaller than both the sum of all incoming rates
    and the sum of outgoing rates\cite{park1989nonequilibrium}. As a result, the time derivative term in the vibronic master equation vanishes and can be expressed as a set of non-linear algebraic equations that can be solved separately from the other flow governing equations \emph{i.e.,} the vibronic master equation can be decoupled from the flow governing equations. This drastically reduces the computational cost, since less expensive two-temperature models \cite{park1993review,park2001chemical,Dunn1973} can be used to compute the flowfield followed by computation of the internal state populations using QSS approximation. Hence, this section assesses the validity of the QSS assumption inside the ICP torch. 

    To compute the QSS populations, flow quantities (pressure, temperatures, and macroscopic compositions) from the StS flowfield are probed at the point of interest. Then the vibronic master equations (\cref{eq:master}) are solved for the given flow quantities:

    \begin{equation}
    \frac{d}{d t} y_i=\frac{M_i \dot{\omega}_i}{\rho} \label{eq:master}
    \end{equation}

    where the index $i$ denotes the vibronic levels. Conventionally, the temporal term is removed under the QSS assumption resulting in a set of algebraic equations to be solved for the populations of the internal states\cite{park1989nonequilibrium}. In this work, however, QSS populations are retrieved by numerically integrating the master equations in time for a 0D isochoric and isothermal reactor with fixed bath temperatures and identifying the plateau region in the internal temperatures evolution plot denoting the QSS region. The population of the ground states for each component (N, $\mathrm{N}_2$, \emph{etc.}) are updated at each time step to have the same total number density for each macroscopic component as in the StS flowfield.  \cref{fig:qss_pop_N,fig:qss_pop_N2} show the population distributions of N and $\mathrm{N}_2$ at various locations for a particle moving along a streamline starting from the inlet and ending at the outlet. At location A, which is cold, the QSS populations of the high-lying states for N and $\mathrm{N}_2$ differ from the full StS populations. However, at such low temperatures, populations of high-lying states are negligibly small and hence it does not make any sense to look at those. Location B is characterized by a sudden jump in plasma temperatures, along with the scarcity of electrons in the region. Because the collisional processes are responsible for the equilibration of internal energy states (in particular the processes involving electrons as the colliding partners), a lack of electrons along with a sudden change in plasma conditions can lead to the failure of QSS. This is what is observed looking at the population distributions of N and $\mathrm{N}_2$ at location B, although the differences are small \emph{i.e.,} less than 1 order of magnitude difference between the QSS and the full time-dependent StS populations. However, as the particle enters the hot plasma core, complete QSS conditions start prevailing as shown by the population distributions at locations C and D. Hence, this analysis shows that the QSS assumption indeed holds true in the plasma core. As a consequence, global rate coefficients can be derived from the vibronic StS rates under quasi-steady-state assumption along with other parameters needed for a conventional two-temperature model, which can drastically reduce the computational cost without compromising the results.

    \subsection{A High-pressure case}\label{sec:high_pr} 
\begin{figure}[!htb]
    \centering
    \subfloat[][]{\includegraphics[scale=0.15,valign=c]{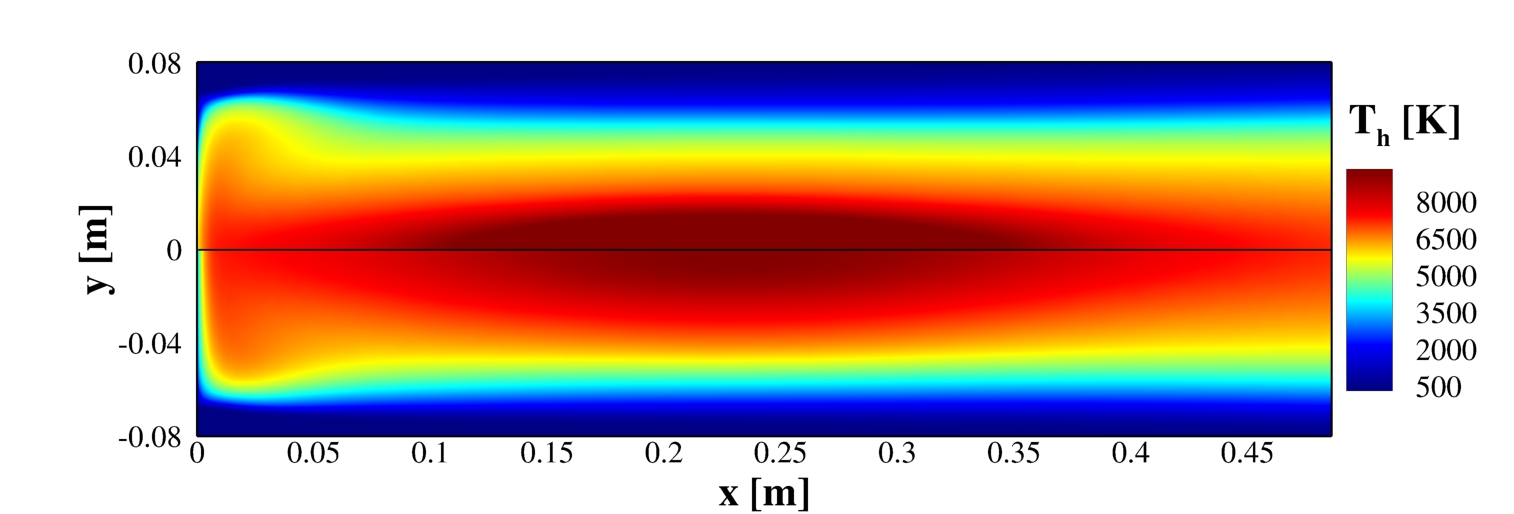}
    \vphantom{\includegraphics[scale=0.4,valign=c]{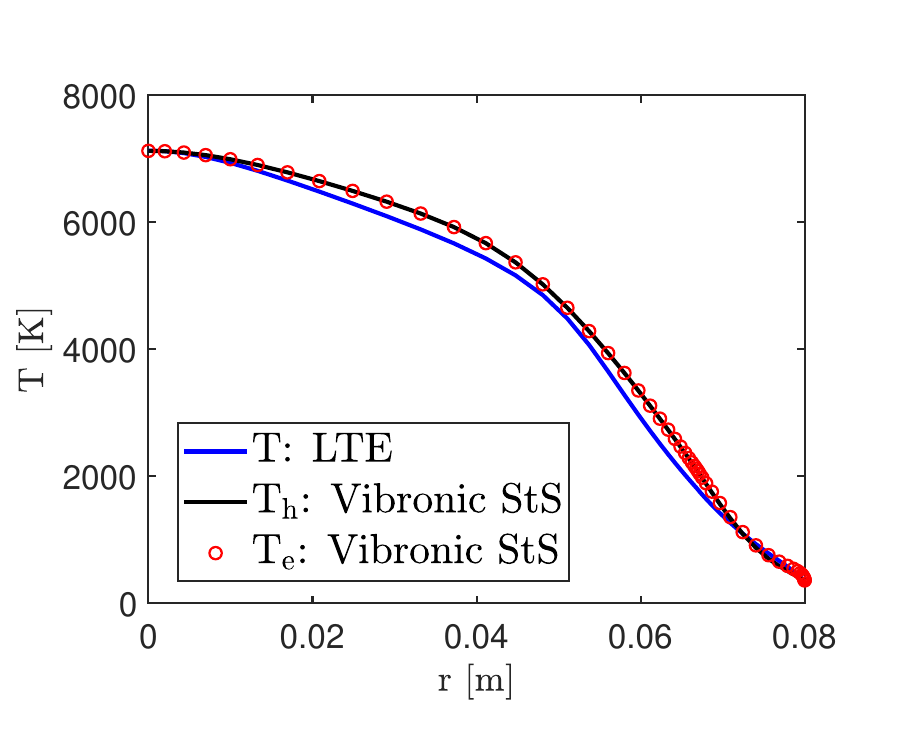}}}
    \subfloat[][]{\includegraphics[scale=0.5,valign=c]{lte_vs_sts_30kPa/T_profiles_x0p485_lte_vs_sts.pdf}}
    \caption{(a) Heavy-species temperature field (top: LTE, bottom: vibronic StS), and (b) radial temperature profiles at the torch outlet. Operating conditions: \SI{30}{kPa}, \SI{50}{kW} and \SI{6}{g/s}. }
    \label{fig:T_profiles_lte_vs_vibronic_30kpa}
    \end{figure}

This section presents the results for a very high-pressure ICP torch simulation, under which LTE assumption should hold and results obtained using a non-LTE model should collapse to LTE results. All the operating conditions remain the same as before except for the pressure which is set to \SI{30}{kPa}. \cref{fig:T_profiles_lte_vs_vibronic_30kpa} compares the radial temperature profiles at the torch outlet obtained from LTE and StS simulations. As expected, at high pressures, LTE conditions start to prevail and hence the temperature profiles obtained from vibronic StS simulations lie very close to the LTE temperature profile. This suggests that a computationally cheaper LTE model can be safely used at high-pressure cases ($>$ \SI{30}{kPa}). This also confirms the ability of the StS model to predict the LTE conditions and ensures the consistency of the model.

\section{Conclusions}\label{sec:conclusions}
This paper presented a vibrational-specific state-to-state model for nitrogen plasma implemented within a multi-physics computational framework for ICP simulations with an aim to study non-equilibrium effects in ICP facilities. The StS model was reduced by coarse-graining where an energy-based binning strategy was used, making it feasible to perform CFD calculations. The reduced StS model was found to capture the dynamics of the full StS model with excellent accuracy. State-to-state simulations were performed for the nitrogen ICP torch where the vibronic master equations are solved in a fully coupled manner with the flow governing equations. Simulations have been presented for the VKI ICP torch for low-pressure operating conditions at which significant non-equilibrium effects are observed:
\begin{itemize}
    \item LTE and StS (NLTE) plasma temperature fields show a large difference for the considered operating condition showing strong non-equilibrium conditions inside the torch. This highlights the importance of using an accurate NLTE model to simulate ICP facilities where strong heating from inductor coils leads to strong non-equilibrium conditions at low-pressure conditions.
    \item The population distributions of all the components show deviation from Boltzmann distribution in the colder recombination regions showing that there exists a strong recombination non-equilibrium in the torch. N atom shows large under-population in the hot plasma core region indicating ionization non-equilibrium in the plasma core. $\mathrm{N}_2$ shows mild deviation from Boltzmann distribution in the hot plasma core, while $\mathrm{N}_2^+$ and $\mathbf{N}^+$ follow Boltzmann distribution in the hot plasma region. 
    \item Further, the validity of the QSS assumption which is widely used in the aerospace community was assessed. Populations obtained from the state-to-state simulation were compared against the QSS population at a selected number of locations along a streamline starting from the cold inlet and exiting at the outlet through the hot plasma core. It was found that the QSS assumption holds well in the hot plasma core, except for a narrow shell surrounding the plasma core. This allows us to derive macroscopic global rates under QSS assumption from the state-to-state kinetics, to be used in comparatively cheaper two-temperature models for plasma simulations.
    \item Further, the StS model was used to simulate a very high pressure (\SI{30}{kPa}) case and it was found to give consistent results with that of LTE simulation. This shows that a computationally cheaper LTE model can be safely used for modeling high-pressure ICPs. This also shows that the StS model implementation is consistent and is able to predict the LTE conditions correctly. 
\end{itemize}

However, uncertainties remain in various state-specific rates and may need further calibration based on experimental data. Future work would focus on comparing the StS simulation results against experiments. Part II of this work presents the derivation of a two-temperature (2-T) model from the current StS model under QSS assumption, followed by a comparison of the StS results against the 2-T results for a wide range of operating conditions.
    
\section*{Acknowledgments}
    This work is funded by the Vannevar Bush Faculty Fellowship OUSD(RE) Grant No: N00014-21-1-295 with M. Panesi as the Principal Investigator. The authors would like to thank Prof. Bodony (UIUC) and Dr. Chiodi for the helpful discussions regarding the coupling of the solvers. 
    
\section*{Competing interest}
The authors declare no competing interests.

\section*{Appendix}
\appendix
    \begin{appendices}    
    \section{Additional details for the grouping strategy}\label{appendix:grouping}
    To reduce the number of energy levels in the vibronic StS model for nitrogen to a more manageable number for CFD calculations, 225 vibronic levels of $\mathrm{N}_2$ and 225 vibronic levels of $\mathrm{N}_2^+$ were grouped to 49 and 56 levels, respectively. For atoms, 131 electronic levels of $\mathrm{N}$ and 81 electronic levels of $\mathrm{N}^+$ were grouped into 7 levels each. The grouping was based on the energy of the levels and the mapping between the grouped levels and corresponding actual levels is given in \cref{tab:mapping_N2,tab:mapping_N2p,tab:mapping_atoms}.

    \begin{table}[H]
    \caption{\label{tab:mapping_N2} Mapping of the grouped levels for N\textsubscript{2}}
    \centering
    \begin{tabular}{lcccc}
    \hline\hline
    Grouped level, $i^{\prime}$ & Actual levels, $i$ & Grouped level, $i^{\prime}$ & Actual levels, $i$  \\
    \hline
    N\textsubscript{2}X(1) & N\textsubscript{2}X(1) & N\textsubscript{2}A(6) & N\textsubscript{2}A(16-18) \\
    N\textsubscript{2}X(2) & N\textsubscript{2}X(2) & N\textsubscript{2}A(7) & N\textsubscript{2}A(19-21)\\
    N\textsubscript{2}X(3) & N\textsubscript{2}X(3) & N\textsubscript{2}A(8) & N\textsubscript{2}A(22-24)\\
    N\textsubscript{2}X(4) & N\textsubscript{2}X(4) & N\textsubscript{2}A(9) & N\textsubscript{2}A(25-27)\\
    N\textsubscript{2}X(5) & N\textsubscript{2}X(5-7) & N\textsubscript{2}A(10) & N\textsubscript{2}A(28-32)\\
    N\textsubscript{2}X(6) & N\textsubscript{2}X(8-10) & N\textsubscript{2}B(1) & N\textsubscript{2}B(1-4)\\
    N\textsubscript{2}X(7) & N\textsubscript{2}X(11-13) & N\textsubscript{2}B(2) & N\textsubscript{2}B(5-8)\\
    N\textsubscript{2}X(8) & N\textsubscript{2}X(14-16) & N\textsubscript{2}B(3) & N\textsubscript{2}B(9-12)\\
    N\textsubscript{2}X(9) & N\textsubscript{2}X(17-19) & N\textsubscript{2}B(4) & N\textsubscript{2}B(13-16)\\
    N\textsubscript{2}X(10) & N\textsubscript{2}X(20-22) & N\textsubscript{2}B(5) & N\textsubscript{2}B(17-20)\\
    N\textsubscript{2}X(11) & N\textsubscript{2}X(23-25) & N\textsubscript{2}B(6) & N\textsubscript{2}B(21-24)\\
    N\textsubscript{2}X(12) & N\textsubscript{2}X(26-27) & N\textsubscript{2}B(7) & N\textsubscript{2}B(25-28)\\
    N\textsubscript{2}X(13) & N\textsubscript{2}X(28-30) & N\textsubscript{2}B(8) & N\textsubscript{2}B(29-33)\\
    N\textsubscript{2}X(14) & N\textsubscript{2}X(31-34) & N\textsubscript{2}W(1) & N\textsubscript{2}W(1-9)\\
    N\textsubscript{2}X(15) & N\textsubscript{2}X(35-38) & N\textsubscript{2}W(2) & N\textsubscript{2}W(10-18)\\
    N\textsubscript{2}X(16) & N\textsubscript{2}X(39-42) & N\textsubscript{2}W(3) & N\textsubscript{2}W(19-27)\\
    N\textsubscript{2}X(17) & N\textsubscript{2}X(43-46) & N\textsubscript{2}W(4) & N\textsubscript{2}W(28-36)\\
    N\textsubscript{2}X(18) & N\textsubscript{2}X(47-50) & N\textsubscript{2}W(5) & N\textsubscript{2}W(37-45)\\
    N\textsubscript{2}X(19) & N\textsubscript{2}X(51-56) & N\textsubscript{2}B\textsuperscript{'}(1) & N\textsubscript{2}B\textsuperscript{'}(1-8)\\
    N\textsubscript{2}X(20) & N\textsubscript{2}X(57-62) & N\textsubscript{2}B\textsuperscript{'}(2) & N\textsubscript{2}B\textsuperscript{'}(9-18)\\
    N\textsubscript{2}A(1) & N\textsubscript{2}A(1-3) & N\textsubscript{2}B\textsuperscript{'}(3) & N\textsubscript{2}B\textsuperscript{'}(19-28)\\
    N\textsubscript{2}A(2) & N\textsubscript{2}A(4-6) & N\textsubscript{2}B\textsuperscript{'}(4) & N\textsubscript{2}B\textsuperscript{'}(29-38)\\
    N\textsubscript{2}A(3) & N\textsubscript{2}A(7-9) & N\textsubscript{2}B\textsuperscript{'}(5) & N\textsubscript{2}B\textsuperscript{'}(39-48)\\
    N\textsubscript{2}A(4) & N\textsubscript{2}A(10-12) & N\textsubscript{2}C(1) & N\textsubscript{2}C(1-5)\\
    N\textsubscript{2}A(5) & N\textsubscript{2}A(13-15) & &\\
    
    \hline\hline
    
    \end{tabular}
    \end{table}

\newpage

    \begin{table}[H]
    \caption{\label{tab:mapping_N2p} Mapping of the grouped levels for $\mathrm{N}_2^+$}
    \centering
    \begin{tabular}{lcccc}
    \hline\hline
    Grouped level, $i^{\prime}$ & Actual levels, $i$ & Grouped level, $i^{\prime}$ & Actual levels, $i$  \\
    \hline
    $\mathrm{N}_2^+$X(1) & $\mathrm{N}_2^+$X(1-3) & $\mathrm{N}_2^+$A(7) & $\mathrm{N}_2^+$A(25-28)\\
    $\mathrm{N}_2^+$X(2) & $\mathrm{N}_2^+$X(4-6) & $\mathrm{N}_2^+$A(8) & $\mathrm{N}_2^+$A(29-32)\\
    $\mathrm{N}_2^+$X(3) & $\mathrm{N}_2^+$X(7-9) & $\mathrm{N}_2^+$A(9) & $\mathrm{N}_2^+$A(33-36)\\
    $\mathrm{N}_2^+$X(4) & $\mathrm{N}_2^+$X(10-12) & $\mathrm{N}_2^+$A(10) & $\mathrm{N}_2^+$A(37-40)\\
    $\mathrm{N}_2^+$X(5) & $\mathrm{N}_2^+$X(13-15) & $\mathrm{N}_2^+$A(11) & $\mathrm{N}_2^+$A(41-44)\\
    $\mathrm{N}_2^+$X(6) & $\mathrm{N}_2^+$X(16-18) & $\mathrm{N}_2^+$A(12) & $\mathrm{N}_2^+$A(45-48)\\
    $\mathrm{N}_2^+$X(7) & $\mathrm{N}_2^+$X(19-21) & $\mathrm{N}_2^+$A(13) & $\mathrm{N}_2^+$A(49-52)\\
    $\mathrm{N}_2^+$X(8) & $\mathrm{N}_2^+$X(22-24) & $\mathrm{N}_2^+$A(14) & $\mathrm{N}_2^+$A(53-56)\\
    $\mathrm{N}_2^+$X(9) & $\mathrm{N}_2^+$X(25-27) & $\mathrm{N}_2^+$A(15) & $\mathrm{N}_2^+$A(57-60)\\
    $\mathrm{N}_2^+$X(10) & $\mathrm{N}_2^+$X(28-30) & $\mathrm{N}_2^+$A(16) & $\mathrm{N}_2^+$A(61-67)\\
    $\mathrm{N}_2^+$X(11) & $\mathrm{N}_2^+$X(31-33) & $\mathrm{N}_2^+$B(1) & $\mathrm{N}_2^+$B(1-4)\\
    $\mathrm{N}_2^+$X(12) & $\mathrm{N}_2^+$X(34-36) & $\mathrm{N}_2^+$B(2) & $\mathrm{N}_2^+$B(5-8) \\
    $\mathrm{N}_2^+$X(13) & $\mathrm{N}_2^+$X(37-39) & $\mathrm{N}_2^+$B(3) & $\mathrm{N}_2^+$B(9-12)\\
    $\mathrm{N}_2^+$X(14) & $\mathrm{N}_2^+$X(40-42) & $\mathrm{N}_2^+$B(4) & $\mathrm{N}_2^+$B(13-16)\\
    $\mathrm{N}_2^+$X(15) & $\mathrm{N}_2^+$X(43-45) & $\mathrm{N}_2^+$B(5) & $\mathrm{N}_2^+$B(17-20)\\
    $\mathrm{N}_2^+$X(16) & $\mathrm{N}_2^+$X(46-48) & $\mathrm{N}_2^+$B(6) & $\mathrm{N}_2^+$B(21-24)\\
    $\mathrm{N}_2^+$X(17) & $\mathrm{N}_2^+$X(49-51) & $\mathrm{N}_2^+$B(7) & $\mathrm{N}_2^+$B(25-28)\\
    $\mathrm{N}_2^+$X(18) & $\mathrm{N}_2^+$X(52-54) & $\mathrm{N}_2^+$B(8) & $\mathrm{N}_2^+$B(29-32)\\
    $\mathrm{N}_2^+$X(19) & $\mathrm{N}_2^+$X(55-57) & $\mathrm{N}_2^+$B(9) & $\mathrm{N}_2^+$B(33-36)\\
    $\mathrm{N}_2^+$X(20) & $\mathrm{N}_2^+$X(58-60) & $\mathrm{N}_2^+$B(10) & $\mathrm{N}_2^+$B(37-39)\\
    $\mathrm{N}_2^+$X(21) & $\mathrm{N}_2^+$X(61-63) & $\mathrm{N}_2^+$D(1) & $\mathrm{N}_2^+$D(1-7)\\
    $\mathrm{N}_2^+$X(22) & $\mathrm{N}_2^+$X(64-66) & $\mathrm{N}_2^+$D(2) & $\mathrm{N}_2^+$D(8-15)\\
    $\mathrm{N}_2^+$A(1) & $\mathrm{N}_2^+$A(1-4) & $\mathrm{N}_2^+$D(3) & $\mathrm{N}_2^+$D(16-23)\\
    $\mathrm{N}_2^+$A(2) & $\mathrm{N}_2^+$A(5-8) & $\mathrm{N}_2^+$D(4) & $\mathrm{N}_2^+$D(24-31)\\
    $\mathrm{N}_2^+$A(3) & $\mathrm{N}_2^+$A(9-12) & $\mathrm{N}_2^+$D(5) & $\mathrm{N}_2^+$D(32-39)\\
    $\mathrm{N}_2^+$A(4) & $\mathrm{N}_2^+$A(13-16) & $\mathrm{N}_2^+$C(1) & $\mathrm{N}_2^+$C(1-4)\\
    $\mathrm{N}_2^+$A(5) & $\mathrm{N}_2^+$A(17-20) & $\mathrm{N}_2^+$C(2) & $\mathrm{N}_2^+$C(5-8)\\
    $\mathrm{N}_2^+$A(6) & $\mathrm{N}_2^+$A(21-24) & $\mathrm{N}_2^+$C(3) & $\mathrm{N}_2^+$C(9-14)\\
    \hline\hline
    
    \end{tabular}
    \end{table}
    
    \begin{table}[H]
    \caption{\label{tab:mapping_atoms} Mapping of the grouped levels for N and N\textsuperscript{+}}
    \centering
    \begin{tabular}{lcccc}
    \hline\hline
    Grouped level, $i^{\prime}$ & Actual levels, $i$ & Grouped level, $i^{\prime}$ & Actual levels, $i$  \\
    \hline
    N(1) & N(1) & N\textsuperscript{+}(1) & N\textsuperscript{+}(1) \\
    N(2) & N(2) & N\textsuperscript{+}(2) & N\textsuperscript{+}(2) \\
    N(3) & N(3) & N\textsuperscript{+}(3) & N\textsuperscript{+}(3) \\
    N(4) & N(4-6) & N\textsuperscript{+}(4) & N\textsuperscript{+}(4) \\
    N(5) & N(7-13) & N\textsuperscript{+}(5) & N\textsuperscript{+}(5-6) \\
    N(6) & N(14-27) & N\textsuperscript{+}(6) & N\textsuperscript{+}(7-17) \\
    N(7) & N(28-131) & N\textsuperscript{+}(7) & N\textsuperscript{+}(18-81) \\
    
    \hline\hline
    
    \end{tabular}
    \end{table}
    
    \end{appendices}

\section*{References}
\bibliographystyle{aiaa}
\bibliography{bibliography}

\end{document}